\documentclass[10pt,final,journal,twocolumn,letterpaper]{IEEEtran}

\IEEEoverridecommandlockouts
\usepackage{graphicx,dblfloatfix}
\usepackage{amsmath}
\usepackage{amssymb}
\usepackage[latin1]{inputenc}
\usepackage{times}
\usepackage{cite}
\usepackage{amsfonts}
\usepackage{color}
\usepackage{colortbl} 
\usepackage{multirow}
\usepackage[table]{xcolor}
\usepackage{longtable}
\usepackage{dropping}
\usepackage{caption}
\usepackage{subcaption}
\makeatletter
\let\saved@longtable\longtable
\long\def\foo#1\LT@err#2#3#4!!{\def\longtable{#1#4}}
\expandafter\foo\longtable!!

\long\def\foo#1\@outputpage#2\@outputpage#3!!{%
\def\LT@output{#1\@opcol#2\@opcol#3}}
\expandafter\foo\LT@output!!
\makeatother
\newcommand{\ket}[1]{| #1 \rangle}

\newcommand{\eqr}[1]{Eq.~(\ref{#1})}
\newcommand{\fref}[1]{Fig.~\ref{#1}}
\newcommand{\tref}[1]{Table~\ref{#1}}
\newcommand{\etal}{\textit{et~al.}}
\newcommand{\figures}{fig/}
\newcommand{\food}{\hspace{-2.3pt}$-$ \hspace{5pt}}
\newcolumntype{C}[1]{>{\centering \arraybackslash}p{#1}}
\newcolumntype{R}[1]{>{\raggedleft\let\newline\\\arraybackslash\hspace{0pt}}m{#1}}

\newsavebox\ltmcbox

\newcommand{\tabitem}{~~\llap{\textbullet}~~}        
\newcommand \flabel

\renewcommand{\flabel}[1]{\ifdefined\flabels\ #1\fi}        

\usepackage{acronym}

\ifCLASSINFOpdf
\else
\fi

\begin{document}
%
\title{Polar Codes and Their Quantum-Domain Counterparts}

\author{Zunaira Babar, Zeynep B. Kaykac Egilmez, Luping Xiang, Daryus Chandra, Robert G.~Maunder, Soon Xin Ng 
and Lajos Hanzo 
\thanks{Z.~Babar, Z.~B.~Kaykac, L.~Xiang, D.~Chandra, R.~G.~Maunder, S.~X.~Ng, and L.~Hanzo are with the School of Electronics and Computer Science, University of Southampton, 
SO17 1BJ, United Kingdom. Email: \{zb2g10,zbk1y15,dc2n14,,rm,sxn,lh\}@ecs.soton.ac.uk.}%
\thanks{ L. Hanzo would like to acknowledge the financial support of the Engineering and Physical Sciences Research Council projects EP/Noo4558/1, EP/PO34284/1, COALESCE, of the Royal Society's Global Challenges Research Fund Grant as well as of the European Research Council's Advanced Fellow Grant QuantCom.}}


%


\maketitle

\begin{abstract} 
Arikan's polar codes are capable of achieving the Shannon's capacity
at a low encoding and decoding complexity, while inherently supporting
rate adaptation. By virtue of these attractive features, polar codes
have provided fierce competition to both the turbo as well as the Low
Density Parity Check (LDPC) codes, making its way into the $5$G New
Radio (NR). Realizing the significance of polar codes, in this paper
we provide a comprehensive survey of polar codes, highlighting the
major milestones achieved in the last decade.  Furthermore, we also
provide tutorial insights into the polar encoder, decoders as well as
the code construction methods.  We also extend our discussions to
quantum polar codes with an emphasis on the underlying
quantum-to-classical isomorphism and the syndrome-based quantum polar
codes.
\end{abstract}

%
\IEEEpeerreviewmaketitle

\section*{Acronyms}
\begin{acronym}[ABCDEFGH] 
\acro{ASIC}{Application Specific Integrated Circuit}
\acro{AWGN}{Additive White Gaussian Noise}
\acro{BCH}{Bose-Chaudhuri-Hocquenghem}
\acro{B-DMC}{Binary-Input Discrete Memoryless Channel}
\acro{BEC}{Binary Erasure Channel}
\acro{BER}{Bit Error Rate}
\acro{BICM}{Bit-Interleaved Coded Modulation}
\acro{BICM-ID} {BICM with Iterative Decoding}
\acro{B-DMC}{Binary-input Discrete Memoryless Channel}
\acro{BEC}{Binary Erasure Channel}
\acro{BLER}{BLock Error Ratio}
\acro{B-MC}{Binary-input Memoryless Channel}
\acro{BP}{Blief Propagation}
\acro{BSC}{Binary Symmetric Channel}
\acro{BP}{Belief Propagation}
\acro{CA-SCL}{Cyclic Redundancy Check-Aided SCL}
\acro{CA-SCS}{Cyclic Redundancy Check-Aided SCS}
\acro{CSS}{Calderbank-Shor-Steane}
\acro{CNOT}{Controlled NOT}
\acro{Cz}{Controlled-$\mathbf{Z}$}
\acro{CRC}{Cyclic Redundancy Check}
\acro{DE}{Density Evolution}
\acro{DMC}{Discrete Memoryless Channel}
\acro{eMBB}{enhanced Mobile BroadBand}
\acro{FEC}{Forward Error Correction}
\acro{FPGA}{Field Programmable Gate Array}
\acro{GA}{Gaussian Approximation}
\acro{HARQ}{Hybrid Automatic Repeat Request}
\acro{IRCC}{IRregular Convolutional Code}
\acro{LDPC}{Low Density Parity Check}
\acro{LLR}{Log-Likelihood Ratio}
\acro{LR}{Likelihood Ratio}
\acro{MERA}{Multi-scale Entanglement Renormalization Ansatz}
\acro{ML}{Maximum Likelihood}
\acro{ML-SSC}{Maximum Likelihood Simplified Successive Cancellation}
\acro{mMTC}{massive Machine Type Communication}
\acro{NR}{New Radio}
\acro{PBCH}{Physical Broadcast CHannel}
\acro{PC}{Parity Check}
\acro{QECC}{Quantum Error Correction Code}
\acro{QPSK}{Quadrature Phase Shift Keying}
\acro{QSCD}{Quantum Successive Cancellation Decoder}
\acro{RM}{Reed-Muller}
\acro{RRNS}{Redundant Residue Number System}
\acro{RS}{Reed-Solomon}
\acro{SC}{Successive Cancellation}
\acro{SCAN}{Soft CANcellation}
\acro{SCH}{Successive Cancellation Hybrid}
\acro{SCL}{Successive Cancellation List}
\acro{SCS}{Successive Cancellation Stack}
\acro{SNR}{Signal-to-Noise Ratio}
\acro{SSC}{Simplified Successive Cancellation}
\acro{SSCL}{Simplified Successive Cancellation List}
\acro{TCM}{Trellis Coded Modulation}
\acro{3GPP}{Third Generation Partnership Project}
\acro{TTCM}{Turbo Trellis Coded Modulation}
\acro{URC}{Unity Rate Code}
\acro{URLLC}{Ultra-Reliable Low-Latency Communication}
\acro{XOR}{eXclusive-OR}
\end{acronym}
\setcounter{figure}{1}
\begin{figure*}[!b]
\centering
\includegraphics[width=0.75\linewidth]{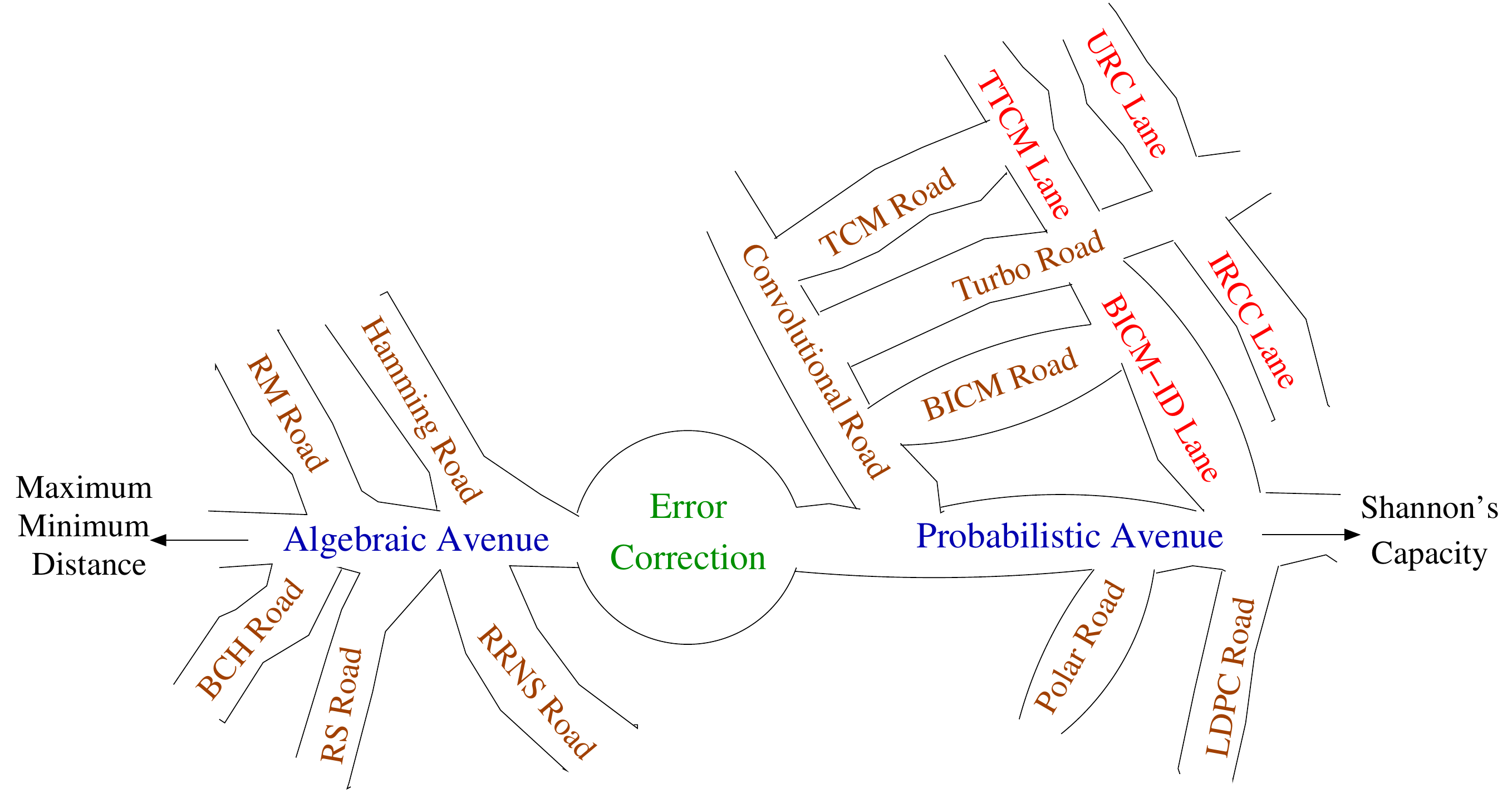}
\caption{Road map portraying the evolution of classical channel coding theory. The algebraic avenue aims at maximizing the minimum distance,
while the probabilistic avenue leads to capacity approaching designs.}
\label{fig:CodeDesignRM}
\end{figure*}
\section{Introduction} \label{sec:intro}

\vspace{0.5cm}

\textbf{\textit{Information is the resolution of uncertainty,}}

\textit{Claude Shannon.}

\vspace{0.5cm}
The inception of classical coding theory dates back to 1948~\cite{shannon1948mathematical}, when Claude Shannon introduced the 
notion of `channel capacity'. 
Explicitly, Shannon predicted in his seminal paper~\cite{shannon1948mathematical} that
virtually error-free transmission over noisy channels can be achieved by invoking error correction codes having coding rate $R$
less than the channel capacity $C$ and having an infinitely long codeword length. The capacity of
an \ac{AWGN} channel having the bandwidth $B$ (Hz) and the noise power spectral density $N_0/2$ (Watts/Hz) per dimension
is quantified by the Shannon-Hartley theorem as follows:
\begin{equation}
 C = B \log_2 \left( 1 + \frac{S}{N_0B} \right),
\end{equation}
when the average transmitted power is $S$ Watts. 
Hence, the maximum permissible coding rate of an error correction code is limited by
the \ac{SNR} ($\frac{S}{N_0B}$) and the bandwidth ($B$) under the idealized support for infinite implementation complexity
and transmission delay. Similarly, the capacities of a \ac{BSC} or a \ac{BEC} are specified by their respective channel characteristics, 
i.e. the cross-over probability of the \ac{BSC} and the erasure probability of the \ac{BEC}, again assuming infinite processing and time resources.
However, practical systems can neither afford an infinite implementation complexity nor an infinite
transmission delay. So, we need optimized codes, which perform close to the Shannon's capacity limit, while guaranteeing the desired performance
metrics, as illustrated in \fref{fig:CodeDesign}. 
\setcounter{figure}{0}
\begin{figure}[t]
\centering
\includegraphics[width=\linewidth]{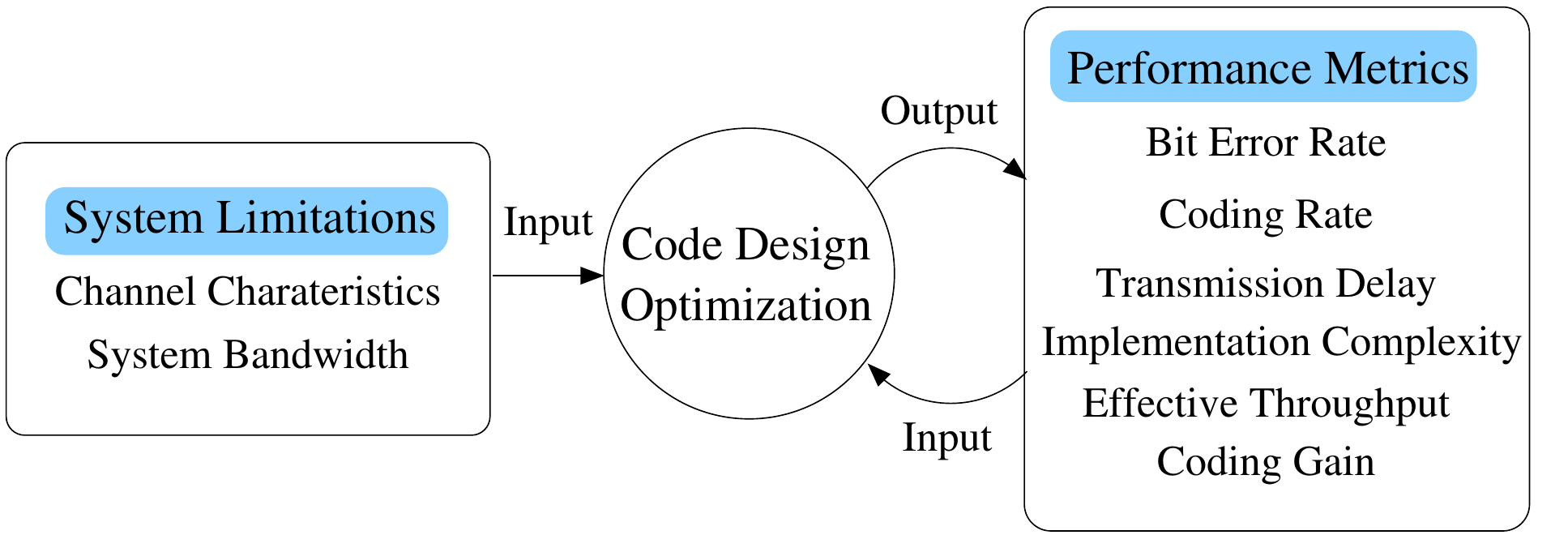}
\caption{Factors driving the code design optimization.}
\label{fig:CodeDesign}
\end{figure}

Shannon quantified the capacity limit and proved the existence of `capacity-achieving' codes based on the random-coding argument.
However, he did not give any recipes for constructing such capacity-achieving codes. Over the last 
seven decades, researchers have endeavored to design optimum codes, which are capable of operating close to the capacity limit, while also offering
the desired performance metrics of \fref{fig:CodeDesign}. Broadly speaking, this quest has taken two avenues: the algebraic coding avenue and 
the probabilistic coding avenue, as portrayed in the stylized road map of \fref{fig:CodeDesignRM}. 
Algebraic coding was the main research avenue for the first few decades.
The aim of this coding paradigm is to design powerful codes by exploiting finite-field arithmetic to maximize the minimum 
Hamming distance\footnote{The Hamming distance between two vectors is equal to the number of positions at which the corresponding elements
(bits or symbols) differ.} between the codewords 
for a given coding rate, or more specifically for the given information word length
$k$ and codeword length $n$. This has given rise to a range of popular coding families, which includes for example
Hamming codes~\cite{hamming1950error}, \ac{RM} codes~\cite{1057465,muller1954application}, 
\ac{BCH} codes~\cite{Hocquenghem_1959,bose1960class}, \ac{RS} codes~\cite{reed1960polynomial}
and \ac{RRNS} codes~\cite{watson1966self,szabo1967residue}. Unfortunately, algebraic avenue does not promise a 
capacity achieving design.
Nonetheless, algebraic codes have found their way into practical applications by virtue of their strong
error correction capabilities (or equivalently low \ac{BER} floors). Explicitly, these codes are useful when the received information is in the 
form of hard decisions. For example, \ac{RS} codes are used in magnetic tape and disk storage as 
well as in several standardized systems, such as the deep-space coding standard~\cite{CCSDS}, where they typically constitute an outer
layer\footnote{Outer layer is with respect to the channel. The inner layer is closer to the channel.} of error correction (known as the outer code) to reduce the BER floor resulting from the inner layer of error correction (called the inner code).

In contrast to the algebraic coding avenue, probabilistic coding avenue has paved
the way to capacity. Explicitly, probabilistic coding avenue of \fref{fig:CodeDesignRM} is inspired by Shannon's random coding philosophy and strives to 
achieve a reasonable trade-off between the performance and the complexity. This design avenue has led to the construction of 
convolutional codes~\cite{elias1955coding}, \ac{LDPC} codes~\cite{gallager62,mackay1995good,mackay1996near}, 
turbo codes~\cite{turbo93, berrou1996near} as well as the polar codes~\cite{5075875}. The probabilistic coding paradigm also includes various `turbo-like'
iterative coding schemes, for example turbo \ac{BCH} codes~\cite{hagenauer1996iterative}, turbo Hamming codes~\cite{nickl1997approaching}, and
\ac{URC}-assisted and \ac{IRCC}-assisted concatenated schemes of~\cite{div:tcm-urc:ip,ircc:tuchler_hagenauer}, as well as the
coded modulation schemes, including \ac{TCM}~\cite{TCM82,1093542,1093541}, 
\ac{BICM}~\cite{zevahi19928,caire1998bit}, \ac{BICM-ID}~\cite{li1997bit}, and \ac{TTCM}~\cite{TTCM98}.
In particular, the turbo and LDPC codes made it possible to operate arbitrarily close to the Shannon limit, while the polar codes finally managed 
to provably achieve the capacity, albeit at infinitely long codeword lengths. Despite being a relatively immature coding scheme, polar coding  
has proved to be a fierce competitor of turbo and LDPC codes, both of which have been ruling for over two decades now. 
Polar codes have already found their way into the 5G \ac{NR} for the control channels of the \ac{eMBB} and the \ac{URLLC} use-cases
as well as for the \ac{PBCH}. Polar codes have also been identified as potential candidates for the data and control channels of the
\ac{mMTC} use-cases.

Polar codes emerged at times when the apprehension that `coding is dead' started looming again\footnote{The notion that `coding is dead'
first officially surfaced in the IEEE Communication Theory Workshop held in St.~Petersburg, Florida, in April 1971, where a group of coding theorists
concluded that there was nothing more to do in coding theory. This workshop became famous as `coding is dead' workshop.}.
Hence, the discovery of polar codes re-energized the coding community and equipped them with a radically different approach for achieving the Shannon's capacity.
In addition to its influence on the classical coding theory, polar codes have also attracted considerable attention within the quantum research
community. \textit{Motivated by the growing interest in polar codes, in this paper we provide a 
comprehensive survey on the classical as well as quantum polar codes, taking the readers through the major
milestones achieved 
and providing a slow-paced tutorial on the related encoding and decoding algorithms.} This tutorial paper  
sets the necessary background for understanding the operation of the \ac{3GPP} $5$G~\ac{NR} polar codes, which have been surveyed 
in~\cite{bioglio2018design, Zeynep2019}. 
It is pertinent to mention here that, to the best of authors' knowledge, only two survey papers~\cite{6852102, 8616900} 
exist on polar codes at the time of writing. In~\cite{6852102}, the author's have provided a succinct overview of the fundamental concepts 
pertaining to polar codes, including the encoding, decoding and construction methods,
while~\cite{8616900} focuses on the \ac{ASIC} implementation of polar decoders. By contrast, this paper has a broader scope, since we provide
in-depth tutorial insights (with explicit examples) as well as a comprehensive survey on polar encoders, decoders as well as polar code construction methods.

\fref{fig:PaperStruct} provides the overview of the paper. We begin our discourse in Section~\ref{sec:ch-polarization},
where we discuss Arikan's channel polarization philosophy, which is the key to Shannon's capacity. We then survey the polar encoding and
decoding algorithms in Section~\ref{sec:encoder} and Section~\ref{sec:decoder}, respectively, with detailed tutorial insights. Continuing further
our discussions, we present the polar code design principles and guidelines in Section~\ref{sec:polarcodedesign}. In Section~\ref{sec:isomorphism},
we detail the transition from the classical to the quantum channel polarization by identifying the underlying isomorphism. Based on the 
isomorphism of Section~\ref{sec:isomorphism}, we proceed to quantum polar codes in Section~\ref{sec:Qpolarcode}. Finally, we conclude our discussions
in Section~\ref{sec:conclusion}.
\setcounter{figure}{2}
\begin{figure}[tb]
\centering
\includegraphics[width=0.7\linewidth]{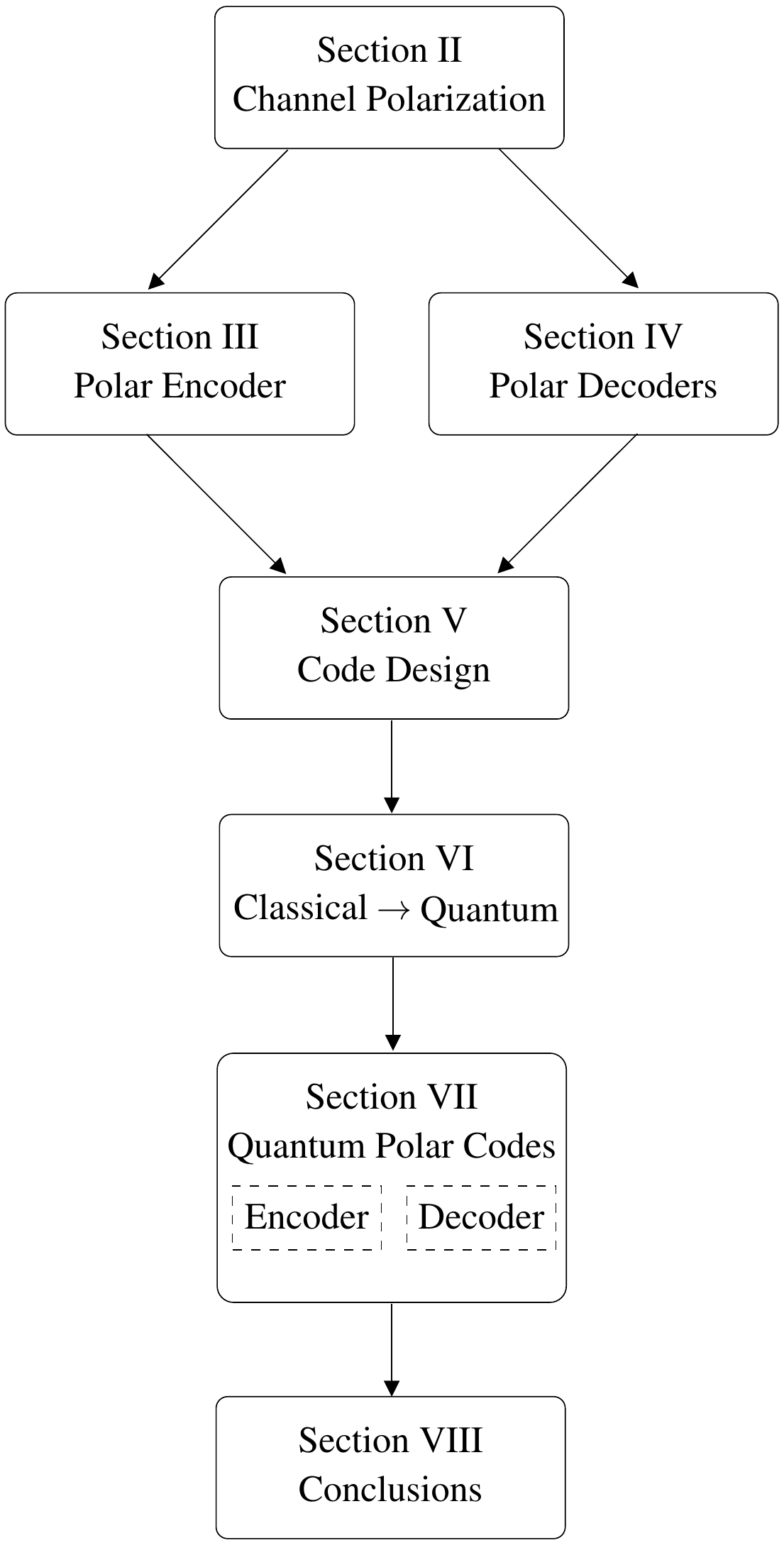}
\caption{Paper structure.}
\label{fig:PaperStruct}
\end{figure}
\section{The Philosophy of Channel Polarization} \label{sec:ch-polarization}
\begin{figure}[tb]
\centering
\includegraphics[width=0.8\linewidth]{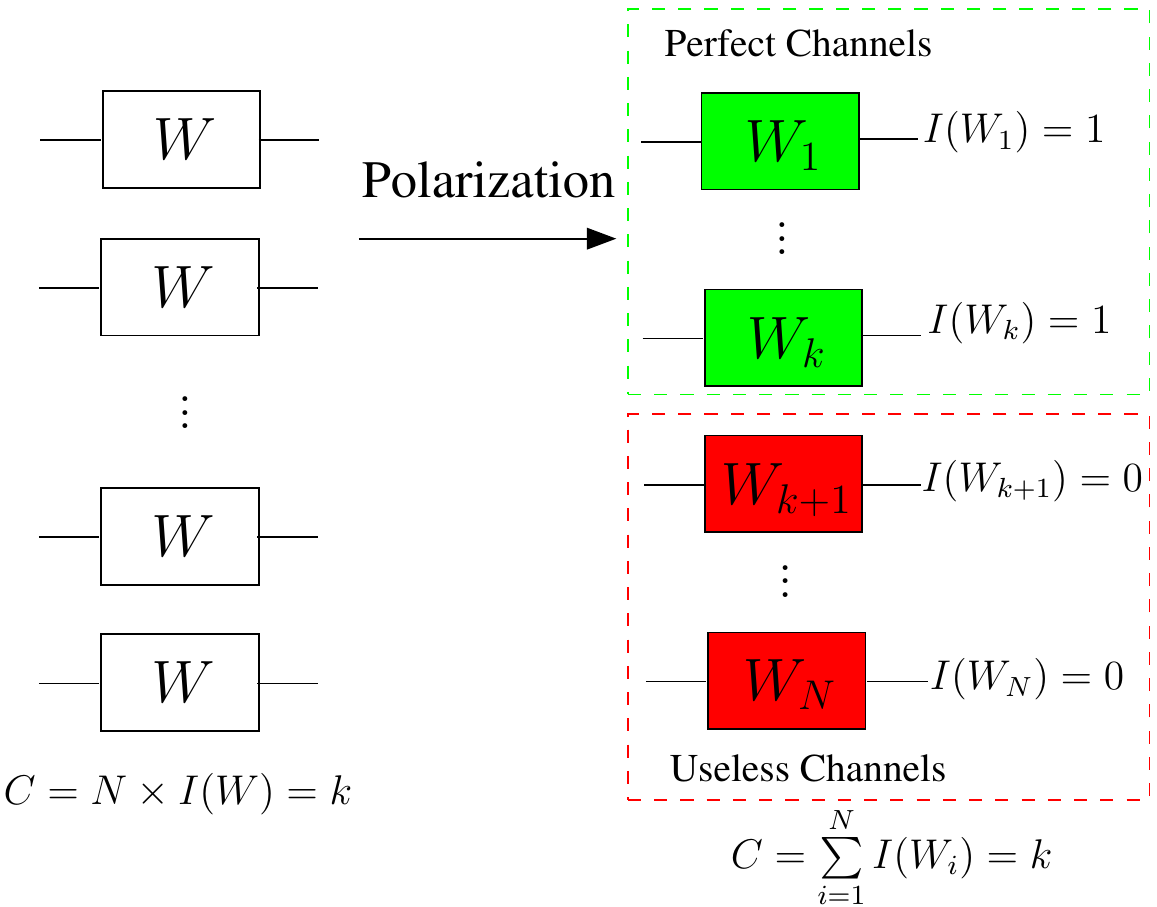}
\caption{The philosophy of channel polarization: the total capacity of $N$ \acp{B-DMC} is redistributed, resulting in 
$k = N \times I(W)$ perfect channels having unit capacity each and $(N-k)$ useless channels having zero capacity, while conserving the 
total capacity. Perfect channels are used for transmitting $k$ information bits, while the useless channels are frozen.}
\label{fig:ch_polar}
\end{figure}

Polar codes rely on the phenomenon of channel polarization, which is the key to unlock Shannon's capacity.
Channel polarization is basically the process of redistributing channel capacities among the various instances, or more precisely
uses, of a transmission channel, while conserving the total capacity, as encapsulated in \fref{fig:ch_polar}. 
Explicitly, channel polarization implies that a set of given channels is polarized into perfect and useless (or completely random) 
channels, having capacities of $1$ and $0$, respectively. This in turn makes the channel coding problem trivial, since the induced perfect
channels may be used for transmitting uncoded information without any errors, while the induced useless channels can be discarded.

Let us consider $N$ uses of a \ac{B-DMC} $W$, each having a capacity of $I(W)$, as exemplified in \fref{fig:ch_polar}. 
In the asymptotic region, i.e. when $N$ is infinitely large, channel polarization induces $k =N \times I(W)$ perfect channels having near-$1$ capacity 
and $(N - k)$ useless channels having near-$0$ capacity.
Thereafter, $k$ information bits are sent uncoded (rate-$1$) through the induced perfect channels, while the $(N - k)$ inputs to the induced useless channels are frozen, implying that known
redundant bits are sent across these channels (rate-$0$)\footnote{An induced bit-channel is a hypothetical end-to-end channel, which consists
of the encoder, the real channel and the decoder, as discussed further in Section~\ref{sec:sec:Arikan}. Hence, the information sent across the 
real channels is encoded. However, the resulting induced bit-channels may be viewed as rate-$1$ and rate-$0$ channels, since they exhibit a capacity of $0$ 
and $1$, respectively.}. Hence, the resulting coding rate is equivalent to the channel capacity, i.e. we have:
\begin{equation}
 R = \frac{k}{N} = \frac{N I(W)}{N} = I(W). 
 \label{eq:polar-cap1}
\end{equation}
\textit{But how do we achieve channel polarization?} This is where Arikan's polar codes come in~\cite{5075875}, which inspired 
researchers to develop more sophisticated coding schemes for achieving channel polarization. 
More explicitly, polar encoders convert the $N$ inputs into bits that can be sent across the real channels at the Shannon's capacity.
\subsection{Arikan's Polar Codes} \label{sec:sec:Arikan}
\begin{figure}[tbp]
        \centering
        \begin{subfigure}[]{\linewidth}
        \begin{center}
                \includegraphics[width=0.6\linewidth]{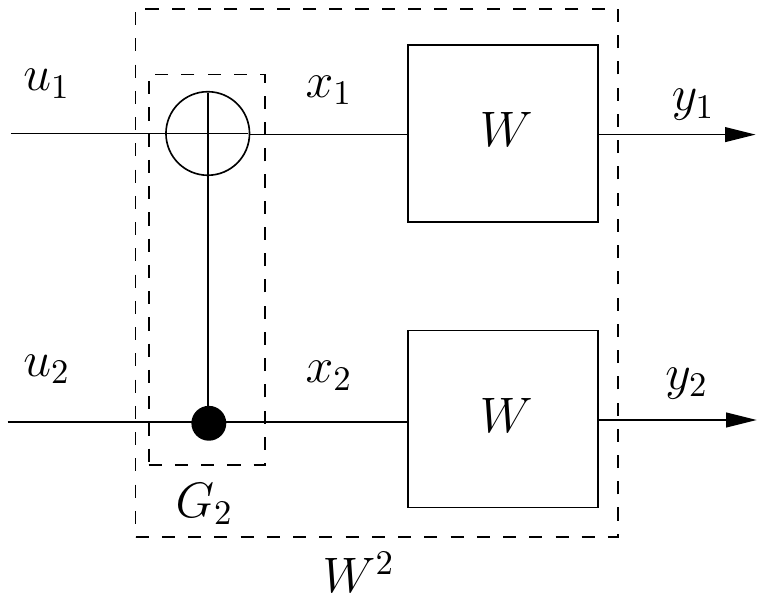}
                \caption{Channel combining.}
                \label{fig:ch_comb}
        \end{center}
        \end{subfigure}%
        \\ 
        \begin{subfigure}[]{\linewidth}
        \begin{center}
                \includegraphics[width=0.6\linewidth]{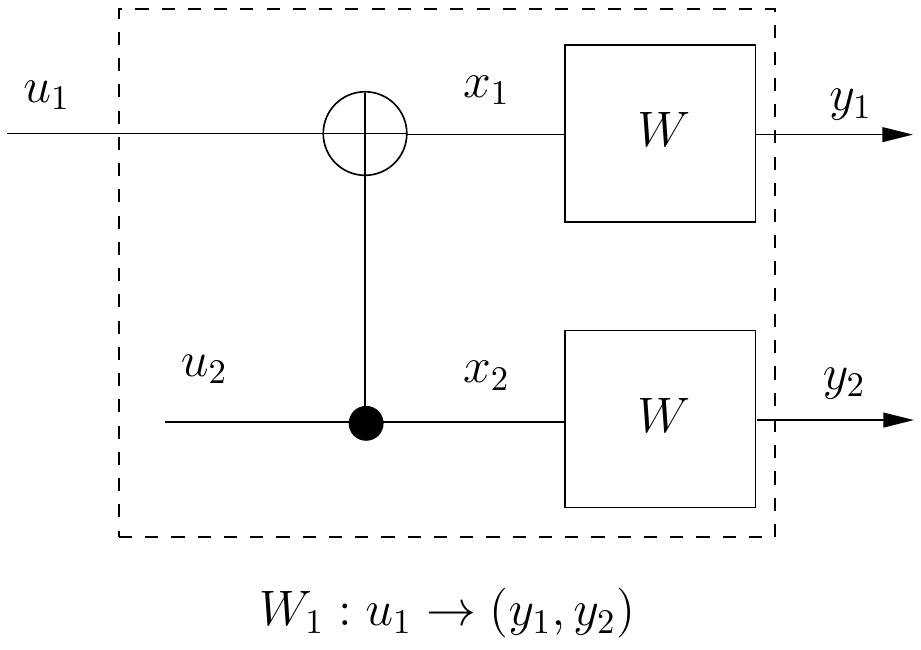}
                \caption{Channel splitting: first bit channel.}
                \label{fig:ch_split_W1}
        \end{center}
        \end{subfigure}
        \\
                \begin{subfigure}[]{\linewidth}
        \begin{center}
                \includegraphics[width=0.6\linewidth]{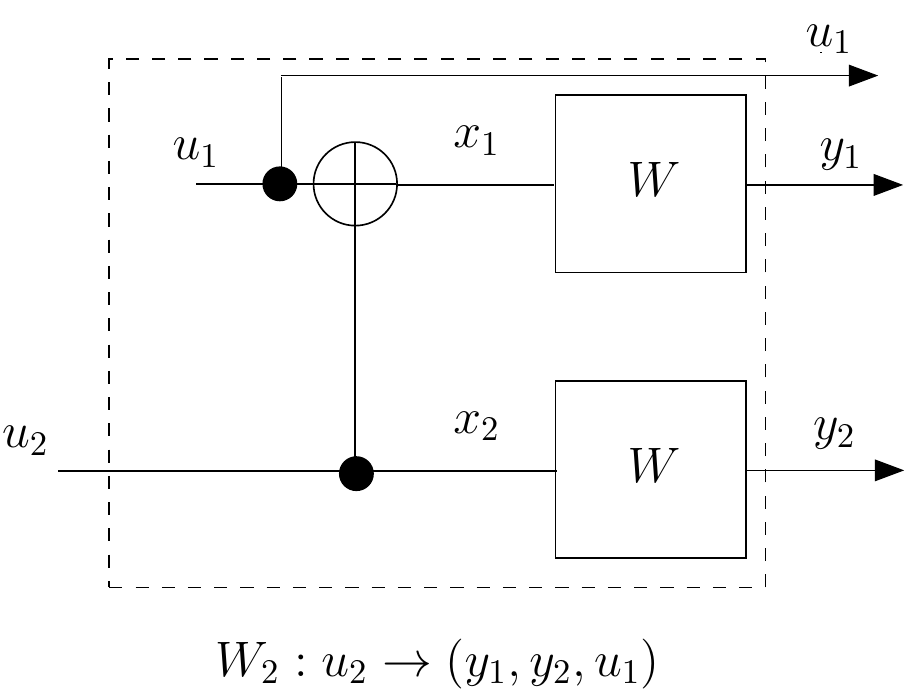}
                \caption{Channel splitting: second bit channel.}
                \label{fig:ch_split_W2}
        \end{center}
        \end{subfigure}
        \caption{Arikan's $2$-bit polar code, relying on channel combining and channel splitting for channel polarization.}
         \label{fig:ch_comb_split}
\end{figure}

Arikan's polar codes~\cite{5075875} achieve channel polarization by recursively invoking the simple $2$-bit encoding kernel of \fref{fig:ch_comb}. 
More specifically, two uses of the \ac{B-DMC} 
are combined using a single \ac{XOR} gate at the encoder. This step is termed as `channel combining' and the resulting compound channel is denoted by $W^2$,
which has a capacity of:
\begin{equation}
 C (W^2) = I (u_1,u_2;y_1,y_2) = I (x_1,x_2;y_1,y_2) = 2 \times I(W).
 \label{eq:bec:2w-1}
\end{equation}
For simplicity, let $W$ be a \ac{BEC} having an erasure probability of $\epsilon$. Hence, the total
capacity of the compound channel $W^2$ is:
\begin{equation}
 C (W^2) = 2 \times I(W) = 2 \times \left(1 - \epsilon \right).
 \label{eq:bec:2w}
\end{equation}
According to the chain rule~\cite{WYNER197851}, \eqr{eq:bec:2w-1} may also be expressed as follows:
\begin{align}
 C (W^2) &= I (u_1,u_2;y_1,y_2) \nonumber \\
 &= I(u_1;y_1,y_2) + I(u_2;y_1,y_2|u_1) \nonumber \\
 &= I(u_1;y_1,y_2) + I(u_2;y_1,y_2,u_1),
 \label{eq:bec:2w-2}
\end{align}
since $u$ is an independently and identically distributed random variable, $u_1$ and $u_2$ are independent.
\eqr{eq:bec:2w-2} implies that the compound channel $W^2$ may be split into two single-bit channels $W_1$ and $W_2$ of
\fref{fig:ch_split_W1} and \fref{fig:ch_split_W2}, respectively, which are defined as follows:
\begin{align}
 W_1 & \buildrel\triangle\over = u_1 \rightarrow \left(y_1,y_2\right)  \label{eq:W1} \\
 W_2 & \buildrel\triangle\over = u_2 \rightarrow \left(y_1,y_2,u_1\right). \label{eq:W2}
\end{align}
This splitting process, which occurs at the decoder, redistributes the total capacity between the two induced bit channels $W_1$ and $W_2$, such
that one channel gets better, while the other gets worse. 

\begin{figure}[tb]
\centering
\includegraphics[width=0.9\linewidth]{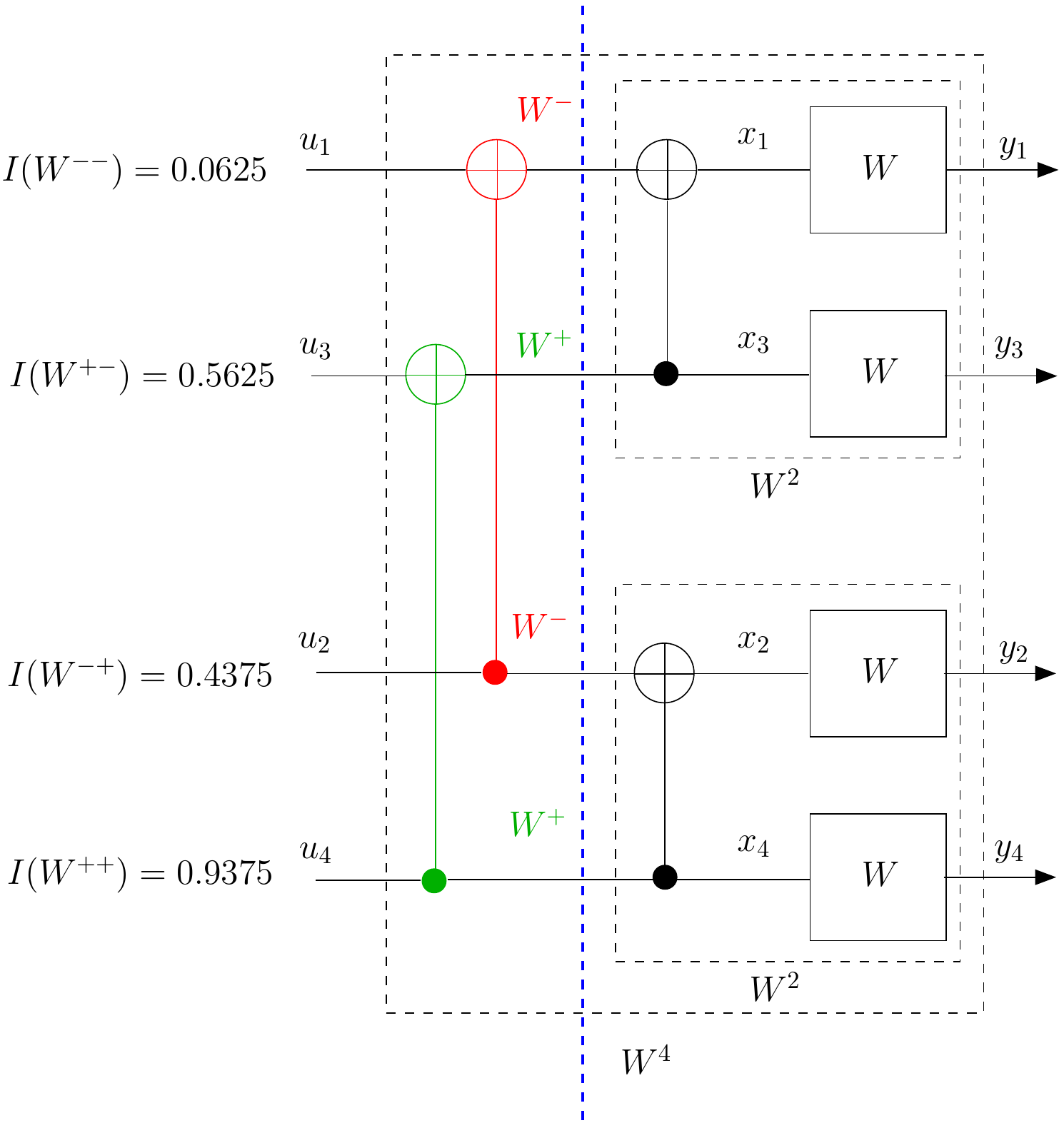}
\caption{$4$-bit compound channel $W^4$ (bit-reversed indexing): it is constructed from two $2$-bit channels $W^2$ by coupling the bad channels $W^-$ using the 
red \ac{XOR} gate and the good channels $W^+$ using the green \ac{XOR} gate (viewing the schematic from right to left). This enhances the strength of polarization yielding two strongly
polarized channels $W^{--}$ and $W^{++}$.}
\label{fig:polar-ex2}
\end{figure}

\begin{figure}[tb]
\centering
\includegraphics[width=0.7\linewidth]{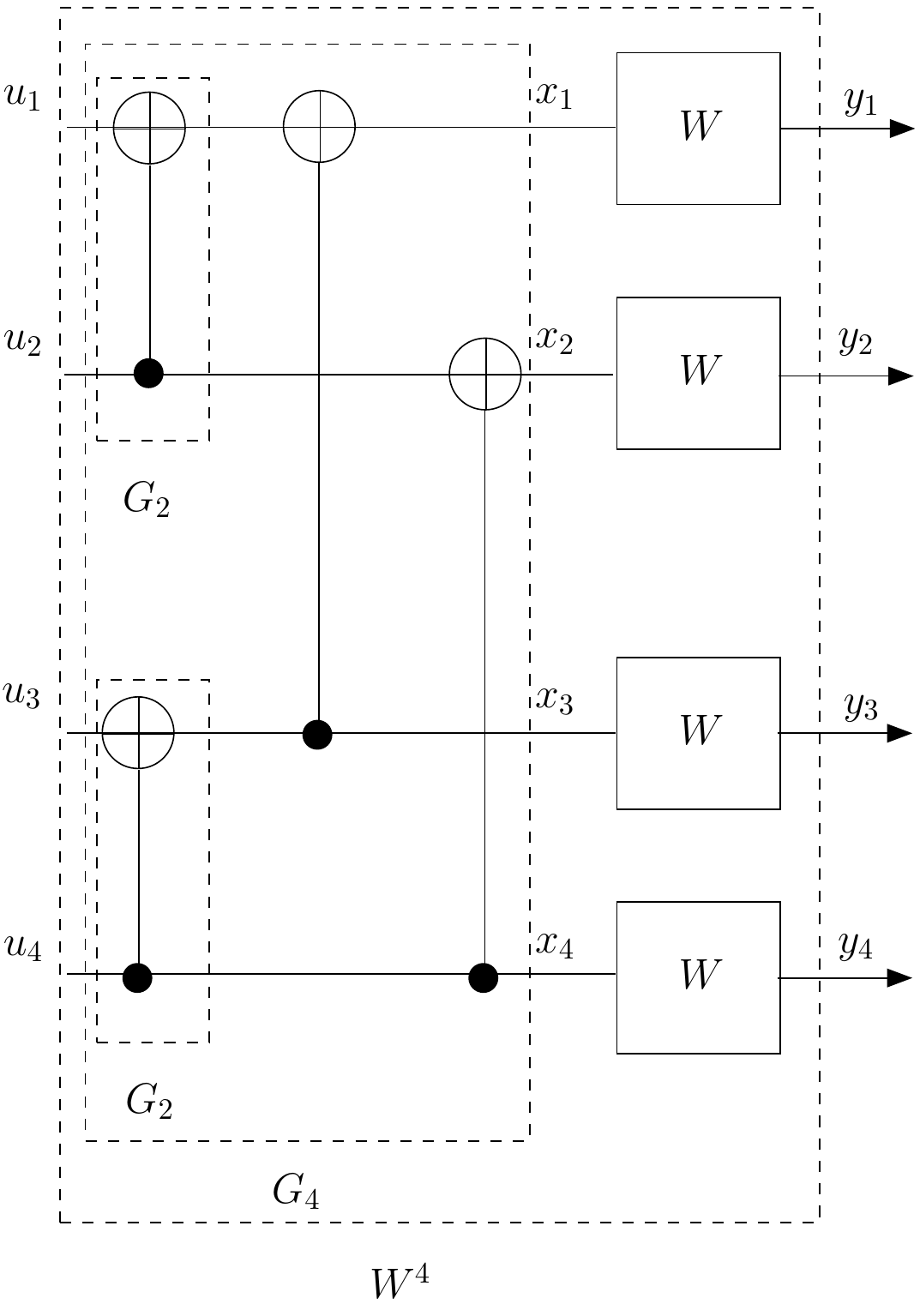}
\caption{$4$-bit compound channel $W^4$ (natural indexing), which is obtained by re-wiring the inputs of \fref{fig:polar-ex2}.}
\label{fig:polar-ex2e}
\end{figure}

Resuming our example of a \ac{BEC}, let us calculate the capacities of the two induced channels.
The first bit-channel $W_1$ of \fref{fig:ch_split_W1} has no information about $u_2$. So, the receiver estimates 
$u_1$ based on the received bits $y_1$ and $y_2$ as follows:
\begin{equation}
 \hat{u}_1 = y_1 \oplus y_2,
\end{equation}
where $\oplus$ denotes modulo-$2$ addition. Hence, $u_1$ can be decoded only when neither $y_1$ nor $y_2$ is erased. Hence, the
erasure probability of the induced bit-channel $W_1$ is:
\begin{equation}
 \epsilon_1 = 1 - (1-\epsilon)^2 = 2\epsilon -\epsilon^2,
\end{equation}
which is worse than that of the original channel $W$. Consequently, $W_1$ is the worse channel, also denoted as $W^{-}$, having a reduced capacity
of:
\begin{equation}
 I(W^-) = 1 - \epsilon_1 = 1-2\epsilon+\epsilon^2.
 \label{eq:c-}
\end{equation}
The second bit channel of \fref{fig:ch_split_W2} outputs $u_1$ as well as $y_1$ and $y_2$. Explicitly, the availability of
$u_1$ implies that we have a genie decoder or a side channel, which reveals the value of $u_1$. Consequently, $u_2$ can be decoded as long as either $y_1$
or $y_2$ is not erased. Hence, the erasure probability of the induced channel $W_2$ is: 
\begin{equation}
 \epsilon_2 = \epsilon^2,
\end{equation}
which is lower than that of the original channel $W$. This implies that $W_2$ is the better channel, also denoted as $W^{+}$, which
has a capacity of:
\begin{equation}
 I(W^+) = 1 - \epsilon_2 = 1-\epsilon^2.
  \label{eq:c+}
\end{equation}
Based on \eqr{eq:c-} and \eqr{eq:c+}, we may conclude that:
\begin{equation}
 I(W^{-}) \leq I(W) \leq I(W^{+}),
 \label{eq:Iw}
\end{equation}
which implies that $W^{-}$ (or equivalently $W_{1}$) tends to polarize towards zero-capacity, while $W^{+}$ (or equivalently $W_{2}$)
tends to polarize towards unit capacity. It is pertinent to mention here that the equality in \eqr{eq:Iw} holds only when $W$ is an extreme
channel having a capacity of either $0$ or $1$.
Furthermore, the total capacity is conserved, since we have:
\begin{equation}
 I(W^{-}) + I(W^{-}) = 2 \times I(W).
\end{equation}
To elaborate further, let us assume that $\epsilon = 0.5$, hence we have:
\begin{align}
I(W) &= 0.5 \nonumber \\
 I(W_1) = I(W^{-}) &= 0.25 \nonumber \\
 I(W_2) = I(W^{+}) &= 0.75,
  \label{eq:cap-ex1}
\end{align}
where we may observe that the value of $I(W^{-})$ gets closer to zero, while that of $I(W^{+})$ gets closer to one. The impact 
of channel polarization may be enhanced by recursively invoking the basic encoding kernel of \fref{fig:ch_comb}, which is labeled `$G_2$'. More specifically,
as exemplified in \fref{fig:polar-ex2}, a compound channel $W^4$ can be constructed by using two copies of the compound channel
$W^2$. 
In the first layer of polarization, two independent copies of $W^2$ are invoked, inducing two good channels $W^+$ and two
bad channels $W^-$, which have the capacities of $0.25$ and $0.75$, respectively, for $\epsilon = 0.5$. In the second layer of polarization, the bad channels
are coupled using the encoding kernel $G_2$ of \fref{fig:ch_comb}, which is simply a \ac{XOR} gate marked in red in \fref{fig:polar-ex2}.
Recall that the encoding transformation of \fref{fig:ch_comb} yields a bad and a good channel, 
whose capacities may be calculated using \eqr{eq:c-} and \eqr{eq:c+}, respectively. Consequently, the red \ac{XOR} gate
of \fref{fig:polar-ex2} polarizes the two $W^-$ channels a the worse channel $W^{--}$ and a better channel $W^{-+}$, having capacities of
$0.0625$ and $0.4375$, respectively. We may notice here that the second layer of polarization polarizes the first channel more towards the
zero capacity, while the capacity of the other channel tends to increase towards one. Similarly, the second layer of polarization couples
the two good channels $W^{+}$ using the \ac{XOR} gate marked in green in \fref{fig:polar-ex2}. This second \ac{XOR} gate induces 
the channels $W^{+-}$ and $W^{++}$ having capacities of $0.5625$ and $0.9375$, respectively. Hence, a two-layered polarization yields
two strongly polarized channels $W^{--}$ and $W^{++}$, which exhibit a higher degree of polarization than the channels $W^-$ and $W^+$.
It is important to point out that the bits in \fref{fig:polar-ex2} are indexed according to their decoding order, which will be discussed
further in Section~\ref{sec:decoder}. Furthermore, it may also be noticed that the bits in \fref{fig:polar-ex2} follow a bit-reversed indexing.
Explicitly, bit reversing implies that a number $i \in \{1,N\}$ having the $n$-bit binary representation $i_1i_2\dots i_n$, for $n = \log_2N$,
is mapped onto its bit-reversed counterpart having the binary representation $i_ni_{n-1}\dots i_1$. 
Consequently, the four inputs of \fref{fig:polar-ex2} $\{1,2,3,4\}$ having the binary representations $\{00,01,10,11\}$ are 
indexed as $\{1,3,2,4\}$ corresponding to their bit-reversed binary representations $\{00,10,01,11\}$. 
This bit-reversed order of the bits may also be obtained from the labels $\{W^{--}, W^{+-}, W^{-+}, W^{++}\}$ of the induced channels 
by mapping $+$ and $-$ onto $1$ and $0$, respectively. Hence, the sequence $\{W^{--}, W^{+-}, W^{-+}, W^{++}\}$ yields the bit-reversed order 
$\{00,10,01,11\}$.
The inputs of \fref{fig:polar-ex2} may be re-wired to follow the natural indexing order, as shown in \fref{fig:polar-ex2e}.
Bit-reversed indexing of \fref{fig:polar-ex2} facilitates hardware implementations, while the natural indexing of \fref{fig:polar-ex2e}
is more apt for software implementations. In this paper, we will follow the the natural indexing of \fref{fig:polar-ex2e}. 

\begin{figure}[tb]
\centering
\includegraphics[width=0.7\linewidth]{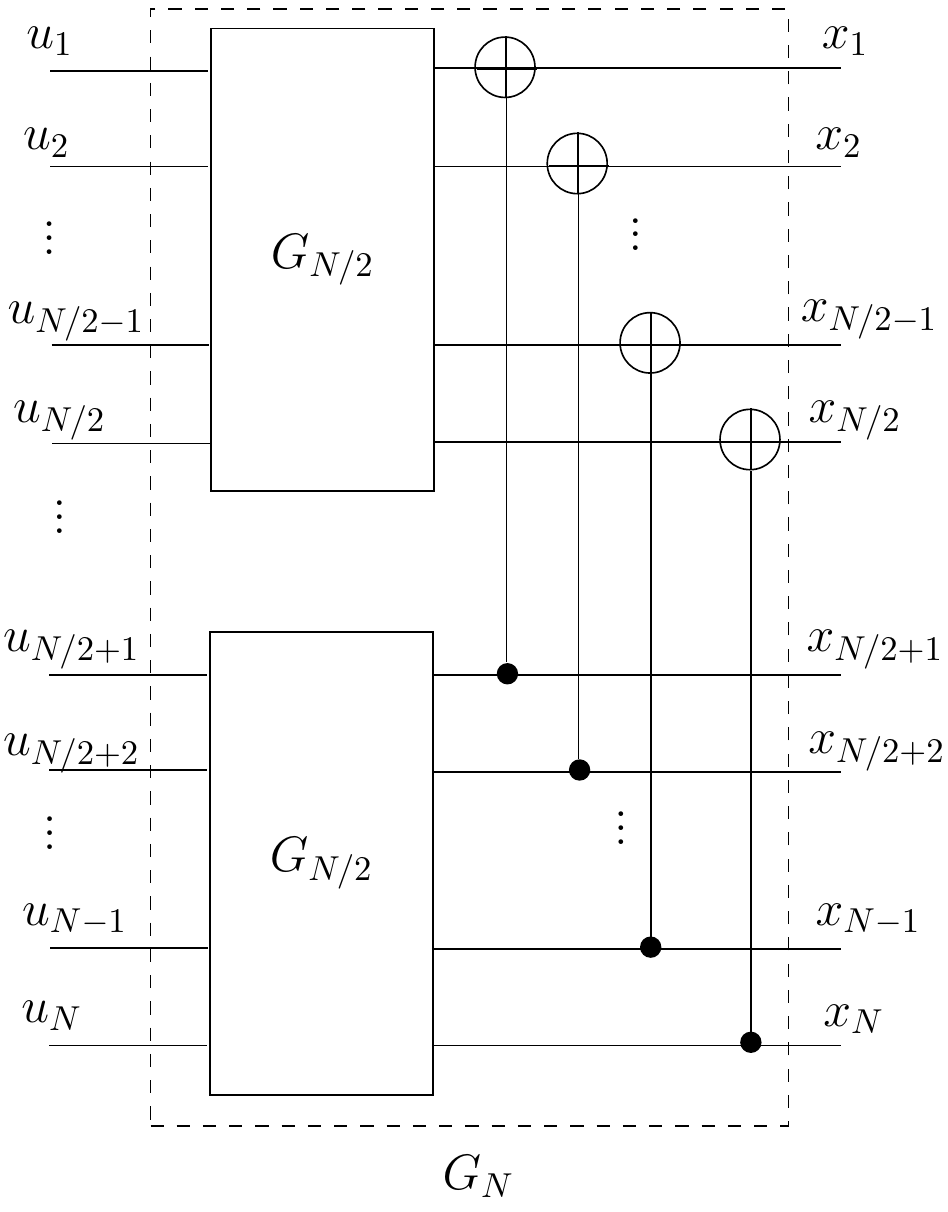}
\caption{Channel combining transformation $G_N$ for constructing an $N$-bit compound channel $W^N$. It is constructed from two ($N/2$)-bit 
transformations by coupling their corresponding outputs using a \ac{XOR} gate (viewing the schematic from left to right).}
\label{fig:polar-exN}
\end{figure}

The polarization of the induced bit-channels of the compound channel $W^4$ may be enhanced 
by invoking a third layer of polarization using two independent copies of $W^4$, resulting in the compound channel $W^8$. The process may be
repeated recursively using the generalized channel combining transformation of \fref{fig:polar-exN} until either the desired length $N$ or the desired polarization strength is achieved. 
\fref{fig:polar-exN} shows the schematic of $G_N$ for constructing an $N$-bit compound channel $W^N$. 
The associated capacities of the induced bit-channels can be recursively calculated as follows:
\begin{align}
 I\left(W^{N}_{2i-1}\right) &=  I\left(W^{N/2}_{i}\right)^2 \nonumber \\
 I\left(W^{N}_{2i}\right) &= 2 I\left(W^{N/2}_{i}\right) - I\left(W^{N/2}_{i}\right)^2,
 \label{eq:recursive-cap}
\end{align}
for $1 \leq i \leq N/2$, where $W^{N}_{i}$ denotes the $i$th  induced bit-channel of an $N$-bit compound channel $W^N$. 
Please note that
the recursive calculation of capacities given in \eqr{eq:recursive-cap} is only valid for \acp{BEC}. 
%

\begin{figure}[tb]
\centering
\includegraphics[width=\linewidth]{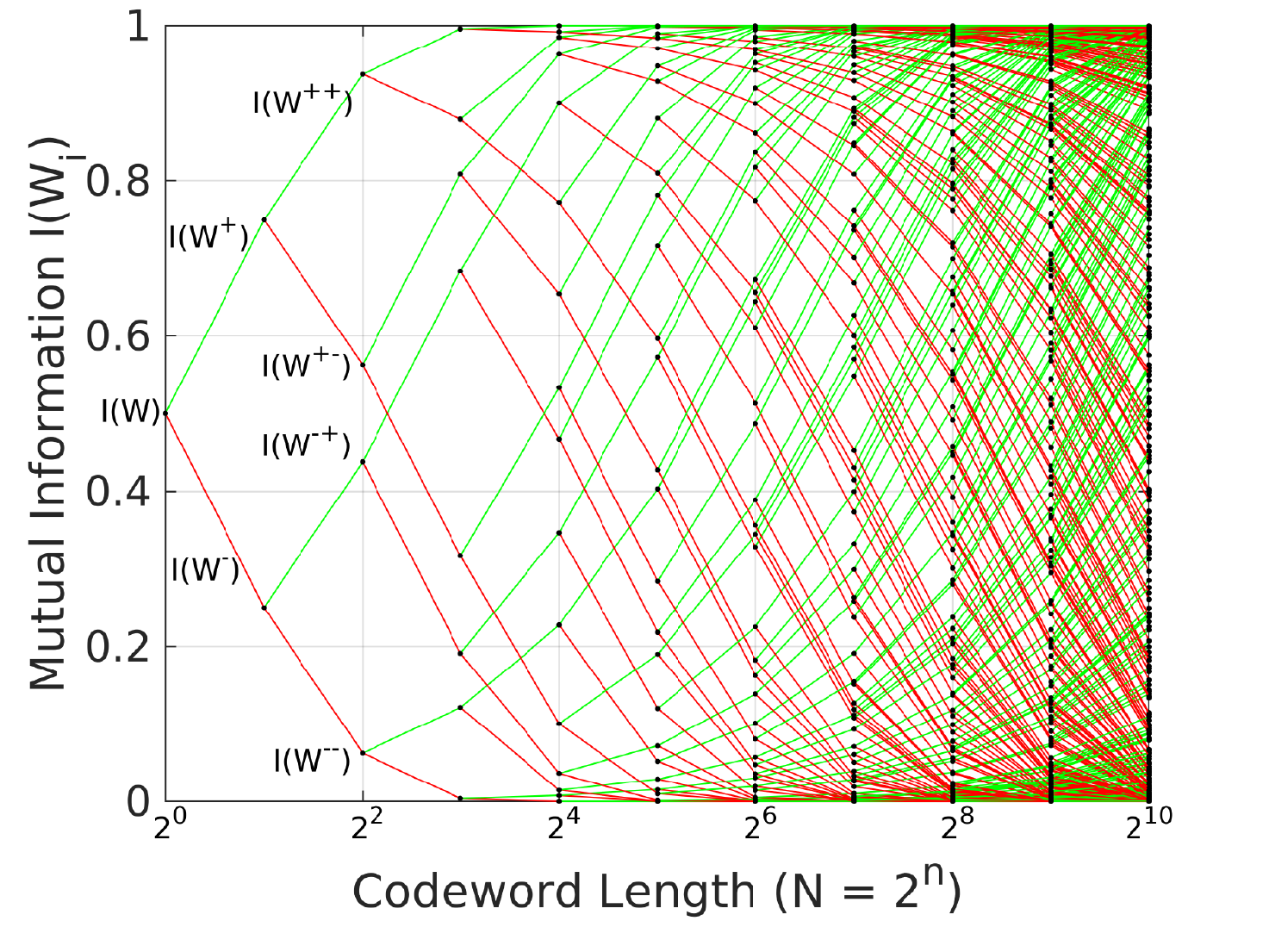}
\caption{$[0,1]$-bounded polarization martingale for a \ac{BEC} $I(W)$ having $\epsilon = 0.5$. The mutual information of the induced bit-channels
$W_i$ converges to $0$ or $1$ upon increasing the codeword length $N$.}
\label{polar-ev}
\end{figure}

\begin{figure}[tb]
\centering
\includegraphics[width=\linewidth]{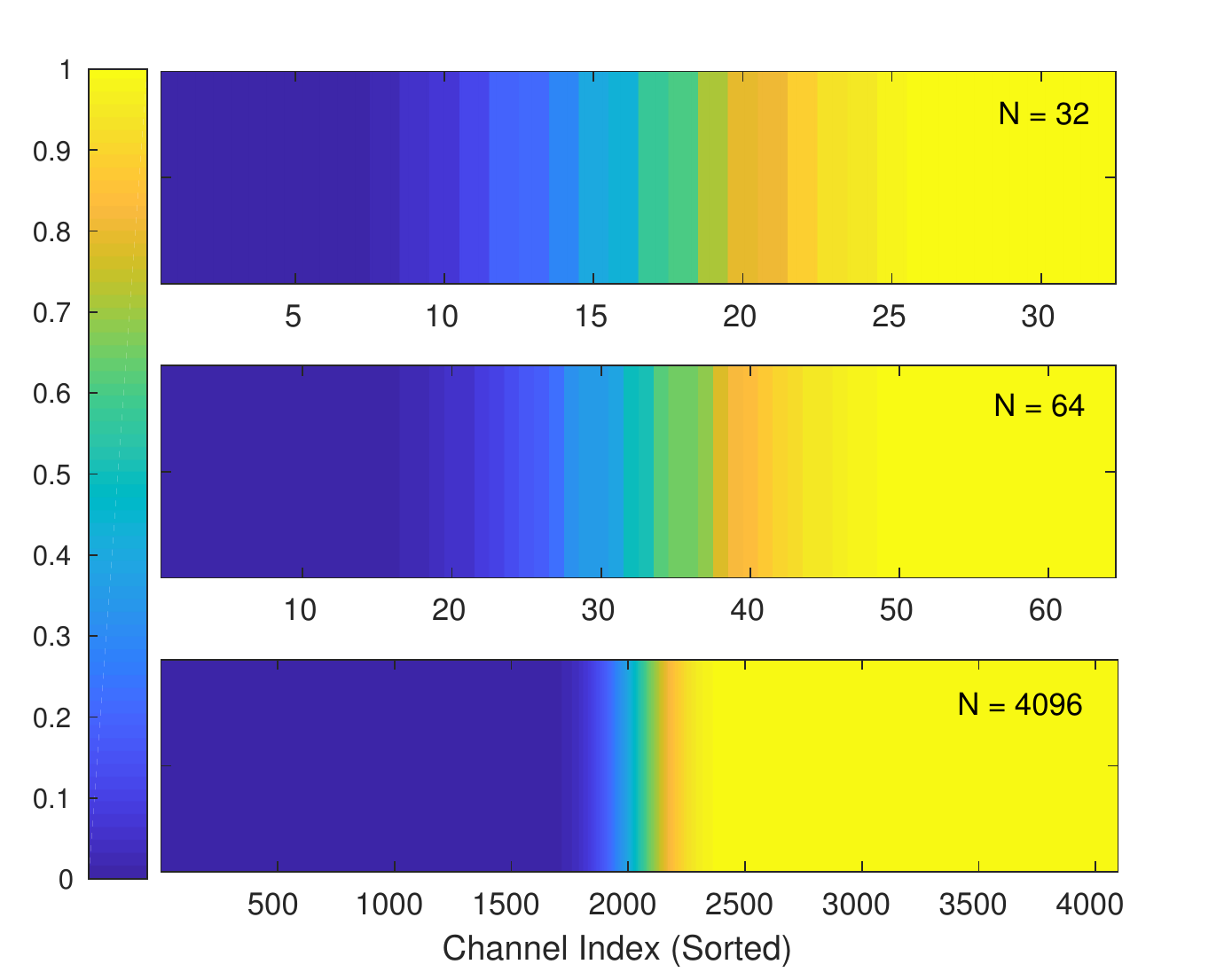}
\caption{Mutual information intensity map for a \ac{BEC} at $\epsilon = 0.5$ and codeword lengths of $32$, $64$ and $4096$. 
The strength of polarization enhances upon increasing the codeword length.} 
\label{IMcolorplot}
\end{figure}

The encoding transformation of \fref{fig:polar-exN} embeds $n$ layers of polarization, when the codeword length is $N = 2^n$.
Each layer of polarization makes the bad channels worse and the good channels better, which is typically called the `Matthew effect', and
demonstrated using the polarization martingale of \fref{polar-ev}, which converges to $0$ and $1$, when $N$ is infinitely long.
To elaborate further, \fref{IMcolorplot} shows the bit-wise mutual information intensity map for increasing codeword length. We may observe in 
\fref{IMcolorplot} that the proportion of near-$0$ and near-$1$ bit-channels increases as the codeword length is increased from $32$ to $4096$.
Please note that the bit-channel indexes in \fref{IMcolorplot} are sorted based on their mutual information values.
Hence, polar codes do not polarize the channels completely at finite codeword lengths. Explicitly, in contrast to the perfect
and useless channels of \fref{fig:ch_polar} having capacities of $1$ and $0$, respectively, finite-length polar codes polarize the 
underlying channels into the set of good and bad channels, which tend to polarize towards the absolute $1$ and $0$ capacity, when 
the codeword length is infinitely long.
\subsection{Non-Arikan Polar Codes} \label{sec:sec:NonArikan}
\begin{figure}[tb]
\centering
\includegraphics[width=0.6\linewidth]{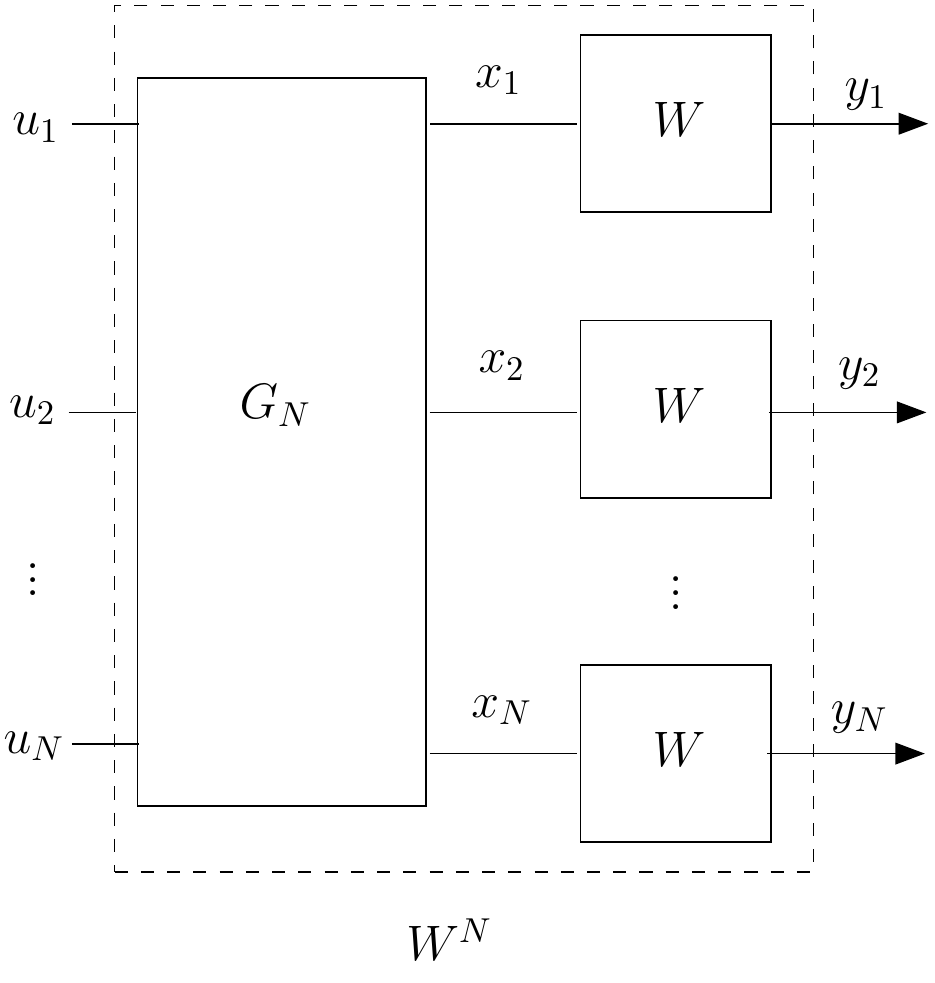}
\caption{Channel combining: the encoder $G_N$ combines the $N$ uses of a \ac{B-DMC} to construct a compound channel $W^N$.}
\label{polar-comb-gen}
\end{figure}

\begin{figure}[tb]
\centering
\includegraphics[width=0.7\linewidth]{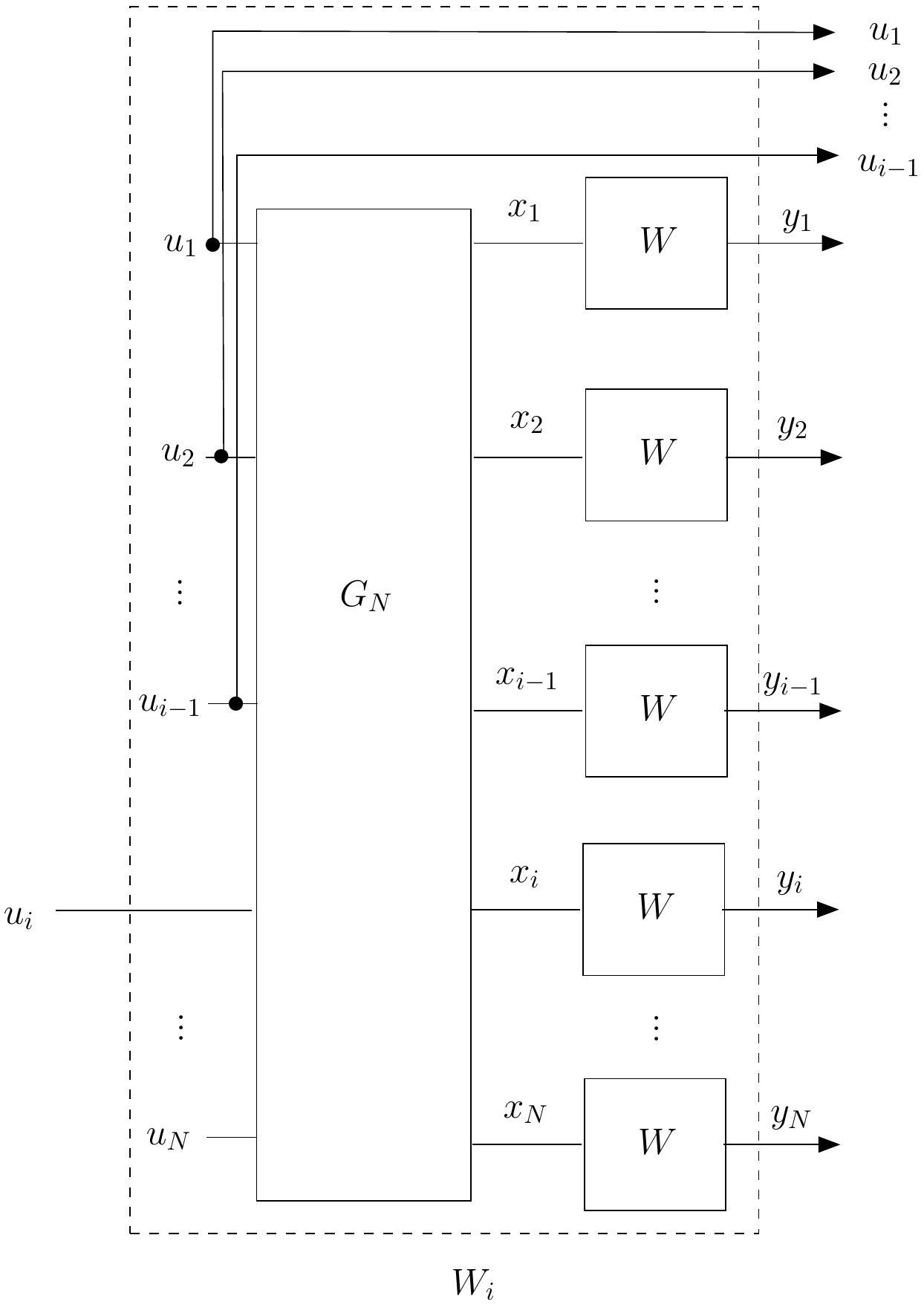}
\caption{Channel splitting: the $i$th induced bit-channel $W_i$.}
\label{polar-split-gen}
\end{figure}

Channel polarization is a general phenomenon and is hence not restricted to Arikan's polar code~\cite{Arikan-course, sasoglu2011polar}. 
The channel combining and channel splitting procedures of \fref{fig:ch_comb_split} may be generalized for an arbitrary encoder $G_N$, as
shown in \fref{polar-comb-gen} and \fref{polar-split-gen}, respectively.
More specifically, the encoder $G_N$ of \fref{polar-comb-gen} is a one-to-one $N$-bit mapper, which constructs a compound channel $W^N$ by 
coupling $N$ uses of a \ac{B-DMC}. 
The 
total capacity of the resulting channel $W^N$ is:
\begin{equation}
 C (W^N) = I (u_1^N;y_1^N) = I (x_1^N;y_1^N) = N \times I(W),
 \label{eq:comb-gen}
\end{equation}
where $a_i^j$ denotes the vector $(a_i, a_{i+1}, \dots, a_j)$. Analogous to \eqr{eq:bec:2w-2}, chain rule may be applied to \eqr{eq:comb-gen},
as follows:
\begin{align}
 C (W^N) &= I (u_1^N;y_1^N)) \nonumber \\
 &= I(u_1;y_1^N) + I(u_2;y_1^N|u_{1}) + \dots I(u_N;y_1^N|u_1^{N-1}) \nonumber \\
 &= \sum_{i=1}^{N} I(u_i;y_1^N|u_1^{i-1})  \nonumber \\
  &= \sum_{i=1}^{N} I(u_i;y_1^N,u_1^{i-1}).
 \label{eq:split-gen}
\end{align}
Hence, the compound channel $W^N$ may be split into $N$ induced bit-channels, such that the $i$th bit-channel is defined as:
\begin{equation}
  W_i \buildrel\triangle\over = u_i \rightarrow \left(y_1^N,u_1^{i-1}\right),
\end{equation}
which is also illustrated in \fref{polar-split-gen}. More specifically, the induced bit-channel $W_i$ takes the input $u_i \in \mathcal{X}$, 
yielding the output $(y_1^N,u_1^{i-1}) \in \mathcal{Y}^N \times \mathcal{X}^{i-1}$, where $\mathcal{X}$
and $\mathcal{Y}$ denote the input and output alphabets of the channel $W$. 
Any random $N$-bit permutation $G_N$ may exhibit good polarization characteristics, as
identified by Shannon's random coding argument. However, Arikan's polar codes bring with them the benefit of a simple recursive structure,
which is easy to implement.

Arikan's polar codes, relying on the $2$-bit kernel $G_2$ of \fref{fig:ch_comb}, exhibit an error exponent of $1/2$. This implies that
the net error probability of the resulting polar code decays exponentially in the square root of the codeword length $N$, when $N$ is sufficiently large.
For the sake of achieving a higher error exponent, higher dimensional kernels, relying on more than $2$ bits, were used by Korada \etal~\cite{korada2010polar}
for polarizing \acp{B-DMC}. Korada \etal~\cite{korada2010polar} also conceived a general formalism for designing \ac{BCH}-based $l$-bit
kernels, which achieve error exponents close to $1$ for large values of $l$. The notion
of code decomposition (also called code nesting) was used in~\cite{6033859,7055274} for increasing the error exponents of higher dimensional kernels,
while \ac{BCH}, \ac{RS} and Golay codes based kernels were explored in~\cite{6979882, 6691213, 7339451, 8258974} for the sake of increasing the minimum distance of polar codes.
Furthermore, the phenomenon of channel polarization was generalized to non-binary channels in~\cite{csacsouglu2009polarization, 6283740},
while explicit \ac{RS}-based non-binary constructions were investigated in~\cite{5513568, mori2010non, 6774879}.
Finally, the idea of mixed kernels, consisting of sub-kernels of arbitrary non-binary alphabet sizes, 
was investigated in~\cite{6034223, 7339419}. The resulting polar codes provided attractive
benefits both in terms of the error correction performance as well as the decoding complexity. 
However, the codeword length was limited to $N=l^n$. For the sake of obtaining arbitrary codeword lengths, 
puncturing and shorterning techniques were exploited in~\cite{6655078, 6936302, 7919040}, while multi-kernel polar codes, invoking multiple $l$-bit kernels, were conceived in~\cite{7962750, 8277949}.
Nonetheless, in this paper, we will focus on Arikan's polar codes.

\section{Polar Encoder} \label{sec:encoder}
Recall from Section~\ref{sec:sec:Arikan} that polar codes induce a set of good and bad channels, so that information bits
can be transmitted through the induced good channels, while the input to the induced bad channels are frozen. When the codeword length $N$ is
infinitely long, the channels are polarized into $k$ good channels (or the so-called perfect channels) and $(N-k)$
bad channels (or the so-called useless channels). Hence, a polar code encodes $k$ information bits into $N$ coded bits using ($N-k$) redundant bits,
which are called the `frozen bits'. It is characterized by the parameters ($N, k, \mathcal{F}, u_{\mathcal{F}}$),
where $\mathcal{F} \subset \{1,2, \dots, N\}$ specifies 
the location of frozen bits, while $u_{\mathcal{F}}$ is an ($N-k$)-bit vector of frozen bits, which are known to the decoder.
The performance of polar codes rely on the parameters $N$, $k$ as well as $\mathcal{F}$. In particular, $\mathcal{F}$ is channel specific and
must be optimized for the channel under consideration, as discussed further in Section~\ref{sec:polarcodedesign}.
However, the performance of polar codes is unaffected by the value of frozen bits, more precisely the vector $u_{\mathcal{F}}$, if the channel is symmetric.
Generally, $u_{\mathcal{F}}$ is assumed to be an all-zero vector. It is important to note here that polar codes 
are intrinsically rate compatible, since the coding rate can be varied by merely changing the number of frozen bits, while using the same
encoder $G_N$. The frozen bits' locations $\mathcal{F}$ can be selected using a sequence that reads the locations.  
The best locations for one coding rate are typically a subset of the best locations for any lower coding rate.

The polar encoding process, which was generalized in \fref{polar-comb-gen} for an arbitrary encoder $G_N$, may be represented
as:
\begin{equation}
 x_1^N = u_1^N G_N.
 \label{eq:PolarEncGn}
\end{equation}
For Arikan's polar code, $G_N$ is the $n$th Kronecker product of the ($2 \times 2$) kernel matrix, which characterizes 
the encoding transformation $G_2$ of \fref{fig:ch_comb}. More specifically, Arikan's kernel $G_2$ may be represented in matrix
form as follows:
\begin{equation}
 G_2 = \left(
\begin{array}{ll}
1 & 0\\
1 & 1
\end{array} \right),
\label{eq:G2}
\end{equation}
while the $N$-bit encoder $G_N$ is defined recursively as:
\begin{equation}
 G_N = G_2^{\otimes n} = \left(\begin{array}{ll}
G_{N/2} & 0\\
G_{N/2} & G_{N/2}
\end{array}\right).
 \label{eq:arikanGn}
\end{equation}
Hence, Arikan's polar code has a recursive structure invoking 
$n = \log_2N$ layers of polarization and each layer of polarization uses $N/2$ \ac{XOR} gates, as previously illustrated in \fref{fig:polar-exN}.
Hence, the encoding operation of \eqr{eq:arikanGn} imposes a complexity of $O(N\log_2N)$. 
We may also notice from \eqr{eq:arikanGn} that polar code assumes a non-systematic structure. Later, 
systematic polar codes were derived in~\cite{5934670, vangala2016efficient, sarkis2016flexible, chen2016low}, which outperformed the classic non-systematic polar codes in terms of the \ac{BER}, 
while retaining the same \ac{BLER} and encoding and decoding complexity. We restrict our discussions to the classic non-systematic polar codes in this paper.

\begin{figure}[tb]
\centering
\includegraphics[width=\linewidth]{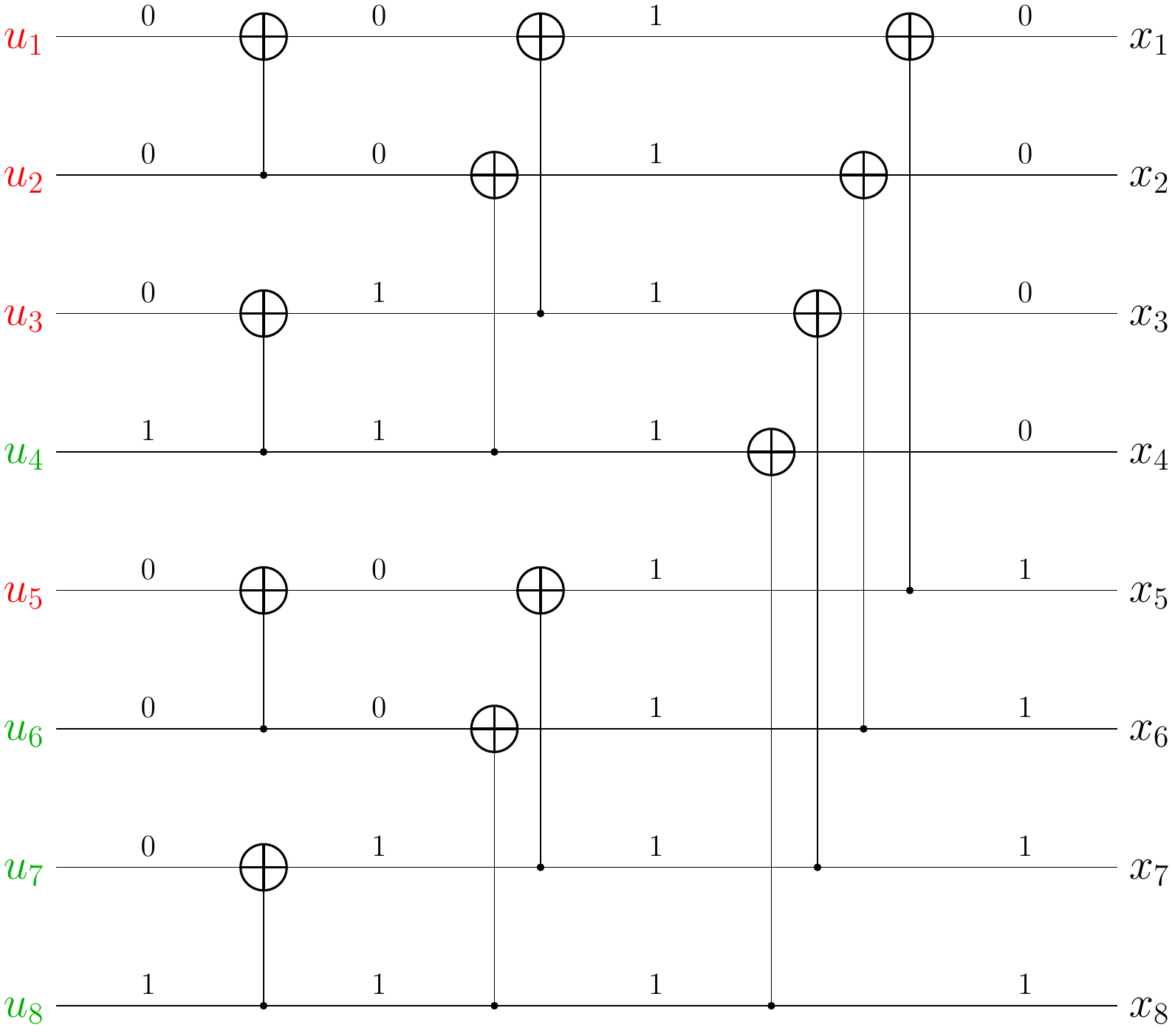}
\caption{Example of polar encoding process: An $N=8$ polar code having $k = 4$, $\mathcal{F} = \{1,2,3,5\}$ and 
$u_{\mathcal{F}} = (0\;0\;0\;0)$ is used to encode the information bit sequence $u_{\mathcal{F}_c} = (1\;0\;0\;1)$. The resulting
codeword is $x_1^8 = (0 \; 0 \; 0\; 0\; 1 \; 1 \; 1\; 1)$, as also shown in \eqr{eq:ex2}.}
\label{example2.fig}
\end{figure}

The $8$-bit polar encoder may be formulated as follows:
\begin{align}
 G_8 = G_2^{\otimes 3}  
&= \left(\begin{array}{llll}
G_2 & 0 & 0 & 0\\
G_2 & G_2 & 0 & 0\\
G_2 & 0 & G_2 & 0\\
G_2 & G_2 & G_2 & G_2
\end{array}\right)  \nonumber \\
& = \left(\begin{array}{llllllll}
1 & 0 & 0 & 0 & 0 & 0 & 0 & 0\\
1 & 1 & 0 & 0 & 0 & 0 & 0 & 0\\
1 & 0 & 1 & 0 & 0 & 0 & 0 & 0\\
1 & 1 & 1 & 1 & 0 & 0 & 0 & 0\\
1 & 0 & 0 & 0 & 1 & 0 & 0 & 0\\
1 & 1 & 0 & 0 & 1 & 1 & 0 & 0\\
1 & 0 & 1 & 0 & 1 & 0 & 1 & 0\\
1 & 1 & 1 & 1 & 1 & 1 & 1 & 1\\
\end{array}\right).
\end{align}
Let us consider an $8$-bit polar code having $k = 4$, $\mathcal{F} = \{1,2,3,5\}$, $u_{\mathcal{F}} = (0\;0\;0\;0)$ and
an information bit sequence $u_{\mathcal{F}_c} = (1\;0\;0\;1)$. Then the encoded output can be computed as follows:
\begin{align}
x_1^8 &= \left(\begin{array}{llllllll}
               0 & 0 & 0 & u_4 & 0 & u_6 & u_7 & u_8
              \end{array}\right) \cdot G_8 \nonumber \\
              &= \left(\begin{array}{llllllll}
               0 & 0 & 0 & 1 & 0 & 0 & 0 & 1
              \end{array}\right) \cdot G_8 \nonumber \\
               &= \left(\begin{array}{llllllll}
               0 & 0 & 0 & 0 & 1 & 1 & 1 & 1
              \end{array}\right).
              \label{eq:ex2}
\end{align}
Please note that $\mathcal{F}_c$ denotes the complementary set of $\mathcal{F}$, which specifies the location of information bits. 
The codeword of \eqr{eq:ex2} may also be directly worked out from the encoding circuit, as exemplified in \fref{example2.fig}.

The encoding process of \eqr{eq:PolarEncGn} may be reformulated as:
\begin{equation}
 x_1^N = u_{\mathcal{F}_c} G_N({\mathcal{F}}_c) + u_{\mathcal{F}} G_N({\mathcal{F}}),
 \label{eq:coset-Gn}
\end{equation}
where $G_N({\mathcal{F}})$ is a submatrix of $G_N$ containing only the rows with indices in $\mathcal{F}$.
When $u_{\mathcal{F}}$ is set to an all-zero bit sequence, \eqr{eq:coset-Gn} reduces to:
\begin{equation}
 x_1^N = u_{\mathcal{F}_c} G_N({\mathcal{F}}_c),
 \label{eq:coset-Gn-2}
\end{equation}
where $G_N({\mathcal{F}}_c)$ is a ($k \times N$) generator matrix. Hence, polar codes are equivalent to linear block codes having
the generator matrix $G_N({\mathcal{F}}_c)$. The encoder of polar codes brings with it the additional benefit of scalability,
both in terms of the codeword length as well as the coding rate. More specifically, the feature of length scalability comes from the recursive
nature of $G_N$, while the rate can be modified by only changing the number of frozen bits. 
It may also be noticed in \eqr{eq:coset-Gn} that when the frozen bits are not set to an all-zero sequence, then the resulting code is a coset
of the linear block code having the generator matrix $G_N({\mathcal{F}}_c)$ and the coset is determined by the 
vector $ u_{\mathcal{F}} G_N({\mathcal{F}})$. 

Polar codes are closely related to the family of \ac{RM} codes~\cite{4542778, arikan2010survey, mondelli2014polar},
since both rely on the encoder $G_N$ of \eqr{eq:arikanGn}. Explicitly, given a pair of integers $0 \leq r \leq n$,
there exists an \ac{RM} code of codeword length $N = 2^n$ and information word length $k = \sum_{i=0}^{r} {n\choose i}$, whose generator matrix 
$G_{\text{RM}}$ is a submatrix of $G_N$ analogous to the polar codes. However, while the generator matrix $G_N({\mathcal{F}}_c)$ of a polar code 
corresponds to the most reliable rows of $G_N$, the generator matrix $G_{\text{RM}}$ of an \ac{RM} code consists of rows of $G_N$ having Hamming
weights $\geq 2^{m-r}$. Equivalently, we may say that we freeze the least reliable channels in polar codes,
while we freeze the the lowest Hamming weight channels in \ac{RM} codes. 
Consequently, polar codes are capacity achieving, while the \ac{RM} codes exhibit a high minimum distance.
It is interesting to point out that the reliability-based selection of frozen bit-channels  
over a \ac{BEC} coincides with the lowest Hamming weight channels for $n=3$ and $n=4$. Hence, the polar code of \eqr{eq:ex2} is equivalent to the $(8,4,4)$ \ac{RM} code. However, the benefits of polar codes begin
to emerge as $n$ increases~\cite{4542778}.
\section{Polar Decoders} \label{sec:decoder}
\begin{figure}[tbp]
\centering
\includegraphics[width=\linewidth]{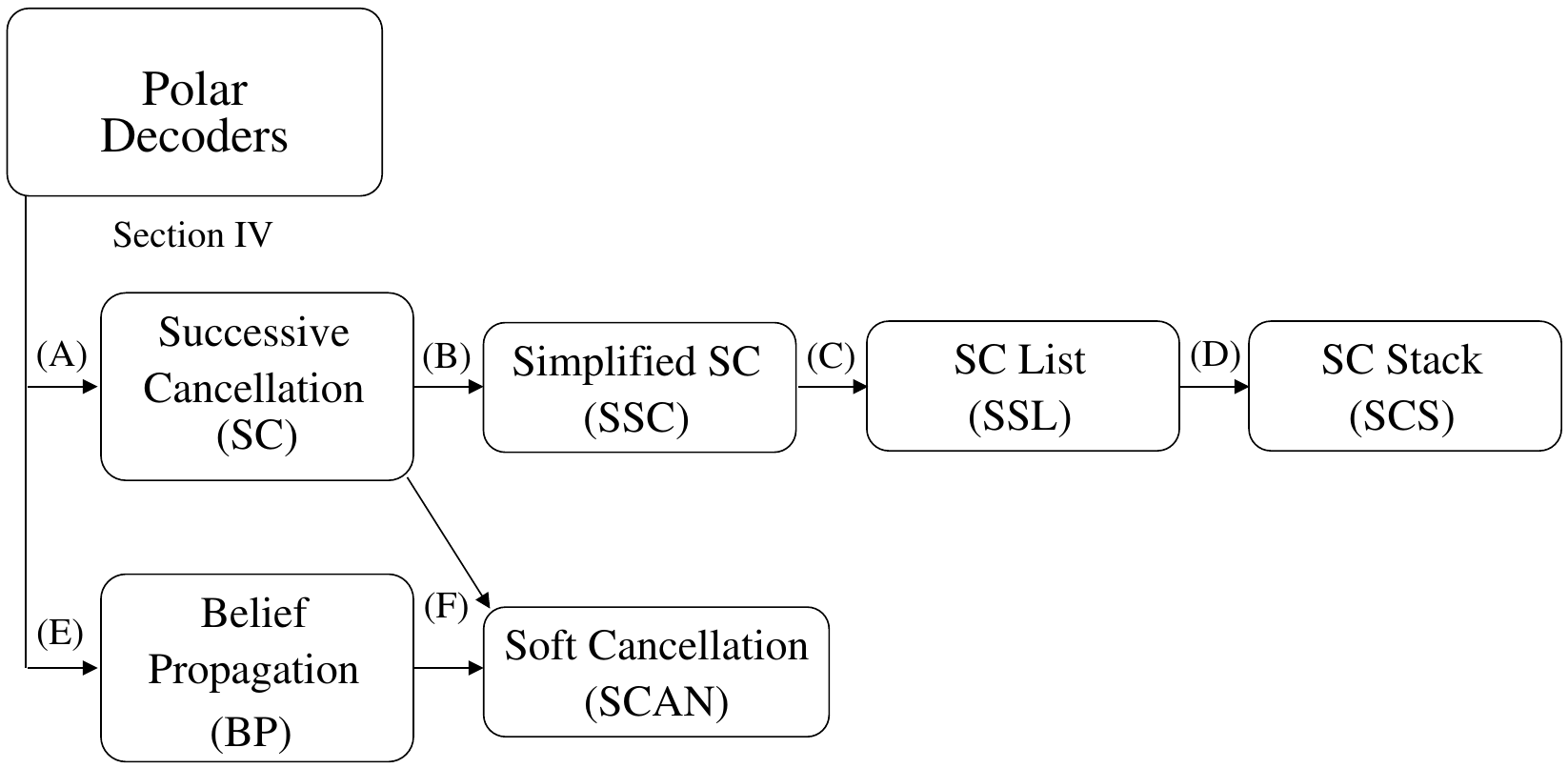}
\caption{Polar decoding algorithms.}
\label{fig:Polar-dec}
\end{figure}

Since the inception of polar codes, intensive research efforts have been invested in improving the polar decoding algorithms from the 
algorithmic perspective as well as from the perspective of hardware implementations. Major contributions in this context are chronologically summarized
in \tref{fig:DecHistory-timeline}.
\begin{table*}[tbp]
\begin{center}
\begin{tabular}[tbp]{R{0.45\linewidth}  p{0.5cm} |@{\food} p{0.45\linewidth}}
\multicolumn{3}{c}{\textbf{Algorithmic Developments} \hfill \textbf{Hardware Implementations}} \\ \\
Belief Propagation (BP) and Successive Cancellation (SC) decoder~\cite{5205856,5075875} &\colorbox{black}{\color{white} \textbf{2009}}&\\
BP decoding improved by exploiting overcomplete factor graphs~\cite{hussami2009performance}&&\\ 
Trellis-based Maximum Likelihood (ML) decoder for short polar codes~\cite{arikan2009performance}&&\\ 
Linear programming based polar decoder for \ac{BEC}~\cite{goela2010lp} &\colorbox{black}{\color{white} \textbf{2010}}     &\\
&&\\
Log Likelihood Ratio (LLR)-based SC decoder~\cite{leroux2011hardware, leroux2012hardware}&\colorbox{black}{\color{white} \textbf{2011}}&Hardware implementations relying on improved scheduling for (LLR)-based SC decoder~\cite{leroux2011hardware, leroux2012hardware}\\
Simplified Successive Cancellation (SSC) decoder~\cite{6065237}&&\\
Successive Cancellation List (SCL) decoder~\cite{6033904,7055304, 6190834}&\\
Successive Cancellation Stack (SCS) decoder~\cite{6215306}&\colorbox{black}{\color{white} \textbf{2012}}     &Pre-computed look-ahead scheduling for reducing the latency of SC decoder~\cite{6364209}\\ 
Cyclic Redundancy Check (CRC)-aided SCL (CA-SCL) and SCS (CA-SCS) decoders~\cite{6297420, 7055304}&&\\
Adaptive CA-SCL decoder\cite{6355936}&&\\
Multistage polar decoder~\cite{6279525} &&\\
ML sphere decoder for short polar codes~\cite{6283643} && \\
Successive Cancellation Hybrid (SCH) decoder~\cite{6560025}&\colorbox{black}{\color{white} \textbf{2013}}&Semi-parallel architecture for SC decoder~\cite{6327689} \\
Maximum Likelihood Simplified Successive Cancellation (ML-SSC) decoder~\cite{6464502}&&Scalable semi-parallel implementation of SC decoder~\cite{6737143, 6876199}\\
Soft CANcellation (SCAN) decoder~\cite{fayyaz2013polar, fayyaz2014low}&&Field Programmable Gate Array (FPGA) implementation of BP decoder~\cite{pamuk2011fpga}\\
&& Two-phase SC decoder architecture having a lower complexity and memory utilization, and a higher throughput~\cite{6620368}\\
&&Overlapped SC decoder architecture~\cite{6475198}\\
ML-SSC improved~\cite{6804939} &\colorbox{black}{\color{white} \textbf{2014}}&Flexible and high throughput architecture of improved ML-SSC decoder~\cite{6804939}\\
Modified BP~\cite{6785962, 6864206}&&Hardware implementation of SCL decoder~\cite{6823099}\\
Early terminated BP decoder~\cite{6942260}: &&Efficient partial-sum network architecture for semi-parallel SC decoder~\cite{6803952}\\
Low-latency CA-SCL decoder~\cite{6986062}&&Architecture of SSC decoder~\cite{6680761}\\
Symbol-based SC and SCL decoders~\cite{6986086,7289453}&&Architecture of $2$-bit SC decoder~\cite{6632947}\\
SC flip decoder~\cite{7094848}&&\\
Error exponent of SCL investigated~\cite{7203252}&\colorbox{black}{\color{white} \textbf{2015}}     &\\
LLR-based SCL decoder~\cite{7114328}&&Hardware architecture of LLR-based SCL decoder~\cite{7114328}\\
&&Architecture of a low-latency multi-bit SCL decoder~\cite{6920050}\\
&&Metric sorter architecture for SCL decoder~\cite{7169066}\\
&&Low-latency SCL decoder relying on double thresholding based list pruning~\cite{7178128, 7339658}\\
Reduced-complexity early terminated BP decoder~\cite{7491324}&\colorbox{black}{\color{white} \textbf{2016}}&Hardware Architecture of low-latency CA-SCL decoder~\cite{7337462} relying on~\cite{6986062}\\
The reduced latency ideas of~\cite{6804939} extended to the SCL decoder~\cite{7339671}&&Implementation of adaptive throughput-area efficient SCL decoder invoking approximate ML decoding components~\cite{7468577}\\
Tree pruning for low-latency SCL decoding~\cite{7348682}&&Sphere decoding based architecture of SCL decoder~\cite{7742998}\\
Reduced-complexity SCL decoder~\cite{7339660} &&Improved metric sorter architecture for SCL decoder~\cite{7407353}\\
Simplified Successive Cancellation List (SSCL) decoder avoiding redundant calculations~\cite{7541412}&&Hardware implementation of CA-SCL based on distributed sorting~\cite{7841865} \\
Parity-check-aided polar decoding~\cite{wang2016parity}&& \\
Reduced latency SSCL decoder~\cite{8010835} &\colorbox{black}{\color{white} \textbf{2017}}&Hardware implementation of SSCL decoder~\cite{8010835}\\
Unsorted SCL decoder~\cite{7867834}&&Hardware implementation of low-latency BP decoder~\cite{7742915}\\ 
Syndrome-based SC decoder~\cite{7990117}&&Two-step metric sorter architecture for parallel SCL decoder~\cite{7756324}\\
Soft SCL decoder for systematic codes~\cite{liu2017parallel,zhou2018performance}&&Memory efficient architectures for SC and SCL decoders~\cite{8070938}\\
Reduced-latency SSC flip decoder~\cite{8369026}&\colorbox{black}{\color{white} \textbf{2018}}     &\\
BP list decoder~\cite{elkelesh2018belief}&&Hardware architecture of multi-bit double thresholding SCL decoder relying on pre-computed look-ahead scheduling~\cite{8361464}\\
 \end{tabular}
  \end{center}
   \caption{Major contributions to the polar decoding paradigm.} 
\label{fig:DecHistory-timeline}
\end{table*}
In this section, we will review the achievements of \tref{fig:DecHistory-timeline} with an emphasis on the major decoding algorithms
identified in \fref{fig:Polar-dec}. 
\subsection{Successive Cancellation (SC) Decoder} \label{sec:sec:SC}
Arikan's seminal paper~\cite{5075875} proposed a \ac{LR} based \ac{SC} polar decoder, which was later modified by 
Leroux~\etal~\cite{leroux2011hardware, leroux2012hardware} to carry out operations in the logarithmic domain; hence reducing the associated
computational complexity. 
Recall from \fref{polar-split-gen} that the compound channel $W^N$ may be split into $N$ polarized bit-channels such that the $i$th bit-channel
$W_i$ takes the input $u_i$ and yields the output $(y_1^N,u_i^{i-1})$. The associated channel transition probabilities 
are denoted by $P_i(y_1^N,u_1^{i-1}|u_i)$, which may be estimated using an \ac{SC} decoder. To elaborate, the channel combining process of
\fref{polar-comb-gen} couples together the input bits $u_i^N$. So, an \ac{SC} decoder reverses this process at the receiver by 
removing the contribution, or more precisely interference, of the bits $u_1^{i-1}$ from the received coded bits $y_1^N$,
hence revealing the value of the $i$th bit $u_i$.

\begin{figure}[tbp]
        \centering
        \includegraphics[width=\linewidth]{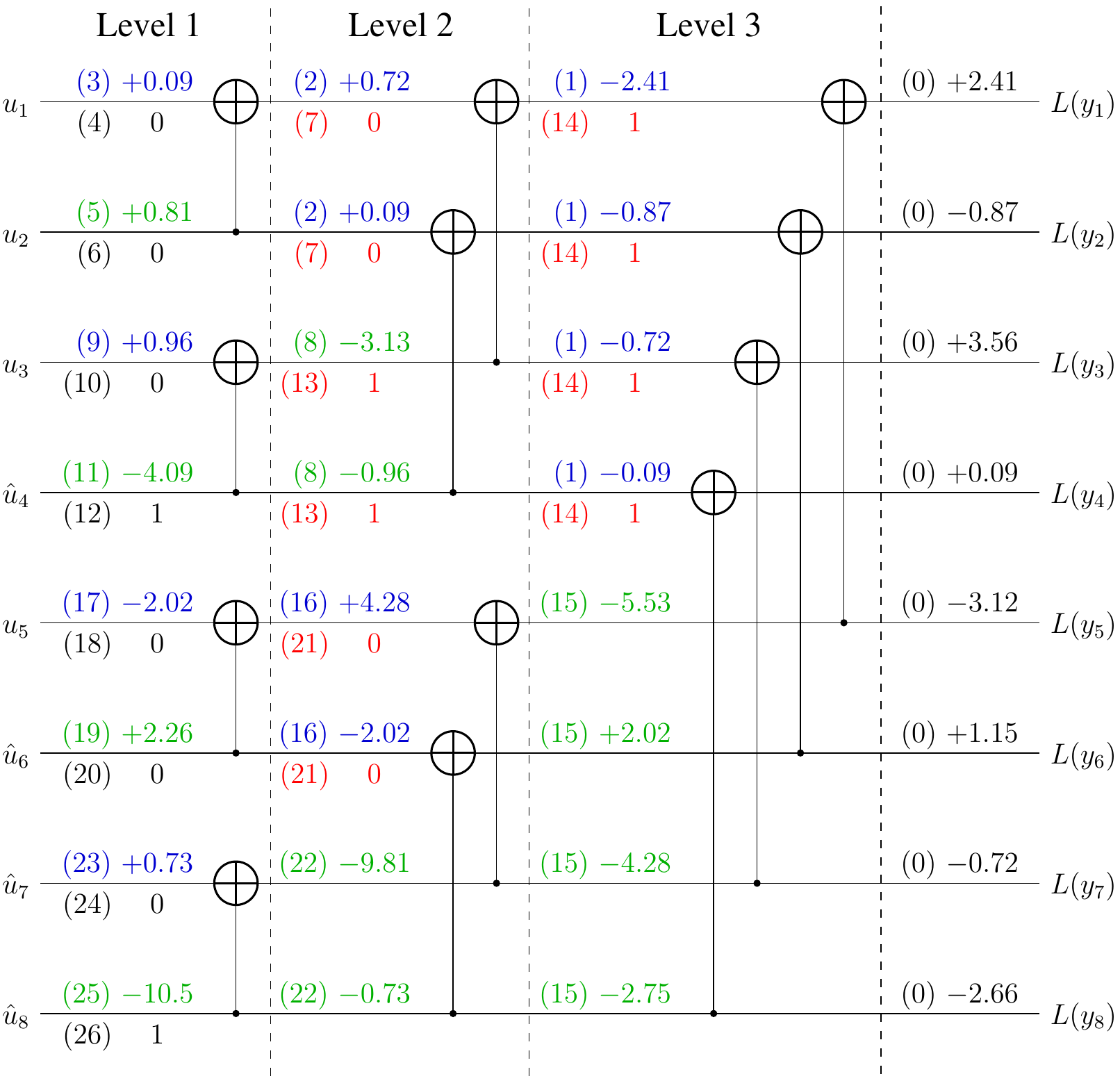}
        \caption{Example of \ac{SC} decoding process: An $N=8$ polar code having $k = 4$, $\mathcal{F} = \{1,2,3,5\}$ and 
$u_{\mathcal{F}} = (0\;0\;0\;0)$ is used to decode the received encoded \acp{LLR} $L(y_i)$ into the $k=4$ recovered information bits 
$\hat{u}_{\mathcal{F}_c} = (1\;0\;0\;1)$. The \acp{LLR} obtained using the $f$ and $g$ functions of \eqr{eq:f} and \eqr{g.eqn}
are shown above each connection in blue and green, respectively. The bits obtained using the partial sum computations of \eqr{PS1.eqn} and 
\eqr{PS2.eqn} are shown below each connection in red. The accompanying numbers in parenthesis identify the step of the \ac{SC} decoding process where the corresponding \ac{LLR} 
or bit becomes available.}
         \label{example.fig}
\end{figure}

An \ac{SC} decoder operates on the same circuit as that of the encoder, as exemplified in \fref{example.fig} for the polar encoder 
of \fref{example2.fig}. 
However, while an encoder always processes the bits from left to right,
an \ac{SC} decoder operates from right to left as well as from left to right. To elaborate,
an \ac{SC} decoder performs computations pertaining to the \acp{XOR} in the circuit according to a sequence that is 
dictated by the availability of data on the left and right hand side of the \ac{XOR}, which introduces data dependencies in the decoding process. 
Hence, the functionality of each \ac{XOR} in the decoding circuit varies, when performing operations on \acp{LLR} at different steps in the 
\ac{SC} decoding process\footnote{The \ac{LLR} $L(b)$ pertaining to the bit $b$ is defined as:
\begin{equation}
 L(b) = \log \left(\frac{P(b=0)}{P(b=1)}\right),
 \nonumber
\end{equation}
where $P(.)$ denotes the probability of occurrence.}. There are three types of computations that can be performed by a particular \ac{XOR} in the decoding circuit, 
depending on the availability of \acp{LLR} provided on the connections on its right-hand side, as well as upon the 
availability of bits provided on the connections on its left-hand side.
Let us exemplify this by considering the $2$-bit elementary kernel of \fref{example.fig}, 
which operates on the $i$th and $(i+2^{j-1})$th bits, where $j \in [1, n]$ 
denotes the level index.
\begin{figure}[tbp]
        \centering
        \begin{subfigure}[]{\linewidth}
        \begin{center}
                \includegraphics[width=0.5\linewidth]{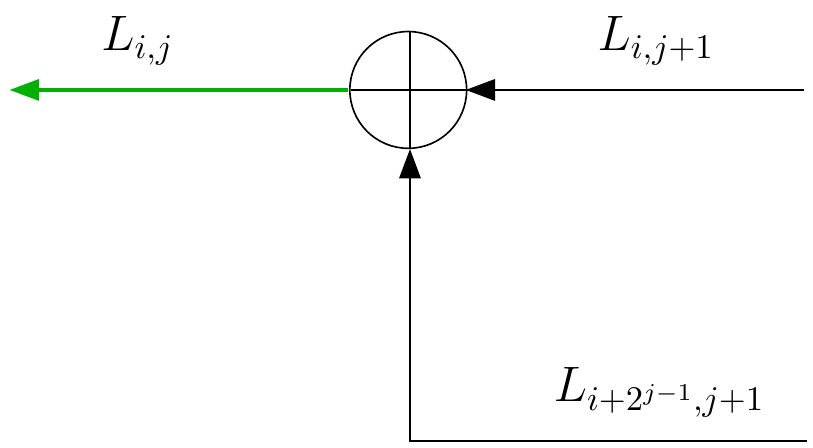}
                \caption{Function $f\left(L_{i,j+1}, L_{i+2^{j-1},j+1}\right)$: \acp{LLR} propagate from right-to-left.}
                \label{fig:SC-1}
        \end{center}
        \end{subfigure}%
        ~ \vfill 
        \begin{subfigure}[]{\linewidth}
        \begin{center}
                \includegraphics[width=0.5\linewidth]{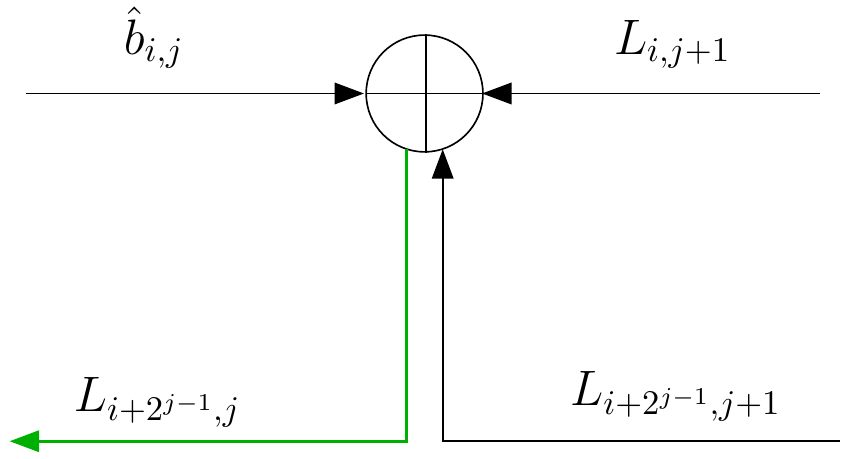}
                \caption{Function $g(L_{i,j+1}, L_{i+2^{j-1},j+1}, \hat{b}_{i,j})$: switch from propagating bits to
propagating \acp{LLR}.}
                \label{fig:SC-2}
        \end{center}
        \end{subfigure}
        ~ \vfill
                \begin{subfigure}[]{\linewidth}
        \begin{center}
                \includegraphics[width=0.5\linewidth]{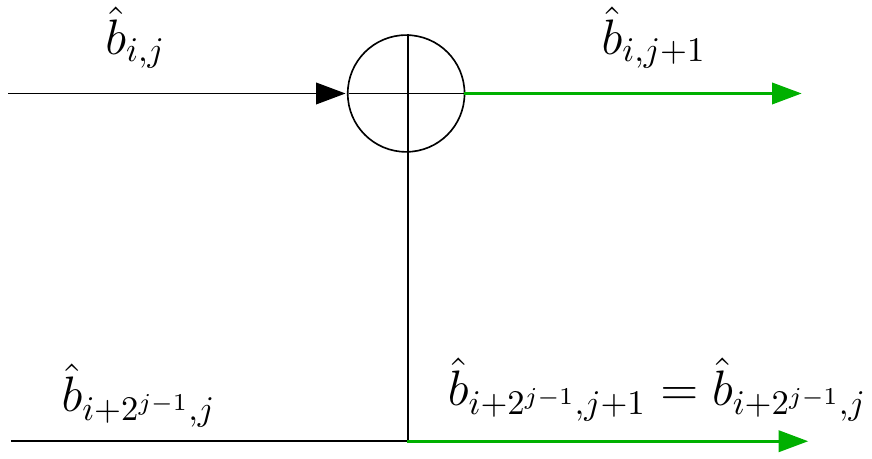}
                \caption{Partial sum calculation $\text{XOR}\left(\hat{u}_1,\hat{u}_2\right)$: bits propagate from left-to-right.}
                \label{fig:SC-3}
        \end{center}
        \end{subfigure}
        \caption{The three computations that can be performed for an \ac{XOR} in the polar decoder. $L_{i,j}$ is the \ac{LLR}
        pertaining to the bit $b_{i,j}$, while $\hat{b}_{i,j}$ is the estimation of bit $b_{i,j}$.}
         \label{fig:SC-123}
\end{figure}

The first occasion when an \ac{XOR} can contribute to the \ac{SC} decoding process is when an \ac{LLR} has been provided by each of the 
connections on its right-hand side, as shown in \fref{fig:SC-1}. Since the \ac{XOR} connects the $i$th and $(i+2^{j-1})$th bits,
we refer to the first and second of these two \acp{LLR} as $L_{i,j+1}$ and $L_{i+2^{j-1},j+1}$, respectively. More specifically, 
$L_{i,j+1}$ and $L_{i+2^{j-1},j+1}$ provide soft-information pertaining to the bits $b_{i,j+1}$ and
$b_{i+2^{j-1},j+1}$, respectively. 
These \acp{LLR} may be generated either by the soft demodulator (for $j=n$) or by the other \acp{XOR} in the circuit (for $j<n$).
Based on the input $L_{i,j+1}$ and $L_{i+2^{j-1},j+1}$, the \ac{XOR} of \fref{fig:SC-1} computes the \ac{LLR} $L_{i,j}$ 
for the first of the two connections on its left-hand side, as follows:
\begin{align}
L_{i,j} &= f\left(L_{i,j+1}, L_{i+2^{j-1},j+1}\right)\nonumber\\
&= L_{i,j+1} \boxplus L_{i+2^{j-1},j+1}, 
\label{eq:f}
\end{align}
where the box-plus operator is defined as~\cite{485714}:
\begin{align}
 &L(b_1) \boxplus L(b_2) \nonumber \\ 
 &= L(b_1 \oplus b_2) \nonumber\\ 
 &= \ln \frac{1 + e^{L(b_1)}e^{L(b_2)}}{e^{L(b_1)} + e^{L(b_2)}}  \nonumber \\
 &= 2\tanh^{-1}(\tanh(L(b_1)/2)\tanh(L(b_2)/2)) \label{boxplus.eqn}\\
&=\mathrm{sign}(L(b_1))\mathrm{sign}(L(b_2))\min(|L(b_1)|,|L(b_2)|) \nonumber \\
&+ \log\left(1+e^{-|L(b_1)+L(b_2)|}\right) - \log\left(1+e^{-|L(b_1)-L(b_2)|}\right) \nonumber \\
& \approx \mathrm{sign}(L(b_1))\mathrm{sign}(L(b_2))\min(|L(b_1)|,|L(b_2)|).\label{min.eqn}
\end{align}
Here, $L(b_1)$ and $L(b_2)$ are the \acp{LLR} pertaining to the bits $b_1$ and $b_2$, respectively.
The $\mathrm{sign}(\cdot)$ of \eqr{min.eqn} returns $-1$ if its argument is negative and $+1$ if its argument if positive. 
Here, \eqr{min.eqn} is referred to as the min-sum approximation.

Later in the \ac{SC} decoding process, the estimated bit $\hat{b}_{i,j}$ is provided on the first of the connections on the left-hand side of 
the \ac{XOR}, as shown in \fref{fig:SC-2}. Together with the \acp{LLR} $L_{i,j+1}$ and $L_{i+2^{j-1},j+1}$ that were previously provided using the 
connections on the right-hand side, 
this enables the \ac{XOR} to compute the \ac{LLR} $L_{i+2^{j-1},j}$ for the second of the two connections on its left-hand side, 
according to the $g$ function as follows:
\begin{align}
L_{i+2^{j-1},j} &= g(L_{i,j+1}, L_{i+2^{j-1},j+1}, \hat{b}_{i,j})\nonumber\\
&= (-1)^{\hat{b}_{i,j}}L_{i,j+1}+L_{i+2^{j-1},j+1}.\label{g.eqn}
\end{align}
We may observe in \eqr{g.eqn} that the $g$ function is analogous to the decoding operation of a repetition node, since the
two \ac{LLR} values are summed together. This is because the information pertaining to the bit $b_{i+2^{j-1},j}$ is contained in  
$L_{i+2^{j-1},j+1}$ as well as $L_{i,j+1}$. Furthermore, the sign of $L_{i,j+1}$ is flipped when $\hat{b}_{i,j} = 1$, since we have 
$b_{i+2^{j-1},j} = b_{i,j+1} \oplus b_{i,j}$. 

Later still, the bit $\hat{b}_{i+2^{j-1},j}$ will be provided on the second of the connections on the left-hand side of the \ac{XOR}, as 
shown in \fref{fig:SC-3}. Together with the bit $\hat{b}_{i,j}$ that was previously provided using the first  
of the connections on the left-hand side, this enables the partial sum computation of bits $\hat{b}_{i,j+1}$ and  $\hat{b}_{i+2^{j-1},j+1}$ 
for the first and second connections on the right-hand side of the \ac{XOR}, where
\begin{align}
\hat{b}_{i,j+1} &= \mathrm{XOR}(\hat{b}_{i,j}, \hat{b}_{i+2^{j-1},j}),\label{PS1.eqn}\\
\hat{b}_{i+2^{j-1},j+1} &= \hat{b}_{i+2^{j-1},j}.\label{PS2.eqn}
\end{align}

As may be appreciated from the discussions above, the $f$ function of \eqr{eq:f} may be used to propagate \acp{LLR} from right-to-left 
within the \ac{SC} decoder, while the partial sum computations of \eqr{PS1.eqn} and \eqr{PS2.eqn} may be used to propagate bits 
from left-to-right and the $g$ function of \eqr{g.eqn} may be used to switch from propagating bits (from left-to-right) 
to propagating \acp{LLR} (from right-to-left). 
The \ac{SC} decoding process begins by processing \acp{LLR} from right to left. However, in order that \acp{LLR} can be propagated from right to left, it is 
necessary to provide \acp{LLR} on the connections on the right-hand edge of the circuit, i.e. right-hand connections at level $3$ of \fref{example.fig}. 
In the example of \fref{example.fig}, this is performed at the start of the \ac{SC} 
decoding process by providing successive \acp{LLR} from a soft demodulator on successive connections on the right-hand edge of the circuit. 
We may also call them channel \acp{LLR}, since they provide soft information pertaining to the channel outputs.
The \ac{SC} decoding process then begins by using the $f$ function of \eqr{eq:f} to propagate \acp{LLR} 
from the right hand edge of the decoding circuit to the top connection on the left-hand edge, allowing the first bit 
to be recovered (steps (0) to (4) in \fref{example.fig}). Explicitly, if the first bit is an information bit, then 
a hard-decision is made based on the resulting \ac{LLR} $L_{1,1}$. By contrast, if the first bit is a frozen bit, then it is set 
equivalent to the known frozen bit.
Then the $g$ function of \eqr{g.eqn} is used to compute the \ac{LLR} pertaining
to the second bit, hence revealing its value (steps (5) and (6) in \fref{example.fig}).
Following this, each successive bit from top to bottom is recovered by using the partial sum computations 
of \eqr{PS1.eqn} and \eqr{PS2.eqn} to propagate bits from left to right, then using the $g$ function of \eqr{g.eqn} for a 
particular \ac{XOR} to switch from bit propagation to \ac{LLR} propagation, before using the $f$ function to propagate \acp{LLR} to the
next connection on the left-hand edge of the circuit, allowing the corresponding bit to be recovered.
It is pertinent to mention here that if bit on the left-hand edge is a frozen bit, then the associated \ac{LLR} is ignored and the value
of the bit is set to the known value of the frozen bit. 
The complexity of this decoding process is $O(N\log_2N)$, since there are  
$n = \log_2N$ levels and each level invokes $N/2$ \ac{XOR} gates. Furthermore, a straightforward implementation of the \ac{SC} decoder incurs
a space complexity $O(N\log_2N)$, which was reduced to $O(N)$ in~\cite{7055304} by exploiting the recursive nature of polar codes
in conjunction  with `lazy-copy' algorithmic techniques.

\begin{figure}[!t]
        \centering
                \includegraphics[width=\linewidth]{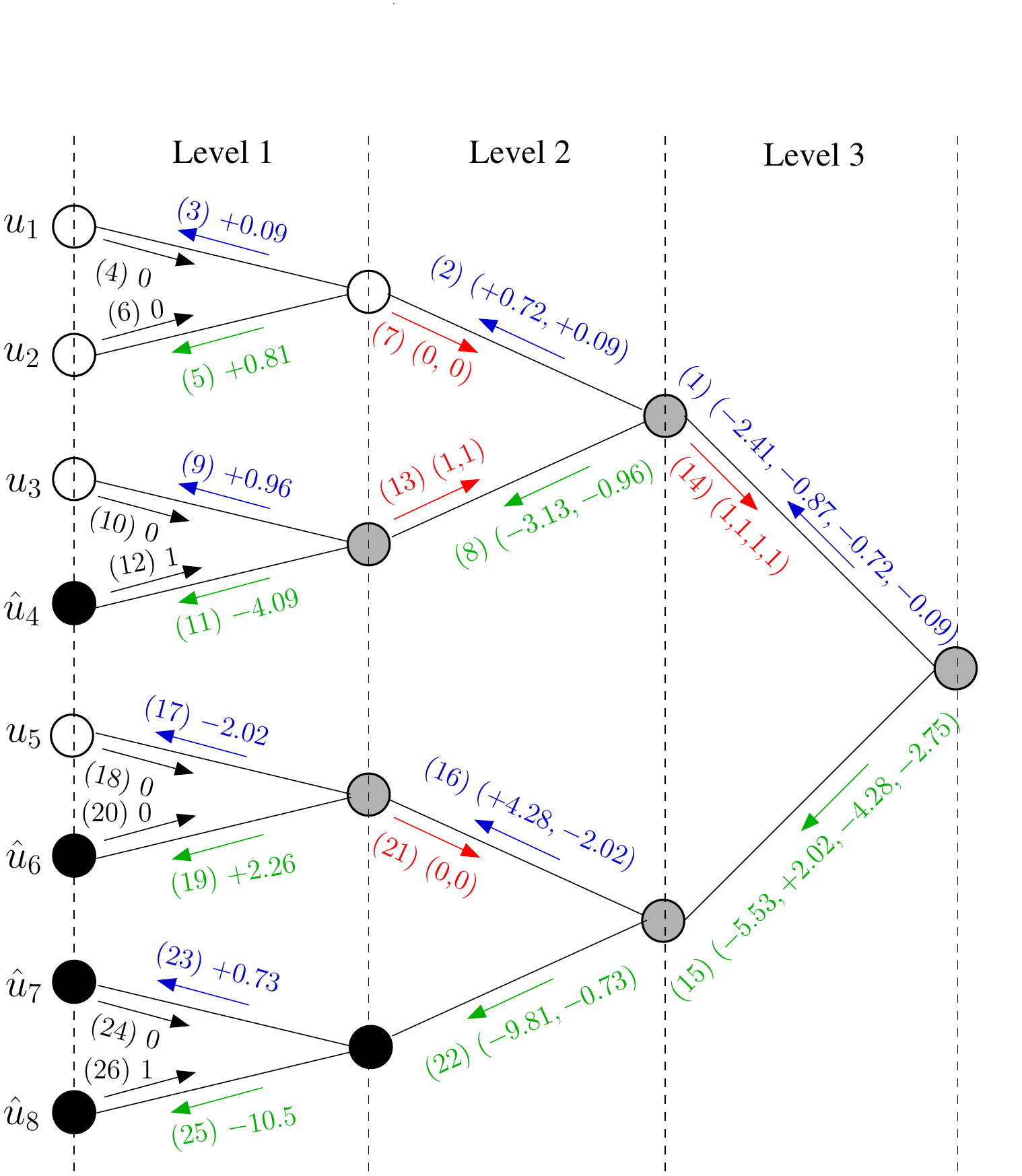}
        \caption{\ac{SC} decoding tree for the \ac{SC} decoding circuit of \fref{example.fig}.
        Rate-$0$ and Rate-$1$ nodes are shown in white and black color, respectively.
The \acp{LLR} obtained using the $f$ and $g$ functions of \eqr{eq:f} and \eqr{g.eqn}
are shown in blue and green, respectively, while the bits obtained using the partial sum computations of \eqr{PS1.eqn} and 
\eqr{PS2.eqn} are shown in red. The accompanying numbers in parenthesis identify the step of the \ac{SC} decoding process 
where the corresponding \ac{LLR} or bit becomes available. }
         \label{exampleTree.fig}
\end{figure}

The \ac{SC} decoding process of \fref{example.fig} may also be visualized over a decoding tree, as shown in \fref{exampleTree.fig}.
The decoding tree of \fref{exampleTree.fig} consists of $n = 3$ levels and each level is composed of $2^{n-i}$ parent nodes and $2^{n-i+1}$ child nodes; hence resulting in $2^{3} = 8$ leaf nodes at level $1$, which correspond to the bits $u$. To elaborate, the number of child nodes at each level corresponds to the number of distinct polarized channels created at that level. Recall from Section~\ref{sec:ch-polarization} that we get two types of polarized channels $W^{-}$ and $W^{+}$ at level $3$, which are further polarized into $W^{--}$, $W^{-+}$, $W^{+-}$ and $W^{++}$ at level $2$ and then into eight types at level $1$. This results in $2^{n-i+1}$ types of polarized channels at each level. Furthermore, the decoding tree at level $3$ starts with a length $N$ parent node, whose length reduces by half at every child node, hence adopting a recursive divide-and-conquer approach.

Let us now elaborate on the flow of \acp{LLR} and bits through the decoding tree of \fref{exampleTree.fig}, where each node acts as a local decoder executing the \ac{XOR} operations of \fref{fig:SC-123}. The \ac{SC} decoding process begins with the parent node at Level $3$, which has the channel \acp{LLR} from the soft demodulator. This parent node generates \acp{LLR} for the top child node using the function $f$ of \fref{fig:SC-1}. It then waits until it receives the hard-decoded bits from the top child node and transfers the parental control to the top child node, which now acts as the next parent node. This process continues in a recursive manner until we reach the leaf nodes at level $1$. At this point, hard decision is made pertaining to the uncoded bits $u$ and sent to the parent node at level $1$. On receiving the hard-decoded bits, the parent node at level $1$ calculates the \acp{LLR} for the bottom child node using the $g$ function of \fref{fig:SC-2} and waits for the hard-decoded bits from the bottom child node. On receiving the bits, the parent node at level $1$ executes the \ac{XOR} operation of \fref{fig:SC-3} and sends the resulting bits to its parent node at Level $2$. The process continues recursively until all the uncoded bits $u$ have been recovered.

\ac{SC} decoders are favored for having a low decoding complexity. However, this is achieved at the cost of a high decoding latency,
since there are several data dependencies associated with the \ac{SC} decoding process. Explicitly, the $f$ operations have to wait for the \acp{LLR} to be made available on 
their right-hand connections, while the $g$ operations have to wait for the availability of the estimated bit values on their 
left-hand connections in addition to availability of the \acp{LLR} on their right-hand connections. Similarly, 
it is necessary to provide bits on the left-hand in order to facilitate the propagation of bits from left to right. 
Hence, owing to these data dependencies, the information bits on the left-hand edge of the circuit are serially recovered from top to 
bottom. This in turn makes the hardware implementation of \ac{SC} decoders challenging. More specifically, 
the data dependencies allow different numbers of operations to be 
completed in parallel at different times, as illustrated in the example of \fref{example.fig}. In order to minimise the 
number of steps required to complete the decoding process, a large amount of hardware may used so that a single processing step is 
sufficient to complete the largest number of parallel operations that are supported by the decoder data dependencies. However, the 
data dependencies will prevent much of this hardware from being used throughout the rest of the decoding process, which may motivate 
the use of a smaller amount of hardware and a greater number of steps. However, either way, the ratio of hardware resource usage to the 
latency required to complete the decoding process may be unfavourable, unless sophisticated alternative techniques can be developed and 
utilised. In this context, Leroux~\etal~\cite{leroux2011hardware, leroux2012hardware} proposed hardware architectures for \ac{SC} decoders, 
which rely on improved scheduling for enhancing resource sharing and memory management. In~\cite{6364209}, pre-computed look-ahead techniques
were invoked for reducing the latency of \ac{SC} decoding process, while semi-parallel implementations of SC decoder
were presented in~\cite{6327689, 6737143, 6876199}. A two-phase 
\ac{SC} decoder architecture was conceived in~\cite{6620368}, which exhibits a lower complexity, memory utilization and latency,
while an overlapped \ac{SC} decoder architecture was presented in~\cite{6475198} for the sake of reducing the latency.
Furthermore, Fan~\etal~\cite{6803952} developed an efficient partial-sum network architecture for semi-parallel \ac{SC} decoder.
In the spirit of further reducing the latency, Yuan~\etal~\cite{6632947} proposed a $2$-bit decoding architecture for \ac{SC} decoders, which concurrently processes two bits during the last stage of the \ac{SC} decoding process. Look-ahead techniques were also invoked in~\cite{6632947} and recently in~\cite{8268893}, while memory-efficient hardware implementations were presented in~\cite{8070938, 8110014}.

The main characteristics of an \ac{SC} polar decoder are summarized in \tref{tab:SC-summary}.
\begin{table}[tbp]
 \centering
 \begin{center}
\begin{tabular}[tbp]{|l|l|}
\hline 
&\\
\textbf{Complexity} & \tabitem Time complexity = $O(N\log_2N)$\\
                    & \tabitem Space complexity = $O(N)$\\ &\\ \hline
& \\
\textbf{Advantages} & \tabitem Low decoding complexity \\
& \tabitem Asymptotically capacity achieving \\ &\\ \hline
&\\
\textbf{Disadvantages} & \tabitem Sub-optimal finite-length performance \\
& \tabitem Serial processing, resulting in high latency (or low throughput) \\
& \tabitem Fully-parallel implementation not feasible \\ &\\ \hline
\end{tabular}
 \end{center}
 \caption{Main characteristics of an \ac{SC} polar decoder.}
\label{tab:SC-summary}
\end{table}

\subsection{Simplified Successive Cancellation Decoder} \label{sec:sec:SSC}
The \ac{SC} decoding process of Section~\ref{sec:sec:SC} consists of some redundant calculations, which may be discarded without
compromising the \ac{BER} or \ac{BLER} performance. Based on this notion, \ac{SSC} decoder was conceived in~\cite{6065237}. As compared to
the classic \ac{SC} decoder, the \ac{SSC} decoder provides significant reduction in the computational complexity as well as the latency, while maintaining the same 
\ac{BER} or \ac{BLER} performance. Quantitatively, it was demonstrated in~\cite{6065237} that the \ac{SSC} decoder
reduces the number of computationally intensive box-plus operator of \eqr{boxplus.eqn} by around $20\%$ to $50\%$, while the decoding latency
is reduced by around $75\%$ to $95\%$.  

The \ac{SSC} decoder exploits the tree structure of \fref{exampleTree.fig} for discarding redundant computations. Explicitly, the nodes of \fref{exampleTree.fig} may be classified into three types, as follows:
\begin{enumerate}
 \item Rate-$0$ nodes, whose descendants are all frozen bits (white nodes in \fref{exampleTree.fig});
 \item Rate-$1$ nodes, whose descendants are all information bits (black nodes in \fref{exampleTree.fig});
 \item Rate-$R$ nodes, whose descendants are a mix of frozen bits and information bits (gray nodes in \fref{exampleTree.fig}).
\end{enumerate}
Recall that the value of frozen bits may be directly used at the decoder rather than estimating it based on the computed \acp{LLR}. 
Consequently, the \ac{SSC} decoder de-activates the $f$ and $g$ operations at Rate-$0$ nodes of \fref{exampleTree.fig} without affecting the decoder's performance. Explicitly, $u_1$ and $u_2$ are initialized to the known value of frozen bits, i.e. $0$, in the \ac{SSC} decoder, and steps ($3$) to ($6$) are not required.
Similarly, the \ac{SSC} decoder further reduces the complexity at the Rate-$1$ nodes by making a hard-decision based on the input \acp{LLR} and passing the hard-decoded bits to the child nodes, rather than computing the $f$ and $g$ functions. In the context of \fref{exampleTree.fig}, an \ac{SSC} decoder makes hard-decision based on the \acp{LLR} computed in step ($22$) revealing $\hat{u_8}$, discards steps ($23$) to ($26$), and estimates $\hat{u_7}$ based on the hard-decision values of step ($22$). Since all nodes connected to the Rate-$1$ nodes are information bits, this simplification does not affect the performance of the decoder. Improved versions of the \ac{SSC} decoder were proposed in~\cite{6464502, 6804939} for the sake of further reducing the associated
latencies, while the hardware architectures of \ac{SSC} decoder were presented in~\cite{6804939, 6680761}.

The main characteristics of an \ac{SSC} polar decoder are summarized in \tref{tab:SSC-summary}.
\begin{table}[tbp]
 \centering
 \begin{center}
\begin{tabular}[tbp]{|l|l|}
\hline 
&\\
\textbf{Complexity} & Depends on the underlying polar code \\ &\\ \hline
& \\
\textbf{Advantages} & \tabitem Lower complexity and latency than \ac{SC} \\
& \tabitem Asymptotically capacity achieving \\ &\\ \hline
&\\
\textbf{Disadvantages} & \tabitem Sub-optimal finite-length performance \\
& \tabitem Serial processing, resulting in high latency (or low throughput) \\
& \tabitem Fully-parallel implementation not feasible \\ &\\ \hline
\end{tabular}
 \end{center}
 \caption{Main characteristics of an \ac{SSC} polar decoder.}
\label{tab:SSC-summary}
\end{table}
\subsection{Successive Cancellation List (SCL) Decoder} \label{sec:sec:SCL}
\begin{figure*}[tb]
        \centering
        \begin{subfigure}[]{0.5\linewidth}
        \begin{center}
                \includegraphics[width=\linewidth]{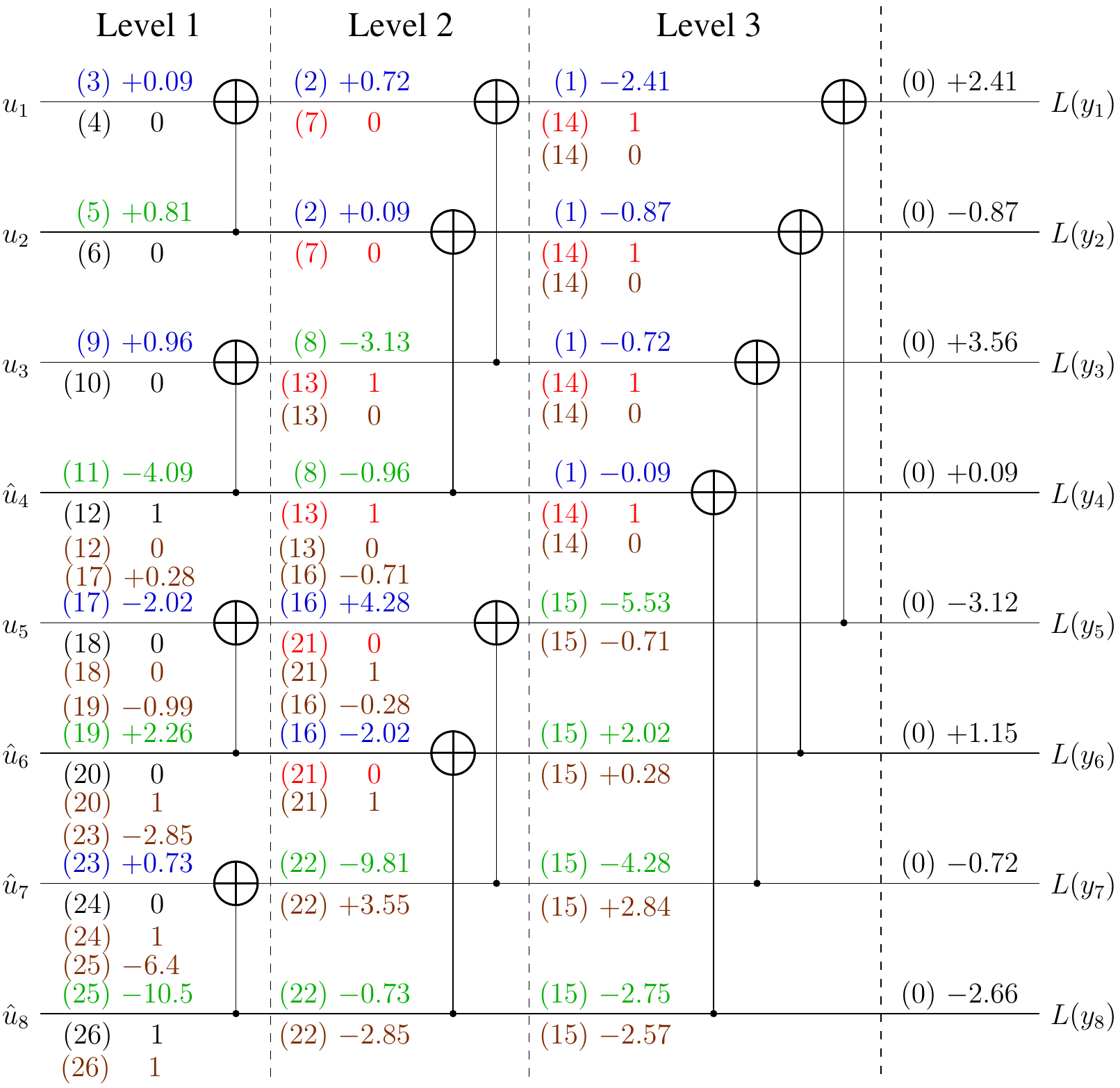}
                \caption{\ac{SCL} ($L=2$) decoding circuit. For the first candidate path, the \acp{LLR} obtained using the $f$ and $g$ functions of \eqr{eq:f} and \eqr{g.eqn}
are shown above each connection in blue and green, respectively, while the bits obtained using the partial sum computations of 
\eqr{PS1.eqn} and \eqr{PS2.eqn} are shown below each connection in red. All the \acp{LLR} and bits pertaining to the second candidate path
are shown in brown. The accompanying numbers in parenthesis identify the step of the \ac{SC} decoding process where the corresponding \ac{LLR} 
or bit becomes available.}
                \label{SCLa.fig}
        \end{center}
        \end{subfigure}%
        ~ 
        \begin{subfigure}[]{0.5\linewidth}
        \begin{center}
                \includegraphics[width=0.9\linewidth]{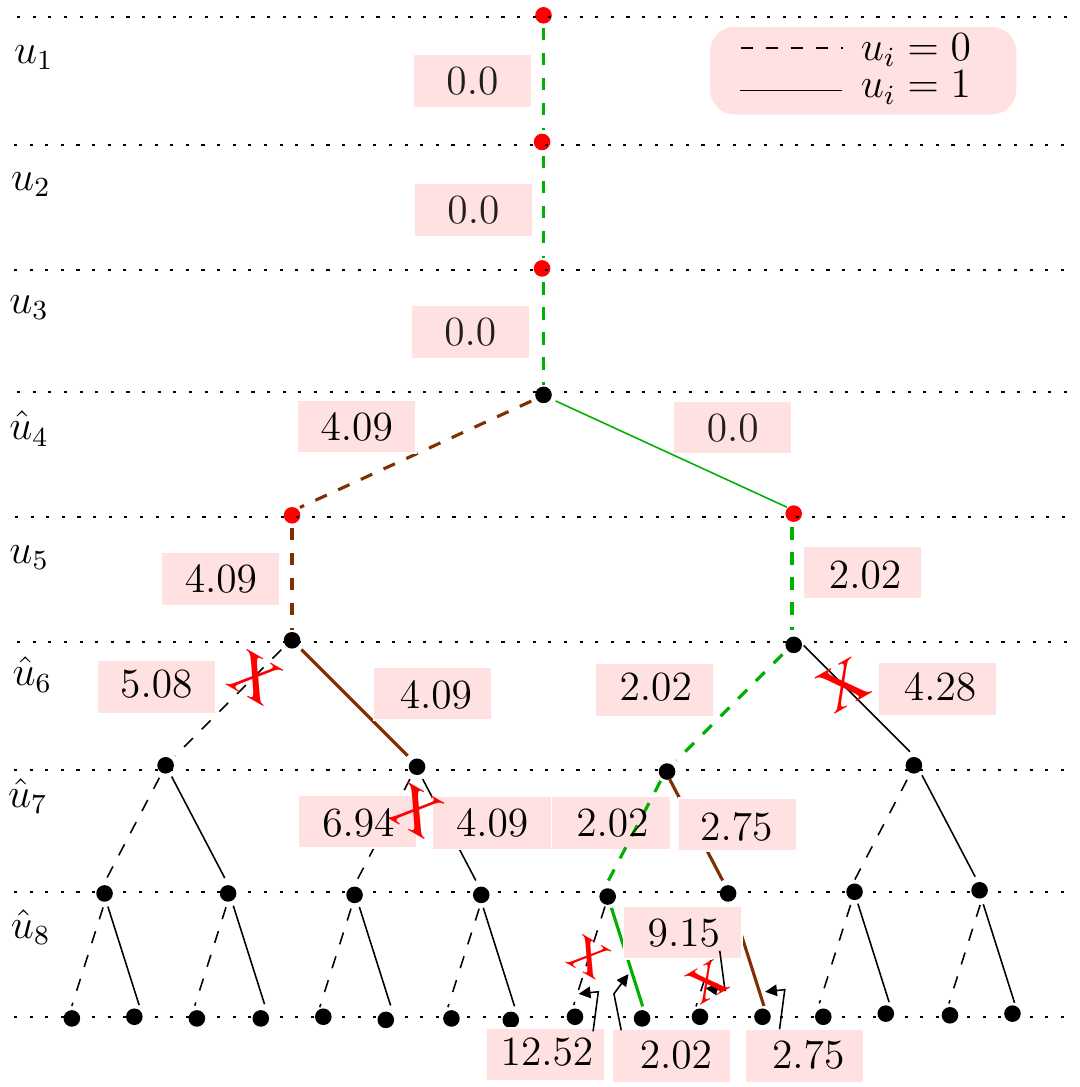}
                \caption{\ac{SCL} decoding over a binary tree. Frozen and information nodes are marked in red and black, respectively, 
                the two candidate paths are shown in green and brown, and the pruned paths
                are marked with a red cross. The values next to the branches are the associated path metrics computed using \eqr{metric.eqn}.
                Please note that the green path is same as the SC decoding path.}
                \label{SCLb.fig}
        \end{center}
        \end{subfigure}
        \caption{Example of \ac{SCL} decoding process ($L = 2$): An $N=8$ polar code having $k = 4$, $\mathcal{F} = \{1,2,3,5\}$ and 
$u_{\mathcal{F}} = (0\;0\;0\;0)$ is used to decode the received encoded \acp{LLR} $L(y_i)$ into the $k=4$ recovered information bits 
$\hat{u}_{\mathcal{F}_c} = (1\;0\;0\;1)$. }
         \label{SCLdecodingexample.fig}
\end{figure*}
The \ac{SC} decoder provably achieves the capacity of a \ac{B-DMC}, when the codeword length is infinitely long. However, it 
does not exhibit good performance at finite codeword lengths, because the channels are not sufficiently polarized. More specifically,
in the \ac{SC} decoding process, the value of each 
recovered information bit $\hat{u}_i$ depends on 
all the previous recovered information bits $\hat{u}_1^{i-1}$. Consequently, if a bit is incorrectly decoder,
it will often catastrophically propagate the error to all subsequent bits. The selection of an incorrect value 
for an information bit may be detected with consideration of the subsequent frozen bits, since the decoder knows that these 
bits should have values of $0$. More specifically, if the corresponding \ac{LLR} has a sign that would imply a value of $1$ 
for a frozen bit, then this suggests that an error may have been made during the decoding of one of the preceding information bits. 
However, in the \ac{SC} decoding process, there is no opportunity to consider alternative values for the preceding information bits. 
Once a value has been selected for an information bit, the \ac{SC} decoding process moves on and the decision is final. 

Inspired by the recursive list decoding of \ac{RM} codes~\cite{dumer2006soft}, Tal and Vardy proposed an \ac{LR}-based \ac{SCL} decoder~\cite{6033904, 7055304}, whose \ac{LLR}-based counterpart
was presented in~\cite{7114328}. In contrast to an \ac{SC} decoder, an \ac{SCL} decoder considers a list of alternative 
values for the information bits, hence improving the finite-length performance of polar codes.
More explicitly,
as the decoding process progresses, an \ac{SCL} decoder considers both options for the value of each successive information bit,
rather than making a hard-decision based on the associated \ac{LLR} value. 
This is achieved by maintaining a list of candidate information bits, where the list 
is built up as the \ac{SCL} decoding proceeds. At the start of the process, 
the list is empty. Whenever the decoding process reaches a frozen bit, a bit value of $0$ is appended to the
list. However, whenever the decoding process reaches an information bit, two replicas of the list are created. 
Here, the bit value of $0$ is appended to the first replica, while the bit value of $1$ is appended to the second replica. 
Hence, the number of lists, or more specifically the number of candidate decoding paths, doubles whenever an information bit is encountered.
This continues until the number of decoding paths reaches a limit $L$, which is known as the list size and is
typically chosen as a power of two. From this point onwards, each time the number of decoding paths is doubled when considering an information bit, 
the worst $L$ amongst the $2L$ candidate paths are identified and pruned from the list. 
In this way, the size of the list is maintained at $L$ until the \ac{SCL} decoding process completes. A straightforward implementation
of the \ac{SCL} algorithm incurs a complexity polynomial in the codeword length $N$. However, Tal and Vardy~\cite{7055304} exploited 
the recursive nature of polar codes together with `lazy-copy' algorithmic techniques to reduce the time complexity to $O(LN\log_2 N)$
and the space complexity to $O(LN)$, both of which are $L$ times the complexities of the classic \ac{SC} decoder.

The \ac{SCL} decoding process may be viewed as a path search in a binary tree of depth $N$, as illustrated in \fref{SCLb.fig}
for the decoding circuit of \fref{SCLa.fig}.
The input bits $u_i$ may be successively recovered as we move down the binary tree.
Explicitly, the nodes of the binary tree (except for the leaf nodes) of \fref{SCLb.fig} correspond to the input $u_i$, 
while the branches represent the possible values $0$ and $1$ of the bit $u_i$. Consequently, all the information nodes (marked in black in
\fref{SCLdecodingexample.fig}) have two branches, while the frozen nodes (marked in red in \fref{SCLdecodingexample.fig}) have a single branch,
since their values are known to the decoder. The $l$th branch at depth $i \in [1,N]$ is identified with a path metric $\phi_{l,i}$, which is calculated as follows:
\begin{align}
&\phi_{l,i} = \phi_{l,i-1}+\ln(1+e^{-(1-2\hat{u}_{l,i})L(u_{l,i})})\\
&\approx \left\{\begin{array}{ll}\phi_{l,i-1}& \textrm{if } \hat{u}_{l,i} = \frac{1}{2}(1-\mathrm{sign}(L(u_{l,i})))\\ \phi_{l,i-1} + |L(u_{l,i})|& \textrm{otherwise}\end{array}\right.,
\label{metric.eqn}
\end{align}
where $\phi_{l,i-1}$ is the parent path's metric at depth $(i-1)$, $\hat{u}_{l,i}$ is the value of $u_i$ associated with the
$l$th branch and $L(u_{l,i})$ denotes the corresponding \ac{LLR} $L_{i,1}$
obtained on the left-hand edge of the polar decoding circuit of \fref{SCLa.fig}.
These \acp{LLR} are obtained throughout the \ac{SCL} decoding process by using separate replicas of the partial sum 
computations of \eqr{PS1.eqn} and \eqr{PS2.eqn} to propagate the bits of each candidate path
from left to right. Similarly, separate replicas of the $f$ and $g$ computations of \eqr{eq:f} and \eqr{g.eqn}
are used to propagate corresponding replicas of the \acp{LLR}, as shown in \fref{SCLa.fig} in brown color for the 
second candidate path. Here, \eqr{metric.eqn} is referred to as the min-sum approximation. Intuitively, \eqr{metric.eqn}
implies that if the value of bit $\hat{u}_{l,i}$ corresponding to the branch at depth $i$ complies with the \ac{LLR} $L(u_{l,i})$, then the 
path metric is left unchanged, otherwise it is penalized by $|L(u_{l,i})|$. Furthermore, \eqr{metric.eqn}
accumulates across all bit positions $i \in [1,N]$. So, it must be calculated for all $L$ candidate paths 
whenever a frozen bit value of $0$ is appended, as well as for all $2L$ candidates when both possible values of an 
information bit are considered. In the latter case, the $2L$ metrics are sorted and the $L$ candidates having the highest values 
are identified as being the worst and are pruned from the list, as shown in \fref{SCLb.fig} for $L = 2$. 
Following the completion of the \ac{SCL} decoding process, 
the candidate path having the lowest metric may be selected as the most likely decoding path. 
It is pertinent to mention here that an SCL decoder having $L = 1$ is same as an \ac{SC} decoder and it
becomes equivalent to \ac{ML} decoder, when $L = 2^k$, implying a search through all possible $2^k$ codewords.

\begin{figure}[tbp]
\centering
\includegraphics[width=\linewidth]{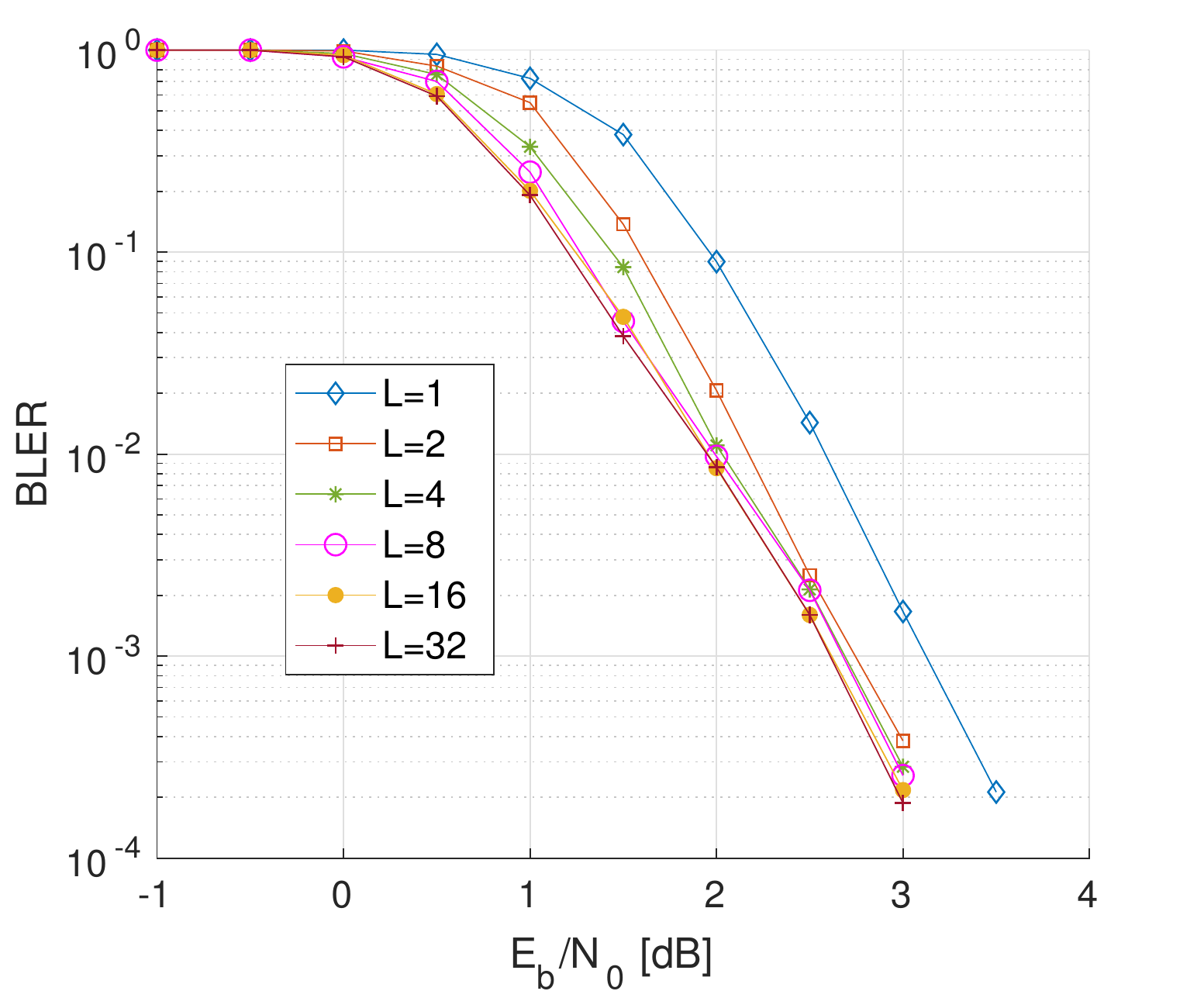}
\caption{Achievable \ac{BLER} performance of the \ac{SCL} decoder for variable list sizes. $(1024,512)$ polar code relying on the $5$G frozen
bit sequence was used in conjunction with \ac{QPSK} transmission over an \ac{AWGN} channel. Min-sum approximation of \eqr{min.eqn} was invoked for
calculating box-plus operations.}
\label{fig:Results:SCL}
\end{figure}

\fref{fig:Results:SCL} records the \ac{BLER} performance of the \ac{SCL} decoder for variable list sizes $L$. 
We may observe in
\fref{fig:Results:SCL} that the \ac{BLER} performance improves upon increasing the value of $L$, providing significant improvement over the
\ac{SC} decoder ($L = 1$). However, the performance improves with diminishing returns at higher values of $L$. More precisely, 
the performance improvement is only marginal for $L > 8$. Hence, a list size of $8$ is deemed sufficient, while a list size of $32$
brings the performance arbitrary close to that of the \ac{ML} decoder, as demonstrated in~\cite{7055304}.

Despite the improved performance of the \ac{SCL} decoder, polar codes failed to outperform the state-of-the-art \ac{LDPC} and turbo codes~\cite{7055304},
which may be attributed to their poor minimum distance. More specifically, the authors of~\cite{6297420} observed 
that when the \ac{SCL} decoder failed to output the correct information sequence, the correct information sequence
was usually present in its list of candidate paths, but with a smaller path metric. This motivated the development of
the \ac{CA-SCL} decoder~\cite{6297420, 7055304}, while an adaptive \ac{CA-SCL} decoding scheme was developed in~\cite{6355936}. 
The \ac{CRC}-aided approach concatenates a \ac{CRC} to the polar code as an outer code, 
so that a portion of the information bits is utilized for carrying the \ac{CRC} bits. More specifically, $(k-r)$ information
bits are first encoded using a \ac{CRC} code, which appends an $r$-bit \ac{CRC} to the $(k-r)$ information bits. 
This \ac{CRC} code can simply be a small random systematic linear block code. The resulting \ac{CRC} encoded
bits are then passed to the inner polar code. At the \ac{CA-SCL} decoder, all candidates paths that do not satisfy the \ac{CRC} are  
pruned during the last stage of the classic \ac{SCL} decoding process, before the candidate having the lowest metric is selected as the 
most likely decoding path. The \ac{CA-SCL} decoder brought the performance of polar codes at par with the state-of-the-art
turbo and \ac{LDPC} codes, while retaining the complexity of the classic \ac{SCL} decoder. However, this is achieved at the cost of
a small loss in the coding rate, since the coding rate of the resulting concatenated code is $(k-r)/N$.
The notion of \ac{CRC}-aided concatenation was generalized to parity check codes in~\cite{wang2016parity}.

In pursuit of approaching the perfromance of the \ac{SCL} decoder with the \ac{SC} decoder's space requirements,
bit-flip decoding strategy was adopted in~\cite{7094848}. The proposed \ac{SC} flip decoder of~\cite{7094848}
enhances the performance of the classic \ac{SC} decoder by exploiting \ac{CRC} in conjunction with the bit-flipping algorithm. More specifically, the \ac{SC} flip decoder commences the decoding process by running a single iteration of the classic \ac{SC} decoder for estimating the codeword $\hat{u}$, as exemplified in \fref{example.fig}. If the \ac{CRC} passes with the estimated codeword $\hat{u}$, the decoding process is terminated. Otherwise, upto $T$ additional \ac{SC} decoding iterations are invoked, sequentially flipping the estimated value of one of the $T$ least reliable bits in each iteration. The process is repeated until the
\ac{CRC} passes or the maximum number of attempts $T$ is reached. While the \ac{SC} flip decoder preserves the space requirements of the classic \ac{SC} decoder as well as improves the decoding performance, it fails to outperform the \ac{SCL} decoder, since the bit-flip algorithm is only capable of correcting a single additional error. Furthermore, the computational complexity is substantially high at low \acp{SNR} and approaches that of the \ac{SC} decoder at high \acp{SNR}. Improvements to the \ac{SC} flip decoder were proposed in~\cite{8368991, 8259253, 8478832}, while the use of the bit-flip algorithm was extended to the \ac{SSC} and the \ac{SCL} decoders in~\cite{8369026, 8684882} and~\cite{8705213}.

Analogous to the classic \ac{SC} decoder, the \ac{SCL} decoding scheme also suffers from a high latency due to the data dependencies
associated with the decoding process. In this context, various low-latency \ac{SCL} decoding schemes have been presented 
in~\cite{6986062, 7178128, 7339658, 7339671, 7348682, 7742998, 8010835}. Furthermore, the complexity of an \ac{SCL} decoder is $L$
times higher than that of a classic \ac{SC} decoder, since it processes $L$ candidate paths. For the sake of reducing 
the complexity of \ac{SCL} decoders, reduced-complexity techniques were explored in~\cite{7339660, 7541412}.
Another key challenge in the \ac{SCL} decoding process is imposed by metric sorting, which is required to identify
the worst $L$ candidate paths among the merged list of $2L$ paths for pruning. Simultaneously comparing
all the $2L$ path metrics requires large amount of hardware resources, while successively comparing the paths increases the latency.
Hence, a suitable compromise must be reached between the required hardware resources and the imposed latency. Hardware implementations
of the \ac{SCL} decoder have been dealt 
in~\cite{6823099, 7114328, 6920050, 7169066, 6986062, 7468577, 7407353, 7841865, 8010835, 7756324, 8070938, 8361464},
while an unsorted \ac{SCL} decoder was conceived in~\cite{7867834}.

The main characteristics of an \ac{SCL} polar decoder are summarized in \tref{tab:SCL-summary}.
\begin{table}[tbp]
 \centering
 \begin{center}
\begin{tabular}[tbp]{|l|p{6.5cm}|}
\hline 
&\\
\textbf{Complexity} & \tabitem Time complexity = $O(LN\log_2N)$\\
                    & \tabitem Space complexity = $O(LN)$\\ &\\ \hline
& \\
\textbf{Advantages} & \tabitem Capable of achieving the \ac{ML} performance\\
& \tabitem Performance of \ac{CA-SCL} is at par with turbo and \ac{LDPC} codes  \\ &\\ \hline
&\\
\textbf{Disadvantages} & \tabitem Serial processing, resulting in high latency (or low throughput) \\
& \tabitem Fully-parallel implementation not feasible \\ 
& \tabitem Higher complexity than \ac{SC} \\
& \tabitem Higher latency than \ac{SC} due to the `metric sorting' operation \\
&\\ \hline
\end{tabular}
 \end{center}
 \caption{Main characteristics of an \ac{SCL} polar decoder.}
\label{tab:SCL-summary}
\end{table}
\subsection{Successive Cancellation Stack (SCS) Decoder} \label{sec:sec:SCS}
The \ac{SC} decoding may be viewed as a greedy search over the binary tree of \fref{SCLb.fig}, since it only considers the path with the lowest
metric, hence making a bit-by-bit decision. By contrast, the \ac{SCL} decoding adopts a breadth-first approach, doubling the number
of paths for each information bit and selecting the best $L$ paths for further processing. The breadth-first approach of \ac{SCL}
provides attractive performance benefits at the cost of an increased computational complexity, because $L$ paths have to be processed in
contrast to the single decoding path of \ac{SC}. For the sake of achieving a reasonable compromise between the performance and computational
complexity, \ac{SCS} decoding was proposed~\cite{6215306} inspired by the stack decoding of convolutional codes~\cite{jelinek1969fast} and
\ac{RM} codes~\cite{stolte2000sequential}. 

Analogous to \ac{SCL} decoding, \ac{SCS} decoding also operates over the binary tree of \fref{SCLb.fig}. However, while an \ac{SCL} decoder
records the $L$ best candidate paths (having the lowest path metric) of the same length, an \ac{SCS} decoder records the $D$ best candidate paths 
of variable lengths in an ordered stack. Furthermore, in contrast to the \ac{SCL} decoding, which processes all the $L$ candidate paths in parallel,
\ac{SCS} decoding only processes the path at the top of the stack at a time, hence reducing the computational complexity. Let us elaborate on this by revisiting
the decoding example of \fref{SCLdecodingexample.fig} with the \ac{SSC} algorithm.

\begin{table*}[tbp]
 \centering
 \begin{center}
\begin{tabular}[tbp]{|l|l|l|l|l|l|l|l|}
\hline
$t = 1$ & $t = 2$ & $t = 3$ & $t = 4$ & $t=5$ & $t=6$ & $t=7$ & $t=8$ \\ \hline \hline
$\hat{u}_{1} \; \; \; \phi_{1,1} = 0.0$ & $\hat{u}_{1}^2 \; \; \; \phi_{1,2}=0.0$ & $\hat{u}_{1}^3 \; \; \; \phi_{1,3}=0.0$ & $\hat{u}_{1}^4 \; \; \; \phi_{2,4}=0.0$&$\hat{u}_{1}^5 \; \; \; \phi_{2,5}=2.02$&$\hat{u}_{1}^6 \; \; \; \phi_{3,6}=2.02$&$\hat{u}_{1}^7 \; \; \; \phi_{5,7}=2.02$&\cellcolor{green}$\hat{u}_{1}^8 \; \; \; \phi_{10,8}=2.02$ \\ \hline
&&&$\hat{u}_{1}^4 \; \; \; \phi_{1,4}=4.09$&$\hat{u}_{1}^4 \; \; \; \phi_{1,4}=4.09$&$\hat{u}_{1}^4 \; \; \; \phi_{1,4}=4.09$&$\hat{u}_{1}^7 \; \; \; \phi_{6,7}=2.75$&$\hat{u}_{1}^7 \; \; \; \phi_{6,7}=2.75$ \\ \hline
&&&&&$\hat{u}_{1}^6 \; \; \; \phi_{4,6}=4.28$&$\hat{u}_{1}^4 \; \; \; \phi_{1,4}=4.09$&$\hat{u}_{1}^4 \; \; \; \phi_{1,4}=4.09$ \\ \hline
&&&&&&\cellcolor{red}$\hat{u}_{1}^6 \; \; \; \phi_{4,6}=4.28$& $\hat{u}_{1}^8 \; \; \; \phi_{9,8}=12.52$ \\ \hline 
\end{tabular}
 \end{center}
 \caption{Example of \ac{SCS} decoding process ($D = 4$) corresponding to the \ac{SCL} decoding example of \fref{SCLdecodingexample.fig}. 
 Each column records the stack outputs at time instant $t$. The deleted path is highlighted in red, while the final optimal path is
 marked in green.}
\label{tab:SCS-ex}
\end{table*}
Analogous to \ac{SCL} decoding, the \ac{SCS} decoding algorithm begins from the root node at depth $i=1$ of the binary tree of 
\fref{SCLb.fig}, computing the path metric $\phi_{l,1}$ according to \eqr{metric.eqn} for all branches at depth $1$.
The resulting path, which is identified by the estimated information bit $\hat{u}_1$ and the path metric $\phi_{l,1}$, is stored in a stack,
as shown in the first column of \tref{tab:SCS-ex}. The stack is sorted in the order of increasing path metrics. 
However, since we only have a single path at time instant $t = 1$ in our example, sorting is not required. Thereafter, the algorithm moves along the binary
tree of \fref{SCLb.fig} by recursively invoking the following operations:
\begin{itemize}
 \item \textbf{Pulling:} The top path from the stack, having the lowest path metric, is pulled out for further processing.
 \item \textbf{Extension:} The pulled path, having the estimated information bits $\hat{u}_1^i$ and the path metric $\phi_{l,i}$,
 is extended along the binary tree of \fref{SCLb.fig} to include the next bit. If the next bit is a frozen bit, 
we get a single extended path by appending a $0$ to the original path $\hat{u}_1^i$. By contrast, if the next bit is an information bit, 
we get two extended paths $\hat{u}_1^{i+1}$: one is obtained by appending a $0$ to $\hat{u}_1^i$, while the other is obtained by 
appending a $1$ to $\hat{u}_1^i$. Path metrics are calculated for the extended paths according to \eqr{metric.eqn}.
\item \textbf{Deletion:} A stack can store at most $D$ paths. Consequently, if $(D-1)$ paths are already stored in the stack, the path at 
the bottom of the stack is deleted in order to make space for the two extended paths. This step may be omitted, if there is a single extended
path.
\item \textbf{Pushing:} The extended paths $\hat{u}_1^{i+1}$ are pushed in the stack along with their path metrics $\phi_{l,i+1}$.  Please note that, consistent with the notation 
of Section~\ref{sec:sec:SCL}, $l$ denotes the $l$th branch in the binary tree of \fref{SCLb.fig}.
\item \textbf{Sorting:} The stack is sorted in order of increasing path metrics, so that the most reliable path having the lowest
path metric appears at the top.
\end{itemize}
The aforementioned five operations are repeated until the top path of the stack reaches the leaf node of the binary tree of \fref{SCLb.fig}. The resulting
stack outputs are recorded in \tref{tab:SCS-ex} for a stack size of $D = 4$. 

The time and space complexities of an \ac{SCS} decoder are $O(DN\log_2N)$ and $O(DN)$, respectively. 
However, the actual complexity of an \ac{SCS} decoder depends on the channel \ac{SNR}. More specifically, at high \acp{SNR}, when the received
information is less noisy, the \ac{SCS} decoder converges faster, approaching the complexity of the classic \ac{SC} decoder. By contrast,
at low \acp{SNR}, the complexity of an \ac{SCS} decoder approaches that of an \ac{SCL} decoder having $L=D$. Nonetheless, \ac{SCS} decoders
are shown to have a lower complexity than the \ac{SCL} decoder at the desired \ac{BLER} of $10^{-3}$~\cite{6215306}, but this is achieved
at the cost of high space complexity. More specifically, a significantly high value of $D$ is required to match the performance of 
the \ac{SCL} decoder of a given list size $L$. For example, a depth size of $D=100$ was used in~\cite{6215306} for the \ac{SCS} decoder 
to match the performance of an \ac{SCL} decoder having $L = 20$. In pursuit of combining the benefits of the \ac{SCL} and \ac{SCS} decoding
schemes, a hybrid scheme called \ac{SCH} was proposed in~\cite{6560025}. The notion of \ac{CRC}-aided decoding is also
readily applicable to \ac{SCS} decoding, resulting in the \ac{CA-SCS} decoder~\cite{6297420}. Furthermore, an efficient software implementation of the \ac{SCS} decoder, relying on the \acp{LLR}, was presented in~\cite{8351832}, which imposes a reduced time and space complexity. The decoding performance of the \ac{LLR}-based \ac{SCS} decoder of~\cite{8351832} was further improved in~\cite{xiang2019crc}. In particular, the improved \ac{SCS} decoder of~\cite{xiang2019crc} incurs a substantially lower time complexity, while maintaining the same space complexity.

The main characteristics of an \ac{SCS} polar decoder are summarized in \tref{tab:SCS-summary}.
\begin{table}[tbp]
 \centering
 \begin{center}
\begin{tabular}[tbp]{|l|p{6.5cm}|}
\hline 
&\\
\textbf{Complexity} & \tabitem Time complexity = $O(DN\log_2N)$\\
                    & \tabitem Space complexity = $O(DN)$\\ &\\ \hline
& \\
\textbf{Advantages} & \tabitem Capable of achieving the \ac{ML} performance when $D=2^k$\\
& \tabitem Performance of \ac{CA-SCS} is at par with turbo and \ac{LDPC} codes  \\ 
& \tabitem Lower complexity than \ac{SCL} at moderate and high \acp{SNR} \\
&\\ \hline
&\\
\textbf{Disadvantages} & \tabitem Serial processing, resulting in high latency (or low throughput) \\
& \tabitem Fully-parallel implementation not feasible \\ 
& \tabitem High time complexity at low \acp{SNR} \\
& \tabitem High space complexity \\
& \tabitem Higher latency than \ac{SC} due to the `metric sorting' operation \\
&\\ \hline
\end{tabular}
 \end{center}
 \caption{Main characteristics of an \ac{SCS} polar decoder.}
\label{tab:SCS-summary}
\end{table}
\subsection{Belief Propagation} \label{sec:sec:BP}
All the afore-mentioned polar decoding schemes are derived from the classic \ac{SC} decoder of Section~\ref{sec:sec:SC} and suffer
from the issue of serial processing, which imposes a high latency. Furthermore, all these coding schemes yield a hard decision output. Hence, these
decoders are not suitable for iterative joint detection and decoding schemes, which require soft-in soft-out decoders.
Fortunately, polar codes may also be represented using a factor graph, which permits soft-in soft-out \ac{BP} decoding~\cite{5075875}.
Furthermore, \ac{BP} decoding algorithm is more amenable to parallel implementation.

\begin{figure}[tb]
\centering
\includegraphics[width=\linewidth]{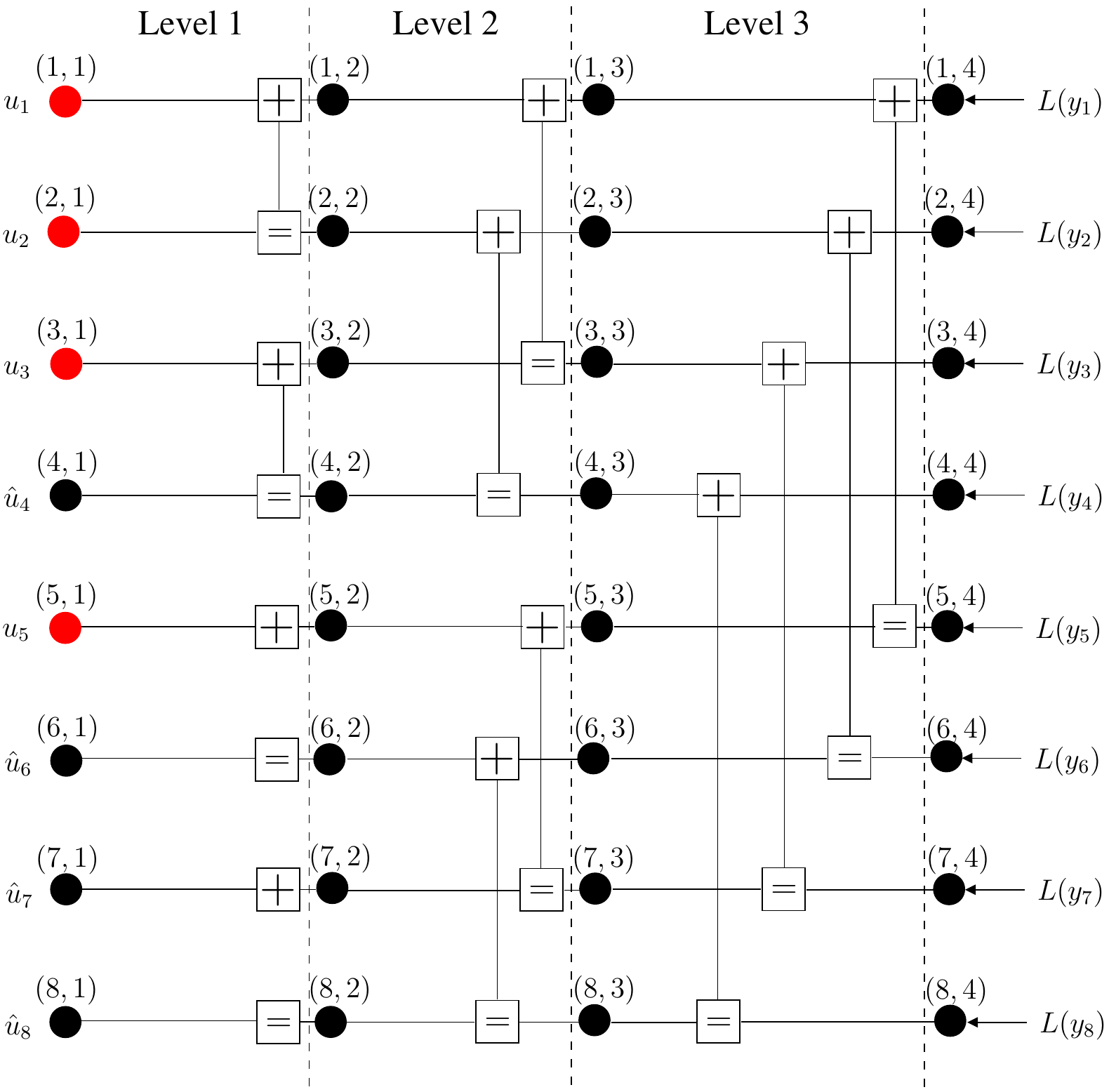}
\caption{Factor graph of an $N=8$ polar code having $k = 4$ and $\mathcal{F} = \{1,2,3,5\}$. Variable nodes and check nodes are denoted
by circle and square, respectively. The frozen variable nodes on the left-hand edge of the factor graph are drawn in red.}
\label{fig:FactorGraphPolar}
\end{figure}

\begin{figure}[tb]
\centering
\includegraphics[width=\linewidth]{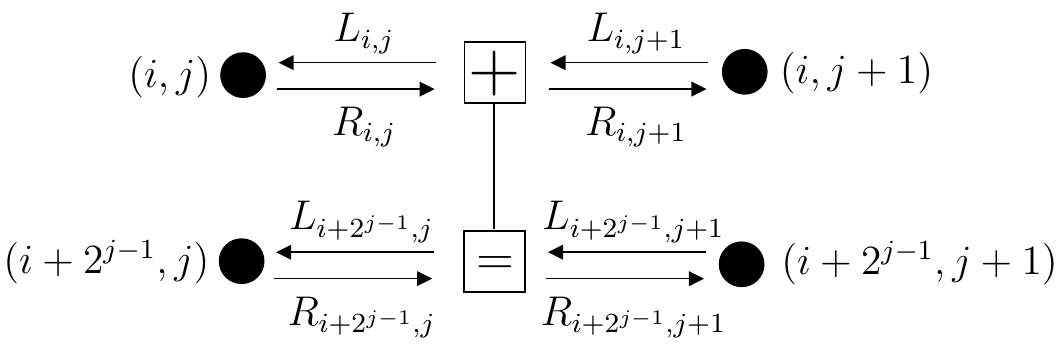}
\caption{Factor graph of the $2$-bit elementary kernel of \fref{fig:FactorGraphPolar} at the $j$th level.}
\label{fig:FactorGraphPolar-a}
\end{figure}

\fref{fig:FactorGraphPolar} shows the factor graph representation of the polar circuit of \fref{example.fig}. We may notice that 
\fref{fig:FactorGraphPolar} is obtained from \fref{example.fig} by replacing each $2$-bit elementary kernel's circuit by 
its factor graph, which is shown in \fref{fig:FactorGraphPolar-a}. 
The resulting factor graph consists of $n = \log_2N$ levels and $N \times (n+1)$ variable nodes. Furthermore, each variable node is identified by the index $(i,j)$ 
where $i$ and $j$ denote the bit index and level index, respectively. 

\begin{figure}[tbp]
        \centering
        \begin{subfigure}[]{0.48\linewidth}
        \begin{center}
                \includegraphics[width=\linewidth]{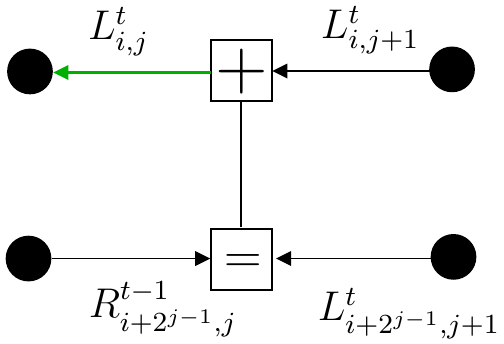}
                \caption{}
                \label{BP-FG-L1.fig}
        \end{center}
        \end{subfigure}%
        ~ 
        \begin{subfigure}[]{0.48\linewidth}
        \begin{center}
                \includegraphics[width=\linewidth]{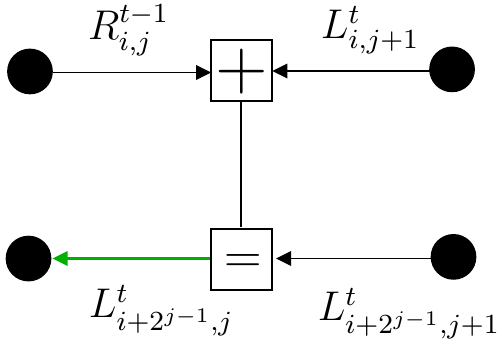}
                \caption{}
                \label{BP-FG-L2.fig}
        \end{center}
        \end{subfigure}
                ~ 
        \begin{subfigure}[]{0.48\linewidth}
        \begin{center}
                \includegraphics[width=\linewidth]{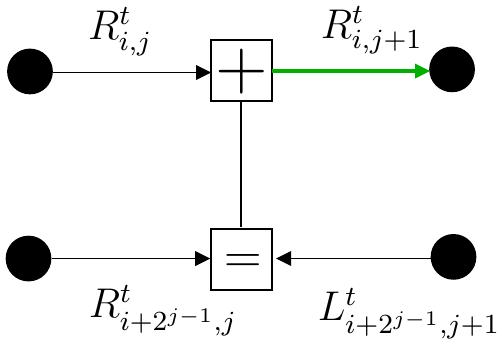}
                \caption{}
                \label{BP-FG-R1.fig}
        \end{center}
        \end{subfigure}
                ~ 
        \begin{subfigure}[]{0.48\linewidth}
        \begin{center}
                \includegraphics[width=\linewidth]{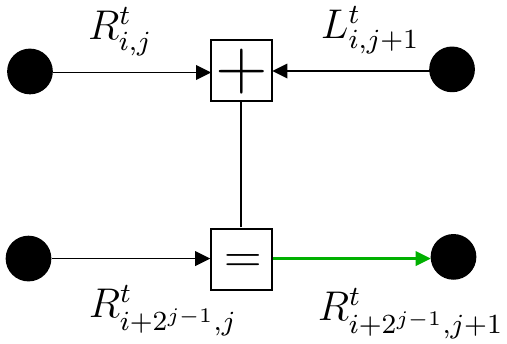}
                \caption{}
                \label{BP-FG-R2.fig}
        \end{center}
        \end{subfigure}
        \caption{Computation of \acp{LLR} at the $j$th level of factor graph during the $t$th iteration of \ac{BP} algorithm:
        (a) The function $f(L^{t}_{i,j+1},L^t_{i+2^{j-1},j+1}+R^{t-1}_{i+2^{j-1},j})$ for calculating the left message $L^t_{i,j}$.
        (b) The function $L^{t}_{i+2^{j-1},j+1}+f(L^t_{i,j+1},R^{t-1}_{i,j})$ for calculating the left message $L^t_{i+2^{j-1},j}$.
        (c) The function $f(R^t_{i,j},L^t_{i+2^{j-1},j+1} + R^t_{i+2^{j-1},j})$ for calculating the right message $R^t_{i,j+1}$.
        (d) The function $R^t_{i+2^{j-1},j} + f(R^t_{i,j},L^t_{i,j+1})$ for calculating the right message $R^t_{i+2^{j-1},j+1}$.}
         \label{BP-FG.fig}
\end{figure}

The \ac{BP} algorithm iteratively exchanges messages over the factor graph of \fref{fig:FactorGraphPolar} until the maximum number of decoding
iterations $I_{\text{max}}$ is reached. Explicitly, as shown in \fref{fig:FactorGraphPolar-a}, two types of messages flow through the factor graph: 
the right-to-left (left) \acp{LLR} $L_{i,j}$ flowing towards the left of the factor graph and the left-to-right (right) \acp{LLR} 
$R_{i,j}$ flowing towards the right of the factor graph. These messages are computed during the \ac{BP} algorithm as follows:
\begin{itemize}
 \item \textbf{Initialization:}  
 The left \acp{LLR} $L^t_{i,n+1}$, pertaining to the variable nodes on the right-hand end of the factor graph, 
 are set equivalent to the channel \acp{LLR} for all decoding iterations $t \in [1,I_{\text{max}}]$.
 Furthermore, all right \acp{LLR} $R^0_{i,j}$ are initialized to zero for the first decoding iteration, except for the \acp{LLR}
pertaining to the frozen nodes on the left-hand edge of the factor graph ($j=1$), which are set to infinity.
 \item \textbf{Right-to-left message exchange:} The \ac{BP} algorithm processes the left \acp{LLR} $L^t_{i,j}$ 
from right to left, starting from the level $j = 3$. Explicitly, the left messages are computed as follows:
 \begin{align}
  &L^t_{i,j} = f(L^{t}_{i,j+1},L^t_{i+2^{j-1},j+1}+R^{t-1}_{i+2^{j-1},j}) \label{eq:L1}\\
  &L^t_{i+2^{j-1},j} = L^{t}_{i+2^{j-1},j+1}+f(L^t_{i,j+1},R^{t-1}_{i,j}) \label{eq:L2},
 \end{align}
 where $f(.)$ is given by \eqr{eq:f},
 for all nodes at the $j$th level and the $t$th decoding iteration. The operation of \eqr{eq:L1} and \eqr{eq:L2} is encapsulated in
 \fref{BP-FG-L1.fig} and \fref{BP-FG-L2.fig}, respectively. We may notice that \eqr{eq:L1} reduces to the \ac{XOR} operation of \fref{fig:SC-1},
 when $R^{t-1}_{i+2^{j-1},j}$ is set to zero, i.e. when we don't have any a-priori information about the bits $b_{i+2^{j-1},j}$. 
 Similarly, \eqr{eq:L2} becomes equivalent to the $g$ operation of \fref{fig:SC-2}, when $R^{t-1}_{i,j}$ of \eqr{eq:L2}
 is replaced by its hard-decision value of plus or minus infinity corresponding to the recovered bit $\hat{b}_{i,j}$. 
 \item \textbf{Left-to-right message exchange:} Following the right-to-left message exchange, the \ac{BP} algorithm
 computes the right \acp{LLR} $R^t_{i,j}$ from left-to-right of the factor graph as follows:
 \begin{align}
 R^t_{i,j+1} = f(R^t_{i,j},L^t_{i+2^{j-1},j+1} + R^t_{i+2^{j-1},j}) \label{eq:R1} \\
 R^t_{i+2^{j-1},j+1} = R^t_{i+2^{j-1},j} + f(R^t_{i,j},L^t_{i,j+1}), \label{eq:R2}
 \end{align}
 for all nodes at the $j$th level and the $t$th decoding iteration, starting from $j=1$. The operation of \eqr{eq:R1} and \eqr{eq:R2} is encapsulated in
 \fref{BP-FG-R1.fig} and \fref{BP-FG-R2.fig}, respectively, which is analogous to the partial sum
 calculation of \fref{fig:SC-1}. To elaborate, the partial sum calculation of \fref{fig:SC-1} operates on the hard-decision values,
 while \eqr{eq:R1} and \eqr{eq:R2} carry out the same operation on the corresponding \acp{LLR}. It is also pertinent to mention here that the values of $R^t_{i,1} = R^0_{i,1}$ for all decoding
 iterations, since they are not updated as the iterations proceed.
 \item \textbf{Iterative message exchange:} One pass of the afore-mentioned `right-to-left' and `left-to-right' message exchange procedures
 constitute one round of \ac{BP} decoding iteration. These procedures are repeatedly invoked during each decoding iteration until the maximum
 number of iterations is reached.
 \item \textbf{Hard-decision:} Finally, the information bits are estimated based on the hard-decision values of the 
 \acp{LLR} $L^{I^{\text{max}}}_{i,1}$ at the left-hand edge of the factor graph, as follows:
 \begin{equation}
  \hat{u}_i =  \left\{
\begin{array}{l l}
 0 & \text{if} \;\;\; L^{I^{\text{max}}}_{i,1} > 0 \\ \\
 1 & \text{otherwise}, \\
\end{array}
\right.
\label{eq:HD}
 \end{equation}
for $i \in \mathcal{F}_c$.
\end{itemize}

The performance of \ac{BP} decoders is comparable to that of the classic \ac{SC} decoder. However, it imposes a higher
computational complexity due to the large number of iterations required for achieving the convergence and higher memory requirements. The time complexity
of the \ac{BP} algorithm is $O(I_{\text{max}}N\log_2N)$, while the space complexity is $O(N\log_2N)$.
Efforts were made in~\cite{6864206, 6942260} to reduce the computational complexity, while the hardware implementations of the \ac{BP}
decoder were presented in~\cite{pamuk2011fpga, 7742915}. Furthermore, improvements were proposed in~\cite{hussami2009performance, 6785962, elkelesh2018belief2}
for the sake of enhancing the performance of the \ac{BP} decoding algorithm. 
More specifically, a different decoding schedule for \ac{BP} was investigated in~\cite{hussami2009performance}, which first completes
the right-to-left message exchange over the polar decoding circuit and then proceeds with the left-to-right message exchange.
An alternative technique was also investigated in~\cite{hussami2009performance}, called the overcomplete representation
(also called permuted factorgraph), which permutes the different levels of the polar factor graph of \fref{fig:FactorGraphPolar} for achieving decoding convergence.
In~\cite{6785962}, the reliability of \ac{BP} messages was improved by incorporated the knowledge of frozen bits,
while a \ac{CRC} was invoked in\cite{elkelesh2018belief2} together with the overcomplete representation of~\cite{hussami2009performance}
for achieving decoding convergence. Despite these efforts, the \ac{BP} algorithm
did not outperform the \ac{SCL} or \ac{SCS} decoders. Recently, a \ac{BP} list decoding algorithm was conceived in~\cite{elkelesh2018belief},
which combines the benefits of the classic \ac{BP} and \ac{SCL} decoders. Explicitly, \ac{BP} list decoder benefits from the good error
correction capabilities of the \ac{SCL} decoder and the soft-in soft-out nature of the classic \ac{BP} algorithm. Furthermore, it also
lends itself to parallel implementation, hence there is a possibility of achieving a lower latency. However, the \ac{BP} list decoder
imposes a high computational complexity and its compatibility with \ac{CRC}-aided polar codes needs to be investigated.

The main characteristics of a \ac{BP} polar decoder are summarized in \tref{tab:BP-summary}.
\begin{table}[tbp]
 \centering
 \begin{center}
\begin{tabular}[tbp]{|l|p{6cm}|}
\hline 
&\\
\textbf{Complexity} & \tabitem Time complexity = $O(I_{\text{max}}N\log_2N)$\\
                    & \tabitem Space complexity = $O(N\log_2N)$\\ 
                    &\\ \hline
& \\
\textbf{Advantages} & \tabitem Soft output, hence compatible with iterative detection\\
& \tabitem Fully-parallel implementation  \\ 
&\\ \hline
&\\
\textbf{Disadvantages} & \tabitem Sub-optimal performance (better than \ac{SC}) \\
& \tabitem High complexity \\ 
& \tabitem High memory requirements \\ 
&\\ \hline
\end{tabular}
 \end{center}
 \caption{Main characteristics of a \ac{BP} polar decoder.}
\label{tab:BP-summary}
\end{table}
\subsection{Soft Cancellation (SCAN)} \label{sec:sec:SCAN}
The \ac{SC} decoder of Section~\ref{sec:sec:SC} is a soft-in hard-out decoder, since hard-decisions are made pertaining to the bits $b_{i,j}$
during the decoding process. In the spirit of extracting soft output from the \ac{SC} decoder, a \ac{SCAN} decoder was proposed~\cite{fayyaz2013polar, fayyaz2014low},
which is in essence a combination of the \ac{SC} decoder of Section~\ref{sec:sec:SC} and the \ac{BP} decoder of Section~\ref{sec:sec:BP}.
Recall that the left-to-right message exchange procedure of \ac{BP} is a soft-valued version of the $f$ and $g$ operations of an \ac{SC}
decoder. Similarly, the right-to-left message exchange of \ac{BP} is a soft-valued counterpart of the partial sum calculation of \ac{SC}.
Hence, we may conclude that all operations of \ac{BP} are soft-valued versions of the operations of \ac{SC}, but they follow a different processing
schedule. A \ac{SCAN} decoder combines the attributes of \ac{SC} and \ac{BP} decoding algorithms by using the \ac{BP} operations of \fref{BP-FG.fig}
in combination with the \ac{SC} processing schedule. This in turn facilitates faster convergence of the resulting decoding algorithm; hence,
drastically reducing the computational complexity. Quantitatively, it was demonstrated in~\cite{fayyaz2014low} that the computational complexity
of the \ac{SCAN} decoder is only $4\%$ of the complexity of the classic \ac{BP} decoder. Another notable contribution in the context of soft-in soft-out polar decoders is the soft counterpart of the \ac{SCL} decoder~\cite{liu2017parallel,zhou2018performance}, which yields soft information for the iterative decoding of concatenated codes. Explicitly, similar to the classic \ac{SCL} decoder, the soft \ac{SCL} decoder maintains a list of $L$ candidate decoding paths. Hence, it outperforms the the \ac{BP} as well as the \ac{SCAN} decoders. However, the soft \ac{SCL} decoder is only applicable to systematic polar codes.

The main characteristics of a \ac{SCAN} polar decoder are summarized in \tref{tab:SCAN-summary}.
\begin{table}[tbp]
 \centering
 \begin{center}
\begin{tabular}[tbp]{|l|p{6cm}|}
\hline 
&\\
\textbf{Complexity} & \tabitem Time complexity = $O(I_{\text{max}}N\log_2N)$\\
                    & \tabitem Space complexity = $O(N\log_2N)$\\ 
                    &\\ \hline
& \\
\textbf{Advantages} & \tabitem Soft output, hence compatible with iterative detection\\
& \tabitem Much lower complexity than \ac{BP}\\
& \tabitem Much lower memory requirements than \ac{BP}\\ 
&\\ \hline
&\\
\textbf{Disadvantages} & \tabitem Sub-optimal performance (better than \ac{SC}) \\
& \tabitem Serial processing, resulting in high latency\\
& \tabitem Fully-parallel implementation not feasible\\
&\\ \hline
\end{tabular}
 \end{center}
 \caption{Main characteristics of a \ac{SCAN} polar decoder.}
\label{tab:SCAN-summary}
\end{table}
\subsection{Comparison of Polar Decoders} \label{sec:sec:DecComp}
\begin{figure}[tbp]
\centering
\includegraphics[width=\linewidth]{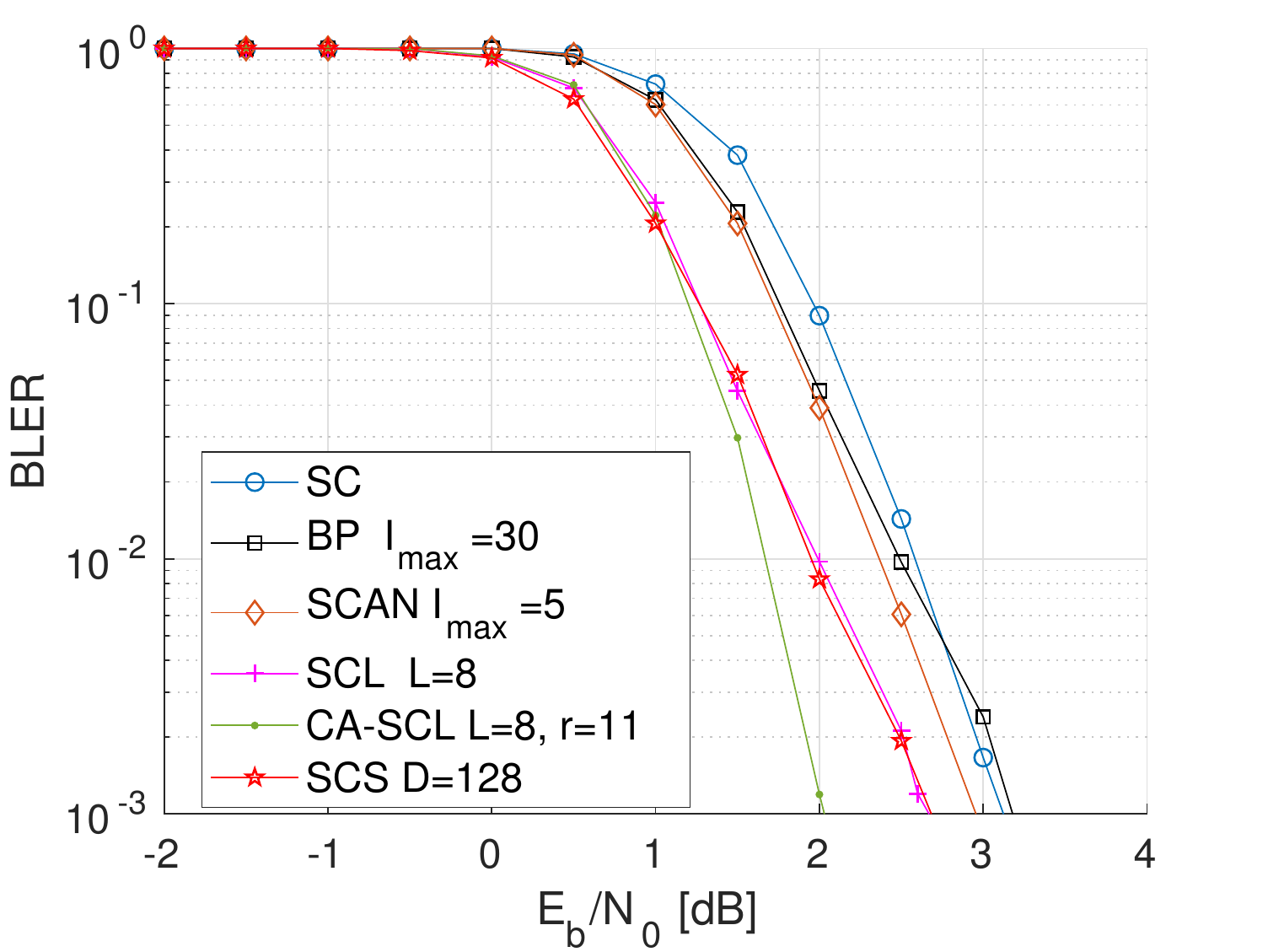}
\caption{Comparison of the achievable \ac{BLER} performance of the polar decoders. $(1024,512)$ polar code relying on the $5$G~\ac{NR} frozen
bit sequence was used in conjunction with \ac{QPSK} transmission over an \ac{AWGN} channel. Min-sum approximation of \eqr{min.eqn} was invoked for
calculating the box-plus operations in all decoders except the \ac{BP} decoder for which the exact computation of \eqr{boxplus.eqn} was used.}
\label{fig:Results:PolarDec}
\end{figure}
\begin{figure}[tbp]
\centering
\includegraphics[width=\linewidth]{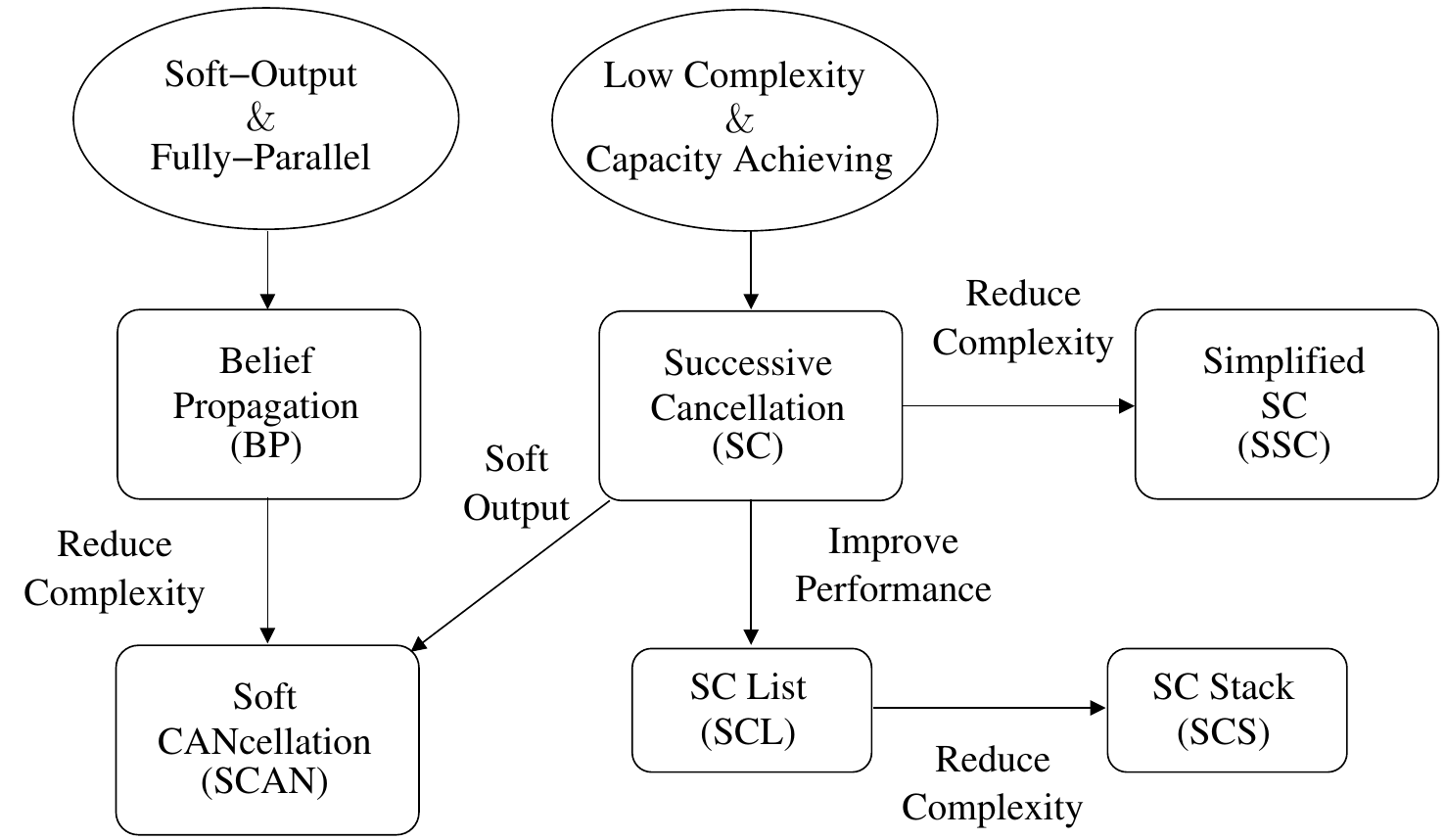}
\caption{Rationale for developing the polar decoders of Sections~\ref{sec:sec:SC} to~\ref{sec:sec:SCAN}.}
\label{fig:dec-motiv}
\end{figure}

In \fref{fig:Results:PolarDec}, we compare the \ac{BLER} performance of the various polar decoders. 
We have used a $1/2$-rate polar code having a codeword length of $1024$ and having the frozen bit sequence specified for the 
$5$G~\ac{NR}~\cite{3gpp38212}. For the sake of ensuring a fair comparison, the $r$ \ac{CRC} bits of the \ac{CA-SCL} scheme are taken from the parity bits of the constituent polar code, rather than from the information bits. More explicitly, the constituent polar code used in the \ac{CA-SCL} scheme has a coding rate of $(k+r)/N=523/1024$; hence, the coding rate of the resultant \ac{CRC}-aided concatenated polar code is $k/N = 1/2$. Furthermore, we have invoked \ac{QPSK} modulation and \ac{AWGN} channel for transmission
in order to allow for comparison with the work that was completed during the development of the $5$G~\ac{NR} standard. We may observe in \fref{fig:Results:PolarDec}
that the \ac{SC} decoder exhibits the worse performance, while the \ac{CA-SCL} decoder has the best performance and that of the
\ac{BP}, \ac{SCAN}, \ac{SCL} and \ac{SCS} lies in between. Furthermore, the performance of \ac{SCS} with a stack size of $D=128$ is same as that of
\ac{SCL} with a list-size of $L=8$.
As discussed in Sections~\ref{sec:sec:SC} to~\ref{sec:sec:SCAN}, the performance improvement
comes at the cost of complexity. 
In \fref{fig:dec-motiv}, we have summarized the rationale for developing the various
polar decoders, while their main characteristics are compared in \tref{tab:DecComp}. 
\begin{table}[tbp]
 \centering
 \begin{center}
\begin{tabular}[tbp]{c||c|c|c|c}
\hline 
&&&&\\
\textbf{Decoder} & \textbf{Complexity} & \textbf{Space Req.} & \textbf{Performance} & \textbf{Fully-parallel}\\ \hline 
\ac{SC} & low & low & sub-optimal & No \\
\ac{SSC} & very low & low & sub-optimal & No \\
\ac{SCL} & medium & low & \ac{ML}& No\\
\ac{CA-SCL}& medium & low & outperform \ac{ML}& No\\
\ac{SCS} & low-medium & high & \ac{ML}& No\\
\ac{BP} & high & high & sub-optimal & Yes\\
\ac{SCAN} & medium & medium& sub-optimal & No \\
\hline
\end{tabular}
 \end{center}
 \caption{Comparison of the polar decoders discussed in Sections~\ref{sec:sec:SC} to~\ref{sec:sec:SCAN}.}
\label{tab:DecComp}
\end{table}
\section{Polar Code Construction: Design Principles, Guidelines \& Examples} \label{sec:polarcodedesign}
Recall from \fref{polar-ev} and \fref{IMcolorplot} that polar codes do not completely polarize at finite block lengths.
Hence, a reliability metric is required to identify the least reliable bit-channels
for transmitting frozen bits. This selection of frozen bit-channels is a very important step in the design of polar codes,
because it directly dictates the resultant \ac{BLER}. Explicitly, the \ac{BLER} of a polar code is upper bounded by the sum of the \acp{BER}
of the individual good bit-channels. So, if a bad channel is inadvertently not frozen, it will deteriorate the 
performance of the resultant polar code. Unfortunately, the set of frozen bit-channels is channel specific and is hence not universal. 
This makes the design of polar codes challenging, as it has to be optimized for the channel under consideration (or equivalently the channel
noise level). However, it must be acknowledged that the capacity changes with changing channel conditions. So, the required optimization is
a natural consequence of any changes in the channel conditions.

The design objective of polar code construction is as follows:

\textbf{Design Objective:} \textit{For a given codeword length ($N$) and channel characteristics, for example \ac{SNR} of an \ac{AWGN} channel, determine
the ($N-k$) least reliable bit-channels (or equivalently the $k$ most reliable bit-channels) - the so-called bad channels.}

\begin{figure}[tbp]
\centering
\includegraphics[width=\linewidth]{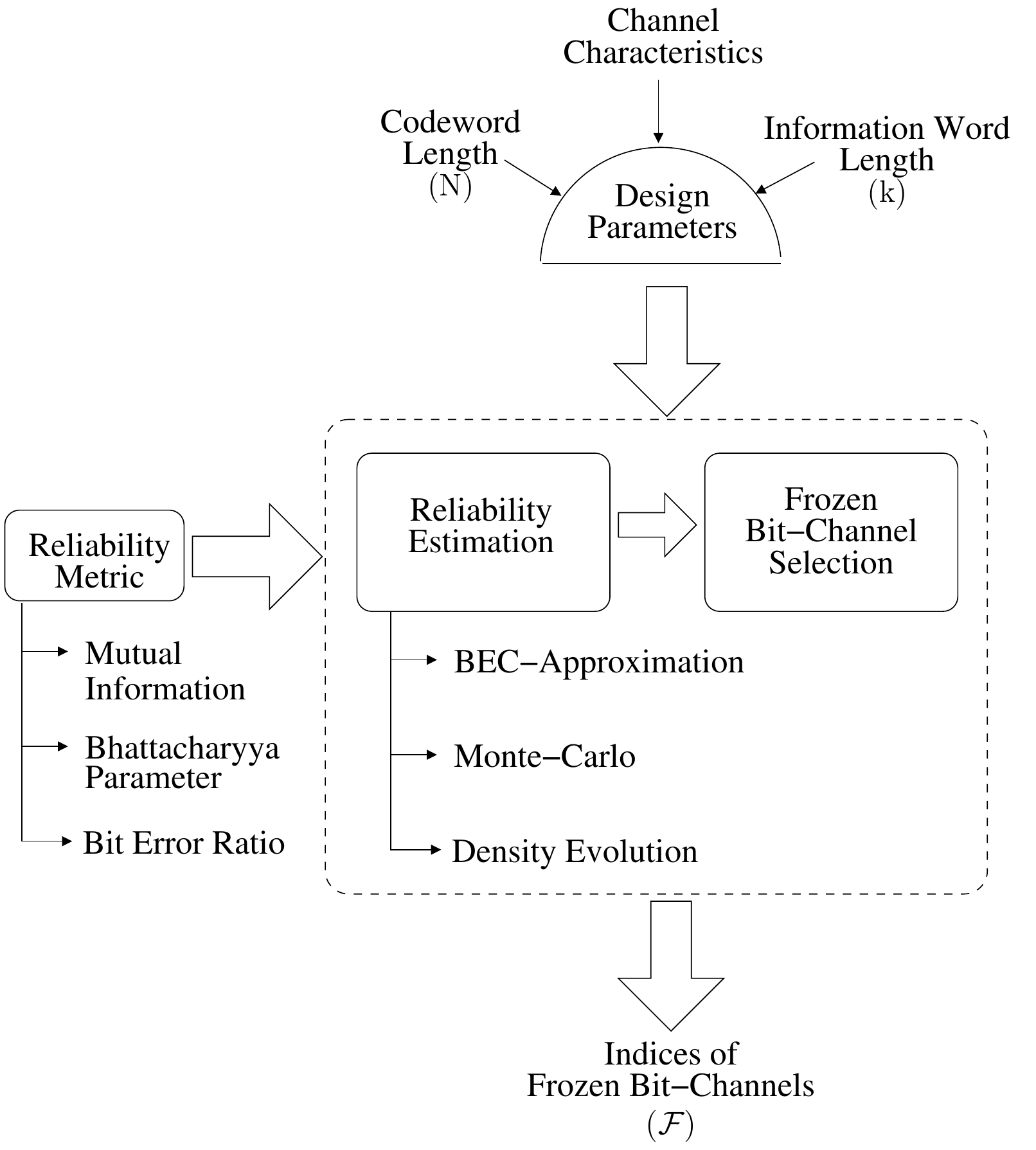}
\caption{Polar code design process.}
\label{fig:PolarDesign}
\end{figure}

The polar code design process relies on the selection of a suitable metric for quantifying the reliability of the induced bit-channels 
and an accurate reliability estimation method, which we will discuss in Sections~\ref{sec:sec:metrics} and~\ref{sec:sec:methods}, respectively. 
Once the reliability of each bit-channel is quantified, the ($N-k$) least reliable bit-channels may be selected as the frozen channels.
Alternatively, a threshold may be defined for rate compatible codes, so that all bit-channels having reliability less than the threshold
are frozen. The overall polar code design process is summarized in \fref{fig:PolarDesign}.
\subsection{Reliability Metrics} \label{sec:sec:metrics}
The reliability of bit-channels can be quantified in terms of the mutual information $I(W_i)$, or more specifically the capacity, of the induced bit-channels $W_i$.
A set of frozen bit-channels $\mathcal{F} \subset \{1, 2, \dots, N\}$ can then be selected based on the mutual information
such that we have:
\begin{equation}
 I(W_i) \leq I(W_j) \;\;\; \forall i \in \mathcal{F}, j \in \mathcal{F}_c.
\end{equation}
Alternatively, as proposed in Arikan's seminal paper~\cite{5075875}, reliability can also be calculated using
the Bhattacharyya parameter, since Bhattacharyya parameter gives an upper bound on the \ac{ML} decision error and is hence a more
accurate representative of the \ac{BLER}. The
Bhattacharyya parameter
is defined as follows: 
\begin{equation}
 Z(W_i) = \sum_{y_1^N,u_1^{i-1}} \sqrt{P_i\left(y_1^N,u_1^{i-1}|u_i=0\right)P_i\left(y_1^N,u_1^{i-1}|u_i=1\right)}.
 \label{eq:Bhattacharyya}
\end{equation}
A higher value of Bhattacharyya parameter indicates a lower reliability, i.e. $Z(W_i) \rightarrow 1$,
$I(W_i) \rightarrow 0$, and vice versa. Hence,
a set of frozen bit-channels $\mathcal{F} \subset \{1, 2, \dots, N\}$ can be selected based on the Bhattacharyya parameter
such that we have:
\begin{equation}
 Z(W_i) \geq Z(W_j) \;\;\; \forall i \in \mathcal{F}, j \in \mathcal{F}_c.
\end{equation}
Reliability can also be directly calculated in terms of the \ac{BER} of the induced bit-channels. All these three metrics, i.e.
mutual information, Bhattacharyya parameter and \ac{BER}, are interchangeably used in the literature as reliability metrics for 
the selection of frozen bit-channels.

\subsection{Reliability Estimation Methods} \label{sec:sec:methods}
The reliability metrics of Section~\ref{sec:sec:metrics} rely on the accurate computation of the transition probabilities 
$P_i(y_1^N,u_1^{i-1}|x_i)$ of the induced bit-channels.
Explicitly, as illustrated in \fref{polar-split-gen}, the $i$th induced bit-channel maps the input $u_i \in \mathcal{X}$
onto the output $(y_1^N,u_1^{i-1}) \in \mathcal{Y}^N \times \mathcal{X}^{i-1}$, where $\mathcal{X}$
and $\mathcal{Y}$ denote the input and output alphabets of the channel $W$. Hence, the cardinality of the output alphabet of the induced channel
is $|\mathcal{Y}^N| \times |\mathcal{X}^{i-1}|$, which grows exponentially with the codeword length $N$.
This in turn implies that the complexity of computing the exact channel transition probabilities grows exponentially
with the codeword length $N$. This is also evident from \eqr{eq:Bhattacharyya}, since the summation
in \eqr{eq:Bhattacharyya} is carried out over all possible values of $y_1^N$ as well as $u_1^{i-1}$.
Therefore, frozen bit-channel selection is deemed intractable. However, 
due to the recursive nature of polar codes, Bhattacharyya parameter as well as the mutual information
may be efficiently computed for \acp{BEC}. The recursive mutual information calculations were shown 
in \eqr{eq:recursive-cap}, while the Bhattacharyya parameter may be recursively calculated as follows:
\begin{align}
 Z\left(W^{N}_{2i-1}\right) &=  2 Z\left(W^{N/2}_{i}\right) - Z\left(W^{N/2}_{i}\right)^2  \nonumber \\
 Z\left(W^{N}_{2i}\right) &= Z\left(W^{N/2}_{i}\right)^2,
 \label{eq:recursive-Z}
\end{align}
with $Z(W_1) = \epsilon$, for the worse and better channels, respectively, where $\epsilon$ is the erasure probability of \ac{BEC}. This algorithm incurs a complexity cost of $O(N\log_2N)$. 

Since it is hard to track the exact  mutual information, Bhattacharyya parameter or \ac{BER} for channels other than \acp{BEC}, various
approximation methods have been proposed over the years. In~\cite{5075875}, Monte-Carlo simulations were invoked for estimating
the bit-channel reliabilities. More specifically, recall from Section~\ref{sec:sec:SC} that \ac{SC} decoder yields the 
channel transitions probabilities $P_i(y_1^N,u_1^{i-1}|x_i)$. Consequently, the Monte-Carlo based method operates by generating random
information and noise sequences for the given channel characteristics, and then estimating the Bhattacharyya parameter
or the mutual information using the output probabilities $P_i(y_1^N,u_1^{i-1}|x_i)$ of the \ac{SC} decoder, assuming that the decoder
knows the bits $u_1^{i-1}$. Equivalently, the \ac{BER} of the induced channels can be estimated based on the output of the \ac{SC} decoder.
This process incurs a complexity of $O(MN\log_2N)$ for $M$ rounds of Monte-Carlo simulations. However, the complexity of the individual
\ac{SC} operations can be reduced for symmetric channels by assuming an all-zero input. Explicitly, since the \ac{SC} decoder is assumed to know correctly the
bits $u_1^{i-1}$ for the purpose of channel reliability estimation, the $g$ operation of \eqr{g.eqn} reduces to:
\begin{equation}
L_{i+2^{j-1},j} = L_{i,j+1}+L_{i+2^{j-1},j+1}.\label{g.eqn.red}
\end{equation}
when the input is an all-zero sequence. This in turn implies that we do not need to carry out the partial sum operations of \eqr{PS1.eqn}.
Hence, the $f$ and $g$ operations of the \ac{SC} decoder can be computed in parallel, which significantly speeds up the process.
This approach is simple, but computationally intensive for long codeword lengths, in particular at high \acp{SNR}, 
because $M$ must be large enough to get reliable estimates. 

Polar codes for arbitrary binary-input channels can also be heuristically designed by considering a 
\ac{BEC} of equivalent capacity~\cite{4542778} or Bhattacharyya parameter~\cite{vangala2015comparative},
while \ac{DE} was invoked in~\cite{mori2009performance, 5166430} to calculate the
bit-channel reliabilities with a linear complexity of $O(N)$. The polar codes of~\cite{mori2009performance, 5166430},
which were customized for \ac{BSC} and \ac{AWGN} channel, outperformed the designs of~\cite{4542778}.
However, the \ac{DE}-based construction method of~\cite{mori2009performance, 5166430}
invokes convolution operations, whose exact implementation imposes exponentially increasing memory requirements.
Alternatively, the memory requirements can be reduced by approximating the convolutional operations using quantization (also called binning).
But this in turn leads to quantization errors. Hence, a suitable compromise has to be reached between the implementation complexity
and the accuracy. Tal and Vardy addressed this issue in~\cite{tal2013construct} by introducing two channel approximations
called the degraded and the upgraded quantizations, which provide a lower and upper bound, respectively, on the error probability
of the underlying channel. Both these approximation methods reduce the cardinality of the channel output based on the parameter $\mu$, so that
the channel outputs become tractable. The complexity of the resulting method is $O(N\mu^2\log_2\mu)$,
where the typical value of $\mu$ is $256$. 
These ideas were further investigated in~\cite{pedarsani2011construction} by exploiting alternate methods for approximating a degraded channel
and generalized to non-binary channels in~\cite{gulcu2018construction}. 

Inspired by the low-complexity \ac{GA}-based \ac{DE} of \ac{LDPC} codes,
Trifonov~\cite{6279525} used \ac{GA}-\ac{DE} for designing polar codes for \ac{AWGN} channels. \ac{GA}-\ac{DE}
tracks the mean value of the \acp{LLR} over the polar decoding circuit, assuming that the \acp{LLR} at all the nodes conform to a 
Gaussian distribution. More explicitly, given that an all-zero codeword is transmitted over an \ac{AWGN} channel, the channel \acp{LLR}
$L(y)$ exhibit a Gaussian distribution with a mean of $2/\sigma^2$ and a variance of $4/\sigma^2$, where $\sigma^2$ denotes 
the noise variance per dimension. The mean of the \acp{LLR} of the polar decoding circuit can then be approximated as follows:
\begin{align}
 \textup{E}\left[L^{N}_{2i-1}\right] &=  \phi^{-1} \left(1 - \left(1 - \phi \left(\textup{E}\left[L^{N/2}_{i}\right]\right) \right)^2\right) \nonumber \\
 \textup{E}\left[L^{N}_{2i}\right] &= 2 \; \textup{E}\left[L^{N/2}_{i}\right],
 \label{eq:GA-DE1}
\end{align}
for the worse and better channels, respectively. In \eqr{eq:GA-DE1}, $\textup{E}$ denotes the expectation (or equivalently mean) operation and the function $\phi$ is defined as follows:
 \begin{equation}
  \phi (x) =  \left\{
\begin{array}{l l}
 1 - \frac{1}{4 \pi x} \int_{-\infty}^{\infty} \tanh \frac{u}{2} \; e^{-\frac{(u-x)^2}{4x}} \; du & \text{if} \;\; x > 0 \\ \\
 1 & \text{if} \;\; x = 0, \\
\end{array}
\right.
\label{eq:phi-GA}
 \end{equation}
which may be approximated as~\cite{chung2001analysis, ha2004rate}:
 \begin{equation}
  \phi (x) \approx  \left\{
\begin{array}{l l}
 e^{-\alpha x^2 + b x} & \text{for} \;\; 0 \leq x < c \\ 
 e^{-\alpha x^\gamma + \beta}  & \text{for} \;\; c \leq x < 10 \\
 \sqrt{\frac{\pi}{x}} e^{-\frac{x}{4}} \left( 1 - \frac{10}{7x}\right) & \text{for} \;\; 10 \leq x,
\end{array}
\right.
\label{eq:phi-GA2}
 \end{equation}
where $a=-0.0564$, $b=0.48560$, $c=0.867861$, $\alpha=-0.4527$, $\beta=0.0218$ and $\gamma=0.86$. Furthermore, the inverse function
$\phi^{-1}$ of \eqr{eq:GA-DE1} can be estimated using numerical analysis techniques, for example the bisection method or the Newton-Raphson method.
The mean \acp{LLR} of \eqr{eq:GA-DE1} may
then be used for approximating the \ac{BER} of the corresponding bit-channels using:
\begin{equation}
 \text{BER}_i \approx Q \left(\sqrt{ \textup{E}\left[L^{N}_{i}\right]}\right).
\end{equation}
Alternatively, the mean \acp{LLR} may be used for calculating the mutual information or the Bhattacharyya parameter.

The aforementioned polar code construction methods rely on the \ac{SC} decoder. Vangala~\etal~\cite{vangala2015comparative} compared these construction methods and demonstrated that all
perform equally well provided that the design \ac{SNR} is carefully chosen. Qin~\etal~\cite{7828081} proposed an improved reliability estimation method tailored for the \ac{BP} decoder, which tracks the evolution \acp{LLR} during the \ac{BP} decoding process. However, only a marginal improvement of upto $0.5$~dB was reported. In~\cite{DBLP:journals/corr/YuanPB17}, the polar code was heuristically optimized for the \ac{SCL} decoder. Inspired by these decoder-specific polar code designs, reinforcement learning techniques and genetic algorithms were invoked in~\cite{DBLP:journals/corr/abs-1904-07511} and~\cite{DBLP:journals/corr/abs-1901-06444, DBLP:journals/corr/abs-1901-10464} to customize the code design for the given polar decoder.

We have summarized the discussions of this section in \tref{tab:frozen-summary}.
\begin{table}[tbp]
 \centering
 \begin{center}
\begin{tabular}[tbp]{|l|l|}
\hline
\textbf{Design Parameters} & Codeword Length ($N$)\\
                           & Information word Length ($k$)\\
                           & Channel characteristics \\ \hline 
\textbf{Reliability Metric} & Mutual Information \\
& Bhattacharyya Parameter \\
& Bit Error Rate \\ \hline
\textbf{Estimation Method} & Monte-Carlo~\cite{5075875} \\
& BEC-approximation~\cite{4542778, vangala2015comparative}\\
& Density Evolution~\cite{mori2009performance, 5166430}\\
& Degraded/Upgraded Quantization~\cite{tal2013construct}\\
& Gaussian Approximation Density Evolution~\cite{6279525} \\ 
& Decoder-specific methods~\cite{7828081,DBLP:journals/corr/YuanPB17,DBLP:journals/corr/abs-1904-07511,DBLP:journals/corr/abs-1901-06444, DBLP:journals/corr/abs-1901-10464} \\ \hline
\end{tabular}
 \end{center}
 \caption{Frozen bit-channel selection procedure for polar codes.}
\label{tab:frozen-summary}
\end{table}
\subsection{Design Examples} \label{sec:sec:ex-costruction}
\begin{figure}[tb]
        \centering
        \begin{subfigure}[]{0.9\linewidth}
        \begin{center}
                \includegraphics[width=\linewidth]{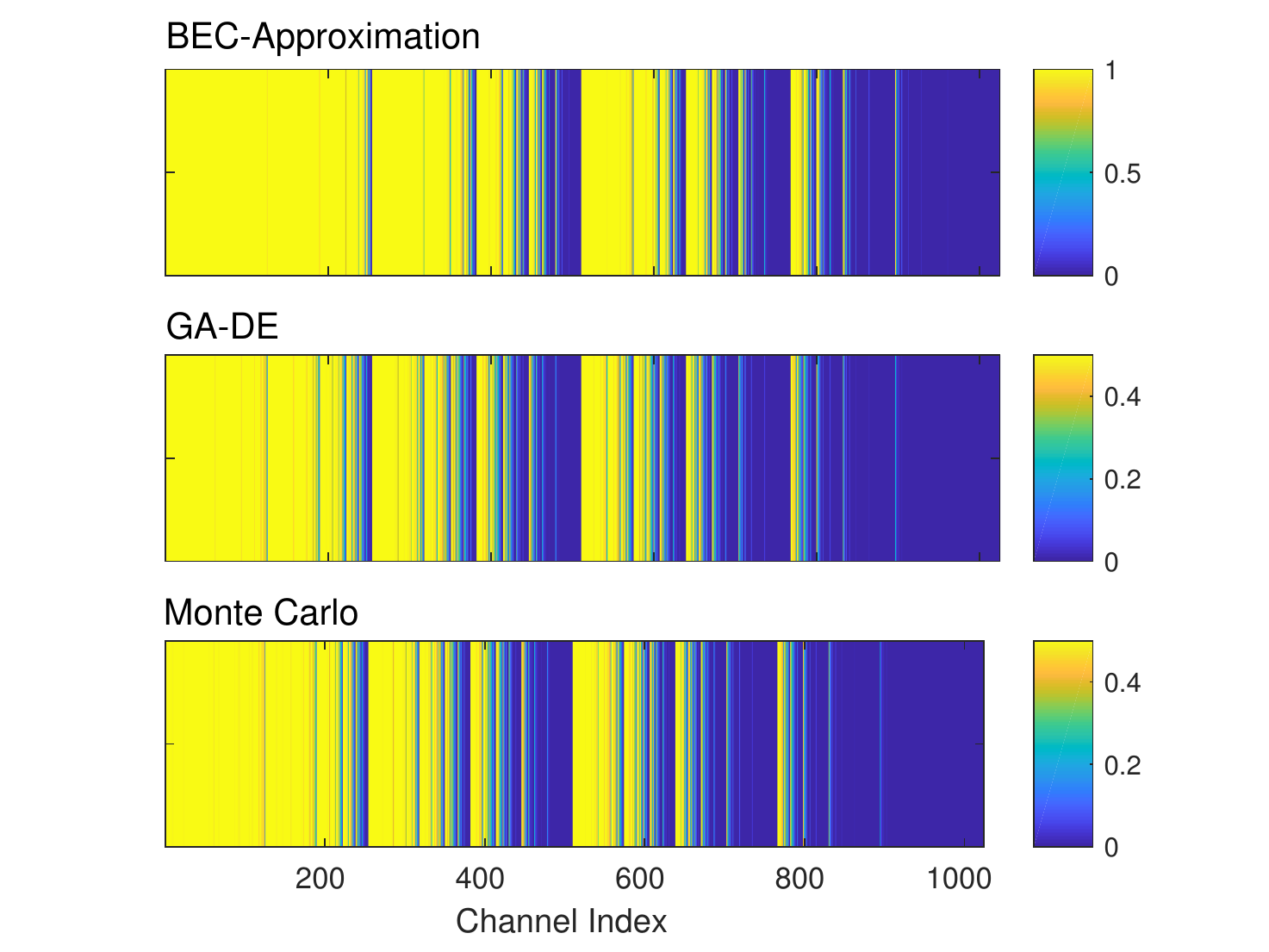}
                \caption{Reliability intensity. Please note that a value of $0$ implies high reliability, since the reliability metric
                is the Bhattacharyya parameter for \ac{BEC}-approximation and the \ac{BER} for GA-DE and Monte Carlo.}
                \label{fig:design0dB-val}
        \end{center}
        \end{subfigure}%
        \vfill 
        \begin{subfigure}[]{0.9\linewidth}
        \begin{center}
                \includegraphics[width=\linewidth]{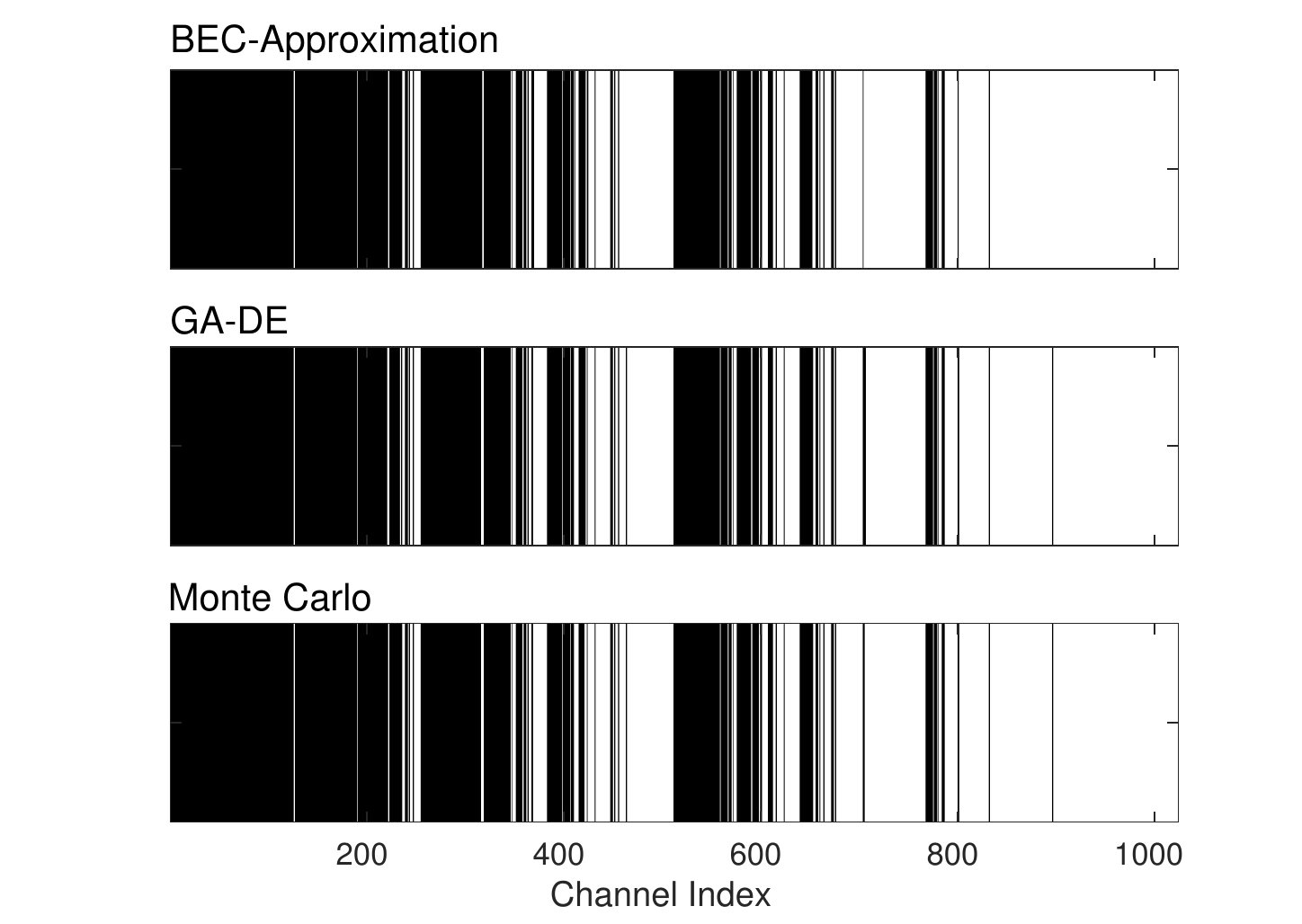}
                \caption{Frozen bit-channel patterns. Black region denotes the frozen indices, while the white one is for information indices.}
                \label{fig:design0dB-bits}
        \end{center}
        \end{subfigure}
        \caption{Comparison of polar code construction methods at $\frac{E_b}{N_0} = 0$~dB: \ac{BEC}-approximation based Bhattacharyya parameter, \ac{GA}-\ac{DE} based \ac{BER}
        and Monte Carlo based \ac{BER}.}
         \label{fig:design0dB}
\end{figure}

In this section, we will compare the different reliability estimation methods of \fref{fig:PolarDesign} by designing a $1/2$-rate
polar code having a codeword length of $N = 1024$ for an \ac{AWGN} channel. In particular, we compare the \ac{BEC}-approximation of~\cite{4542778, vangala2015comparative},
the GA-DE of~\cite{6279525} and the Monte Carlo based method~\cite{5075875}. Furthermore, we chose Bhattacharyya parameter for quantifying
reliability using the \ac{BEC}-approximation method, while \ac{BER} was used for the other two methods. \fref{fig:design0dB-val} records the 
resulting intensity maps for the three methods at $E_b/N_0 = 0$~dB. Explicitly, a value of $0$ in the intensity map of \fref{fig:design0dB-val}
corresponds to the maximum reliability, while a value of $1$ denotes the lowest reliability.
We may notice that there are only slight variations in the three intensity maps, with the GA-DE
and Monte Carlo methods being very similar. 
We next classify the $512$ least reliable bit-channels as frozen bit-channels and plot the frozen bit-channel patterns in \fref{fig:design0dB-bits},
where the frozen bit-channels are colored in black, while the information bit-channels are in white.
Again, the frozen bit-channel patterns are similar for the three methods.

\begin{figure}[tbp]
\centering
\includegraphics[width=\linewidth]{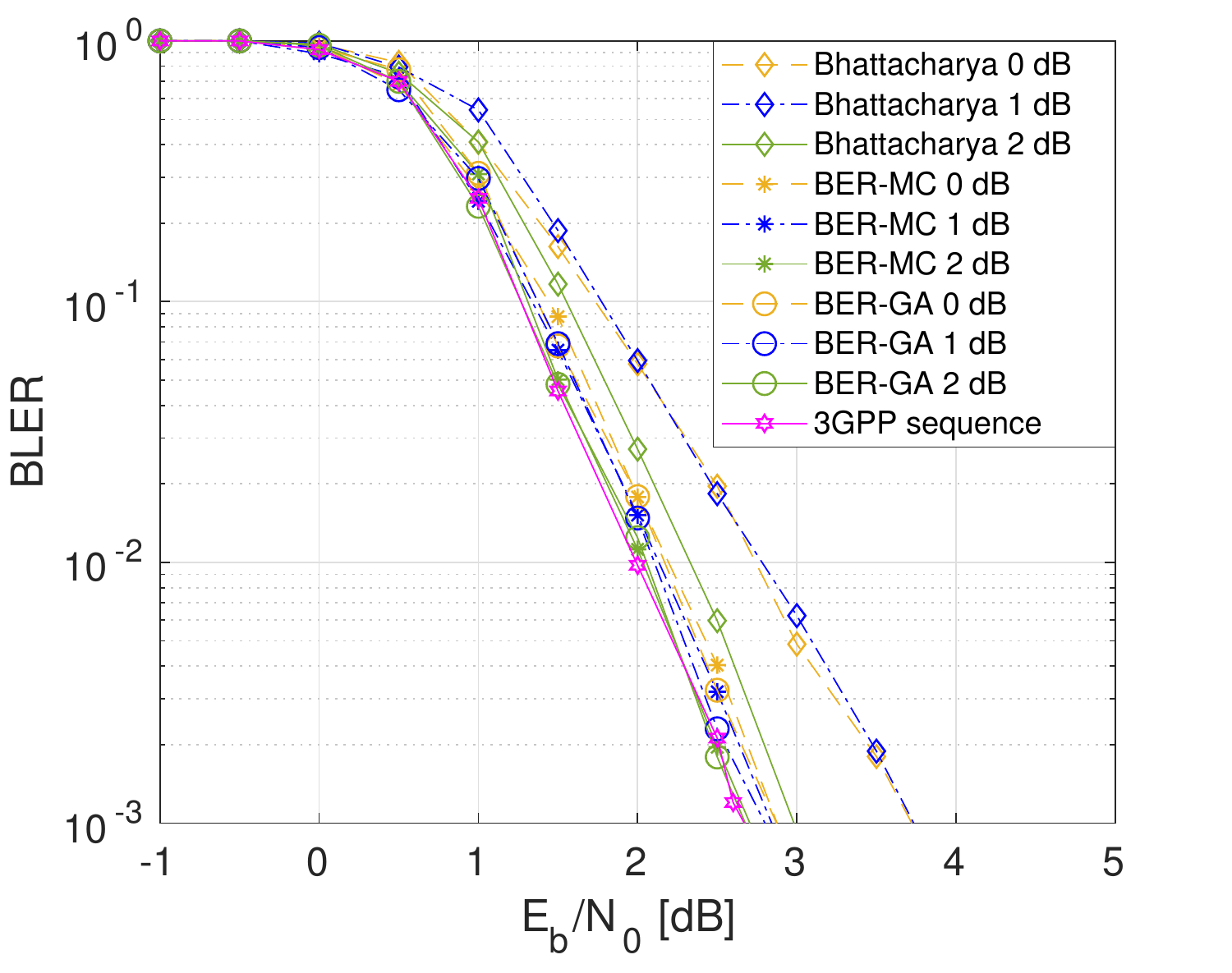}
\caption{Comparison of the achievable \ac{BLER} performance of the $(1024,512)$ polar codes constructed using the various reliability estimation 
methods of Section~\ref{sec:sec:methods} at different design \acp{SNR} per bit. The performance is benchmarked against the \ac{3GPP} $5$G~\ac{NR} frozen bit-channel sequence. \ac{SCL} decoder
having $L = 8$ and \ac{QPSK} transmission over an \ac{AWGN} channel was used.}
\label{fig:Results:PolarConstr}
\end{figure}

In \fref{fig:Results:PolarConstr}, we compare the achievable \ac{BLER} performance of the $(1024,512)$ polar codes constructed using the 
\ac{BEC}-approximation based Bhattacharyya parameter, \ac{GA}-\ac{DE} based \ac{BER} and the Monte Carlo based \ac{BER} at different design 
\acp{SNR} per bit. Inline with \fref{fig:design0dB}, the performance of the polar codes constructed using the Monte Carlo and the 
\ac{GA}-\ac{DE} methods is similar, with the latter being slightly better, while that of the polar codes designed using the \ac{BEC}-approximation
method is worse. Furthermore, \fref{fig:Results:PolarConstr} shows that the \ac{BLER} performance of the 
Monte Carlo and the \ac{GA}-\ac{DE} methods is only sightly affected, when the design \ac{SNR} per bit is 
increased from $0$~dB to $2$~dB.
It is obvious that if we 
choose a very low or high value of the design \ac{SNR} per bit, then the performance will get worse, 
as demonstrated in~\cite{vangala2015comparative}. 
Hence, the polar code construction method is not very sensitive to reasonable discrepancies between the design \ac{SNR} and the operating
\ac{SNR}.
We have also benchmarked the designed polar
codes against the $3$GPP~$5$G~NR polar code in \fref{fig:Results:PolarConstr}. It may be observed that the performance of the 
Monte Carlo and the \ac{GA}-\ac{DE} based polar codes approaches that of the $3$GPP~$5$G~NR when the design $E_b/N_0 = 2$~dB.

\section{Quantum-to-Classical Isomorphism} \label{sec:isomorphism}
In contrast to a classical bit, which can be either $0$ or $1$ at any particular instant, a quantum bit (qubit)\footnote{Please refer to~\cite{babar2018duality} 
for deeper insights into the duality of classical and quantum regime.} exists in superposition of the orthogonal basis states $\ket{0}$
and $\ket{1}$. This superimposed state of the qubit is generally described using the state vector as follows:
\begin{equation}
 \ket{\psi} = \alpha \ket{0} + \beta \ket{1},
 \label{eq:qubit}
\end{equation}
where $\ket{ \cdot }$ is called Ket or Dirac notation~\cite{dirac82}, and $\alpha$ and $\beta$ are complex coefficients conforming to:
\begin{equation}
 |\alpha|^2 + |\beta|^2 = 1.
\end{equation}
Furthermore, unlike a classical bit, which can be `observed' (or `measured') without disturbing its value, any observation of the qubit
perturbs its superimposed state of \eqr{eq:qubit}. To elaborate, if a qubit is observed in the computational basis\footnote{The pair of
orthogonal basis states $\ket{0}$ and $\ket{1}$ is called computational basis.}, it may collapse to the state $\ket{0}$ 
with a probability of $|\alpha|^2$ and the state $\ket{1}$ with a probability of $|\beta|^2$. The quantum superposition of 
\eqr{eq:qubit} makes quantum processing (or computation) systems inherently parallel, while the observation property together with the quantum 
no-cloning theorem\footnote{According to the quantum no-cloning theorem, arbitrary quantum states cannot be cloned 
(or copied)~\cite{citeulike:507853}.} makes quantum transmissions absolutely secure. However, these unusual quantum characteristics,
which have no counterpart in the classical domain, make it challenging to design \acp{QECC}. 
Nonetheless, there exists an underlying isomorphism between the classical and quantum paradigms, which can be exploited for designing efficient
\acp{QECC} 
from the known classical codes~\cite{babar2018duality, zbabar2015_1, zbabar2015_3}.

Environmental decoherence is a major source of noise in quantum systems. It can be modeled using a depolarizing channel, which is considered
the `worst-case scenario'~\cite{mark:book}. Explicitly, a quantum depolarizing channel characterized by the depolarizing probability $p$
independently inflicts an error on each qubit, such that a qubit may experience a bit-flip (Pauli-$\mathbf{X}$), a phase-flip (Pauli-$\mathbf{Z}$)
or a bit-and-phase-flip (Pauli-$\mathbf{Y}$) error with a probability of $p/3$ each\footnote{The 
$\mathbf{I}$, $\mathbf{X}$, $\mathbf{Y}$ and $\mathbf{Z}$ are single-qubit Pauli operators (or gates) defined as:
\begin{equation}
 \mathbf{I}=\begin{pmatrix}
  1& 0 \\
  0& 1
\end{pmatrix}, \;
\mathbf{X}=\begin{pmatrix}
  0& 1 \\
  1& 0
\end{pmatrix}, \;
\mathbf{Z}=\begin{pmatrix}
  1& 0 \\
  0& -1
\end{pmatrix}, \;
\mathbf{Y}=\begin{pmatrix} 
  0& -i \\
  i& 0
\end{pmatrix}. \nonumber
\end{equation}}.
Hence, in contrast to the classical channels, which only inflict bit-flip errors, the quantum depolarizing channel imposes both 
bit-flip as well as phase-flip errors. An interesting point to notice here is that a quantum depolarizing channel may also be viewed as a 
pair of correlated \acp{BSC} inflicting bit-flips and phase-flips respectively.
Fortunately, classical polar codes are capable of concurrently polarizing both the bit-flip as well as the phase-flip channels, 
when the classical \ac{XOR} gates are replaced by the quantum \ac{CNOT} gates\footnote{The \ac{CNOT} gate is a two-qubit gate, having a control qubit and a 
target qubit. When the control qubit is in state $\ket{1}$, the target qubit is flipped; otherwise, the target qubit is left unchanged, as
encapsulated below:
\begin{equation}
 \mathbf{\text{CNOT}} \left(\ket{\psi_0},\ket{\psi_1}\right) = \ket{\psi_0} \otimes \ket{\psi_0 \oplus \psi_1}, \nonumber
\end{equation}
where $\ket{\psi_0}$ is the control qubit, while $\ket{\psi_1}$ is the target qubit. Hence, \ac{CNOT} gate is the quantum analogue 
of the classical \ac{XOR} gate.}; hence achieving the quantum channel capacity, 
as demonstrated in\cite{6472318,6781023}. This is because polar codes merely rely on \ac{CNOT} gates for channel polarization, 
which are capable of concurrently polarizing both the bit-flip as well as the phase-flip channels, but in opposite directions, as further discussed below.

\begin{figure}[tbp]
        \centering
        \includegraphics[width=\linewidth]{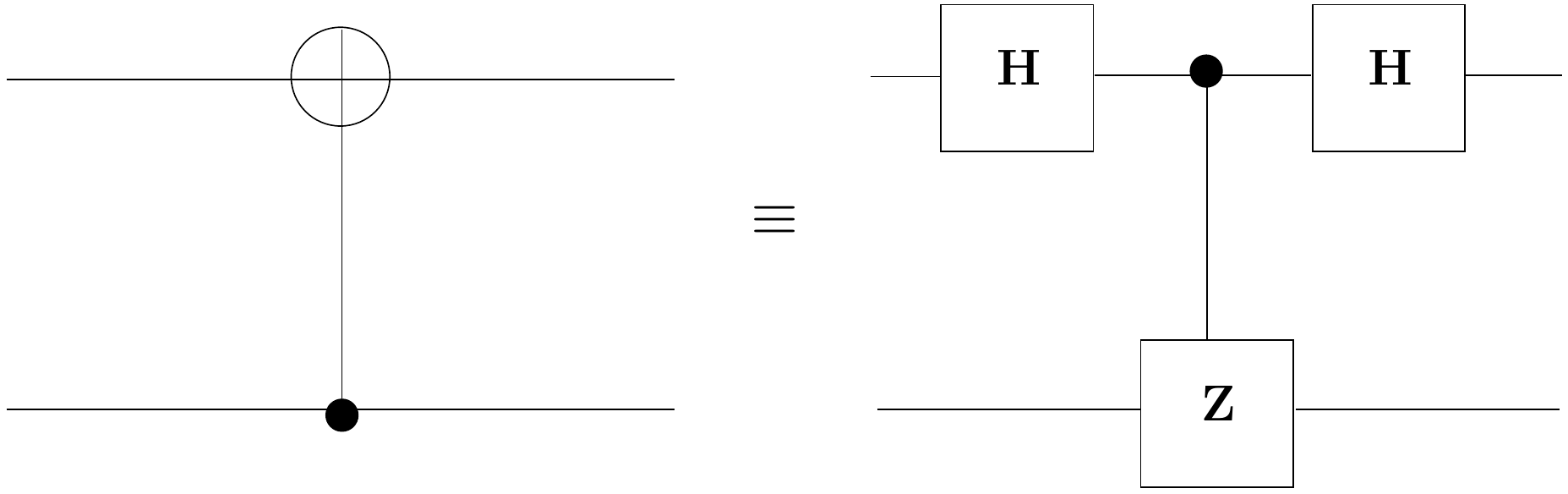}
        \caption{A CNOT gate is equivalent to a Controlled-$\mathbf{Z}$ (Cz) gate, with the control and target qubits swapped, when Hadamard gates are invoked at the
        input and output. The circuit to the left flips the top qubit in computational basis ($\ket{0}$ and $\ket{1}$), when the bottom qubit is in the state $\ket{1}$,
        while the circuit to the right flips the botton qubit in Hadamard basis ($\ket{+}$ and $\ket{-}$), when the top qubit is in the state $\ket{-}$.}
         \label{fig:Cnot_eq}
\end{figure}

\acp{QECC} exploit the computational basis for correcting bit-flip errors,
while the Hadamard basis\footnote{The pair of orthogonal basis states $\ket{+}$ and $\ket{-}$ is called Hadamard basis,
where the basis states $\ket{+}$ and $\ket{-}$ are defined as:
\begin{equation}
 \ket{+} \buildrel\triangle\over = \frac{1}{\sqrt{2}} \left( \ket{0} + \ket{1}\right), \; \;
 \ket{-} \buildrel\triangle\over = \frac{1}{\sqrt{2}} \left( \ket{0} - \ket{1}\right). \nonumber
\end{equation}} is used for phase-flip correction. The action of CNOT on the Hadamard basis is analogous to that on the computational basis
with the role of control and target qubits swapped. Explicitly, let us consider the two-qubit state $\ket{+-}$ in Hadamard basis, which is equivalent to:
\begin{align}
 \ket{+-} &= \frac{1}{\sqrt{2}} \left(\ket{0} + \ket{1}\right) \otimes \frac{1}{\sqrt{2}}\left(\ket{0} - \ket{1}\right) \nonumber \\
 &= \frac{1}{2} \left( \ket{00} - \ket{01} + \ket{10} - \ket{11}\right),
\end{align}
in the computational basis. When a CNOT gate is applied to the second qubit controlled by the first qubit, we get:
\begin{align}
 \ket{+-} \;\; \underrightarrow{\text{CNOT}(1,2)} \;\; &\frac{1}{2} \left( \ket{00} - \ket{01} + \ket{11} - \ket{10}\right) = \ket{--}. 
\label{eq:CNOT-H}
 \end{align}
We may observe in \eqr{eq:CNOT-H} that the operation of CNOT$(1,2)$ on the computational basis is equivalent to that of 
$\overline{\text{CNOT}}(2,1)$ on the Hadamard basis. More explicitly, the classic CNOT$(i,j)$ acting on the computational
basis flips the $j$th qubit (target) between $\ket{0}$ and $\ket{1}$, when the $i$th qubit (control) is in the state $\ket{1}$.
This operation is analogous to that of $\overline{\text{CNOT}}(j,i)$ acting on the Hadamard basis, which 
flips the $i$th qubit (target) between $\ket{+}$ and $\ket{-}$, when the $j$th qubit (control) is in the state $\ket{-}$. 
More specifically, a CNOT gate may also be implemented using a \ac{Cz} gate, with the control and target qubits swapped, as shown in 
\fref{fig:Cnot_eq}.

\begin{figure}[tbp]
        \centering
        \begin{subfigure}[]{0.5\linewidth}
        \begin{center}
                \includegraphics[width=\linewidth]{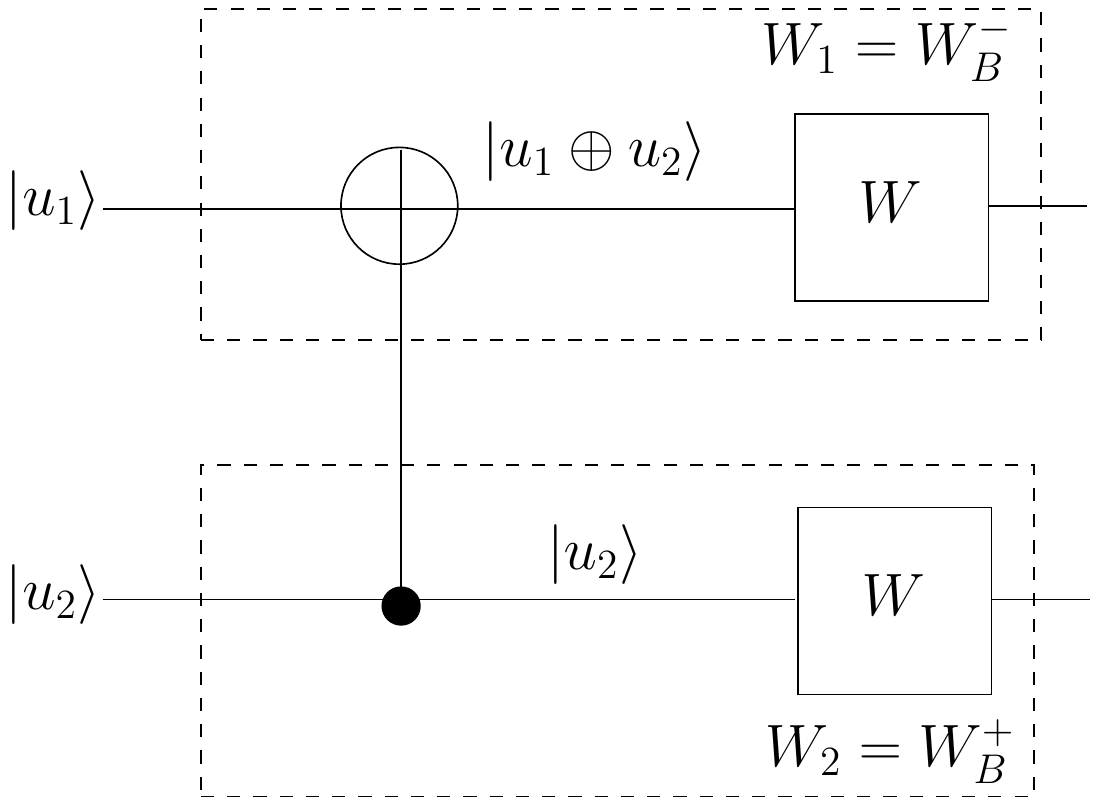}
                \caption{Polarization in computational basis, resulting in polarized bit-flip channels.}
                \label{fig:kernel1}
        \end{center}
        \end{subfigure}%
        ~ 
        \begin{subfigure}[]{0.5\linewidth}
        \begin{center}
                \includegraphics[width=\linewidth]{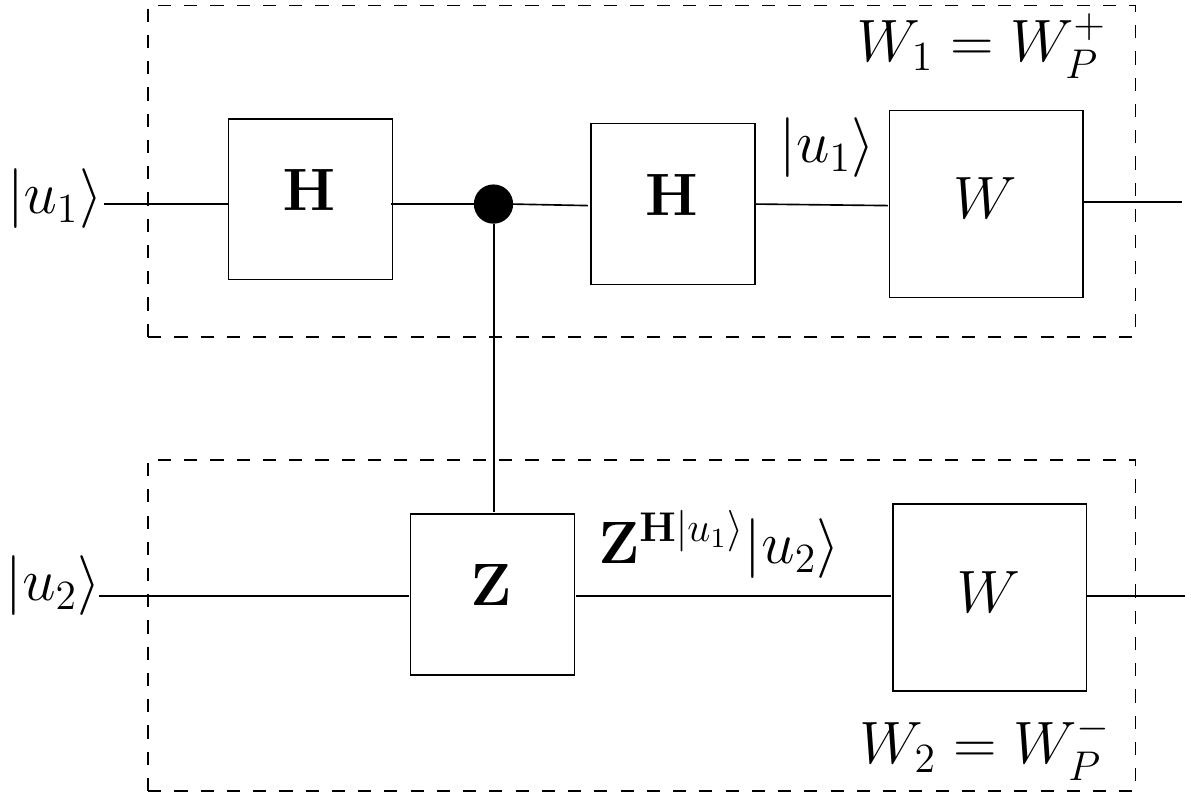}
                \caption{Polarization in Hadamard basis, resulting in polarized phase-flip channels.}
                \label{fig:kernel2}
        \end{center}
        \end{subfigure}
        \caption{The $2$-qubit kernal of a quantum polar code in the computational and Hadamard basis.}
         \label{fig:kernel}
\end{figure}

Let us now consider the $2$-qubit kernel of polar code given in \fref{fig:kernel1}, which is quantum analogue of
Arikan's kernal of \fref{fig:ch_comb}. 
From the perspective of computational basis, the encoder
of \fref{fig:kernel1} combines the information of the two channels in computational basis, so that the computational basis information of the $\ket{u_2}$
is redundantly added to that of $\ket{u_1}$ ($\oplus$ denotes modulo-$2$ addition in the computational basis). Consequently, 
the second qubit-channel $W_2$ becomes more robust against bit-flip errors at the 
cost of deteriorating the robustness first qubit-channel $W_1$. In other words, $W_2$ tends to polarize 
towards a better bit-flip channel $W_B^+$, while $W_1$ tends to polarize towards a worse bit-flip channel $W^-_B$. However, the total capacity
of the two channels is conserved.

\fref{fig:kernel2} shows the polar encoder from the perspective of Hadamard basis; hence, the \ac{CNOT} gate is replaced by the equivalent circuit 
of \fref{fig:Cnot_eq}, which invokes the \ac{Cz} gate. More explicitly, in \fref{fig:kernel2}, the Hadamard basis information of the first qubit $\ket{u_1}$ 
is redundantly incorporated in the Hadamard basis information of the second qubit. Consequently, the information carrying capacity
of the first qubit-channel $W_1$ is enhanced in the phase basis, while that of the second qubit-channel $W_2$ degrades in the phase basis.
Hence, the first qubit-channel $W_1$ tends to polarize towards a better phase-flip channel, while the second qubit-channel $W_2$ tends to polarize towards a worse 
phase-flip channel, denoted by $W^+_P$ and $W^-_P$, respectively. Therefore, the elementary kernal of a polar code is capable of 
polarizing both the bit-flip as well as phase-flip channels, but the direction of polarization is opposite, as illustrated in \fref{fig:kernel}.

Based on the above discussions, a quantum polar code induces four sets of channels,
covering both the bit-flip as well as the phase-flip errors. Explicitly, the polarized channels may belong to one of the following four sets:
\begin{enumerate}
 \item \textbf{Good bit-and-phase channels} ($\mathcal{F}^c$): These induced channels exhibit high information carrying capacity in the computational basis (bit-flip) as well
 as the Hadamard basis (phase-flip). Consequently, these induced channels are used for transmitting the uncoded qubits. 
 \item \textbf{Good bit-only channels} ($\mathcal{F}_P$): These channels have high information carrying capacity in the 
 computational basis (bit-flip), but low capacity in the Hadamard basis (phase-flip). Hence, these channels are frozen in the Hadamard basis
 by transmitting the Hadamard basis states $\ket{+}$ or $\ket{-}$, which are known to the receiver. 
 \item \textbf{Good phase-only channels} ($\mathcal{F}_B$): These channels have low information carrying capacity in the computational basis (bit-flip),
 but hight capacity in the Hadamard basis (phase-flip). Consequently, they are frozen in the computational basis, hence transmitting the 
 computational basis states $\ket{0}$ or $\ket{1}$, which are known to the receiver.
 \item \textbf{Bad bit-and-phase channels} ($\mathcal{F}_{BP}$): These channels have low information carrying capacity in both
 the computational basis (bit-flip) as well as the Hadamard basis (phase-flip). 
Consequently, these channels are frozen in both the computational as well as the Hadamard basis. This is achieved by exploiting pre-shared 
entangled\footnote{ `Entanglement', which Einstein termed as a `spooky action at a distance'~\cite{Born1971}, is the mysterious, 
correlation-like property between two or more qubits, which implies that the entangled qubits cannot be expressed as the tensor product of the 
individual qubits. Furthermore, a strange relationship exists between the two entangled qubits, which entails that measuring one of 
them also reveals the value of the other, even if they are geographically separated.}
qubits, which are referred to as ebits. Explicitly, ebits are created in the Bell state $\ket{\phi^+}$, expressed as: 
\begin{equation}
 \ket{\phi^+} = \frac{\ket{00}^{T_XR_X} + \ket{11}^{T_XR_X}}{\sqrt{2}},
 \label{eq:ebit}
\end{equation}
so that the first qubit is retained at the transmitter ($T_X$), while the associated entangled qubit is sent to the receiver ($R_X$)
before actual transmission commences, for example during off-peak hours, when the channels are under-utilized. It is generally assumed that the 
pre-sharing of ebits takes place over a noiseless quantum channel. Hence, quantum polar codes intrinsically belong to the family
of entanglement-assisted \acp{QECC}~\cite{bowen2002, brun2006}.
\end{enumerate}
\section{Quantum Polar Codes} \label{sec:Qpolarcode}
Inspired by the provably capacity achieving nature of classical polar codes as well as their efficient encoding and decoding structures,
Wilde and Guha~\cite{wilde_polar2013} were the first to demonstrate the existence of the channel polarization phenomenon
for classical-quantum channels, which transmit classical information over quantum channels. These ideas were later extended to the transmission of
quantum information in~\cite{6472318}. The quantum polar codes of~\cite{wilde_polar2013, 6472318} exploit the same encoder as Arikan's
polar codes, except that the classical \ac{XOR} gates are replaced by the quantum \ac{CNOT} gates. Consequently, the quantum polar encoders
of~\cite{wilde_polar2013, 6472318} inherently benefit from the low encoding complexity of Arikan's classic polar codes. Furthermore,
a quantum counterpart of the classical \ac{SC} decoder, named \ac{QSCD}, was conceived in~\cite{wilde_polar2013, 6472318}, which
makes collective measurement on all channel uses. This is achieved by exploiting quantum hypothesis
testing~\cite{helstrom1971quantum, holevo1972analog} in conjunction with Sen's noncommutative union bound~\cite{senachieving}.
The \ac{QSCD} of~\cite{wilde_polar2013, 6472318} failed to match the low decoding complexity of the classical \ac{SC}
decoder. This issue was addressed by Renes~\etal~in~\cite{renes2012}, where an efficient implementation of a quantum polar decoder was given
for quantum Pauli\footnote{A quantum Pauli channel independently inflicts an error on
each qubit, such that a qubit may experience a bit-flip (Pauli-$\mathbf{X}$), a phase-flip (Pauli-$\mathbf{Z}$) and a bit-and-phase-flip 
(Pauli-$\mathbf{Y}$) error with a probability of $p_x$, $p_z$ and $p_y$, respectively. Quantum depolarizing channel
is a special case of a Pauli channel having $p_x = p_z = p_y = p/3$.} and erasure channels. Finally, Wilde and Renes
combined their efforts  in~\cite{wilde_polar2012, 6781023} to present an efficient \ac{QSCD} for arbitrary quantum channels.
The quantum polar codes of~\cite{wilde_polar2013, 6472318, renes2012} rely on 
the sharing of noiseless ebits between the transmitter and the receiver. In this context, the first unassisted quantum polar codes,
consisting of concatenated bit-flip (computational basis) and phase-flip correction (Hadamard basis) polar transformations,
were recently conceived in~\cite{renes2015efficient}. The authors of~\cite{renes2015efficient} also presented efficient encoding and decoding 
implementations for Pauli and erasure channels. In the midst of these advancements, Ferris and Poulin~\cite{ferris2014tensor}
developed a new family of \acp{QECC} based on tensor networks, called the branching \ac{MERA} codes, which is a generalization of quantum polar codes.
In contrast to the quantum polar codes of~\cite{wilde_polar2013, 6472318, renes2012, wilde_polar2012, 6781023},
the tensor network based quantum polar codes of~\cite{ferris2014tensor} invoke syndrome-based classical decoding for estimating channel errors
encountered over the quantum Pauli or erasure channels.

\subsection{Encoder} \label{sec:sec:Qenc}
Recall from Section~\ref{sec:isomorphism} that a quantum depolarizing channel is equivalent to two correlated classical \acp{BSC}. 
The correlation between these two channels can be ignored in the spirit of simplifying the design process, while compromising on the 
achievable performance. Hence, a quantum depolarizing channel having a depolarizing probability
of $p$ may be modeled using two independent \acp{BSC} having a cross-over probability of $2p/3$. This model is widely used for 
constructing \ac{CSS}-type quantum codes~\cite{steane, CS, steane2}, which are designed to independently correct bit-flip and phase-flip errors. 
Consequently, qubits frozen in the computational and Hadamard
basis can be independently determined by finding qubit locations which yield the highest mutual information
for the bit-flip and phase-flip channels, respectively. Furthermore, since the polarization of a quantum bit-flip channel is 
identical to that of a classical channel, as illustrated in \fref{fig:kernel1}, qubits frozen in the computational basis 
can be selected by running the classical frozen bit-channel selection procedure for a \ac{BSC} having a cross-over probability of $2p/3$ (given that
the design depolarizing channel probability is $p$). 
By contrast, we observed in \fref{fig:kernel}, that the polarization of a quantum phase-flip channel is the same as the classical channel polarization (or equivalently
quantum bit-flip channel polarization), but the direction of polarization is opposite. This in turn implies that the pattern of quantum phase-flip
polarization can be obtained by swapping the control and target qubits of the $2$-bit kernel of a classical polar code. The resulting  
$2$-bit kernel for phase-flip channel is given by:
 \begin{equation}
 \tilde{G}_2  = G_2^T = \begin{pmatrix}
        1 & 1\\
        0 & 1
       \end{pmatrix},
       \label{eq:kernal-phase}
\end{equation}
and the equivalent $N$-bit encoder $\tilde{G}_N$ is:
\begin{equation}
\tilde{G}_N = (G_2^T)^{\otimes n} = G_N^T.
 \label{eq:encoder:G:ph}
\end{equation}
Hence, the location of qubits frozen in the computational basis may be determined by invoking the encoder $G_N$, while that of the Hadamard basis
may be selected by using the encoder $G_N^T$. 

\begin{figure}[tbp]
        \begin{center}
                \includegraphics[width=0.8\linewidth]{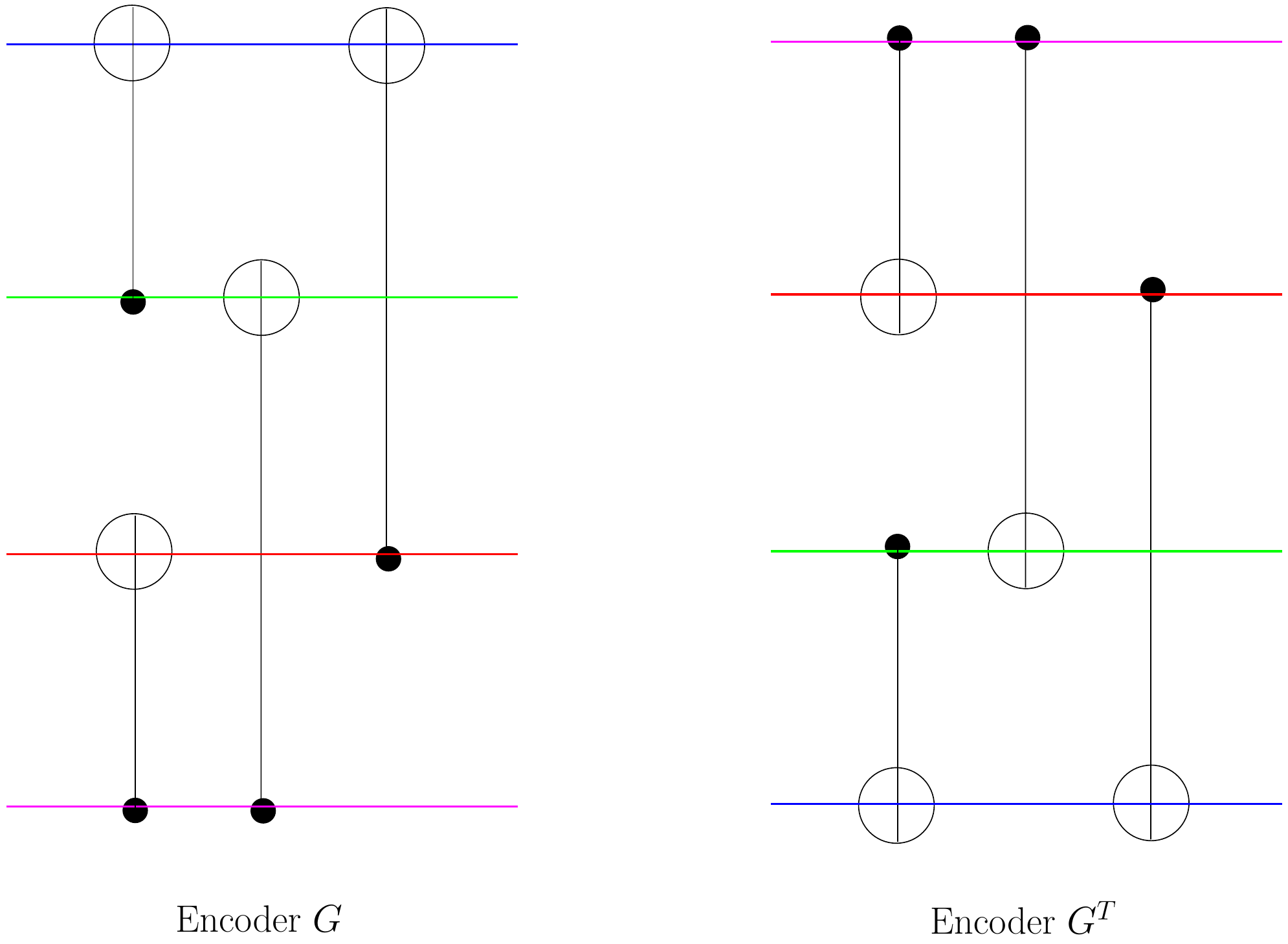}
                \caption{Encoders $G$ and $G^T$ used for selecting frozen channel indices for quantum bit-flip and phase-flip channels, respectively.}
         \label{fig:encoder1}
                \end{center}
\end{figure}

\fref{fig:encoder1} shows the encoders $G$ and $G^T$ for selecting frozen channel indices in the computational and Hadamard basis, respectively, when the codeword length is $4$.
We may observe that the $i$th bit of encoder $G$ is equivalent to the $(4 - i + 1)$th bit of encoder $G^T$. This in turn implies
that the $i$th bit of the encoder $G$ is equivalent to the $(N - i + 1)$th bit of the encoder
$G^T$, where $1\leq i \leq N$.
Therefore, if we freeze the $i$th qubit in computational basis for the bit-flip channel, then we also freeze the $(N - i + 1)$th qubit in the Hadamard basis
for the phase-flip channel.

\begin{figure}[tbp]
        \begin{center}
                \includegraphics[width=\linewidth]{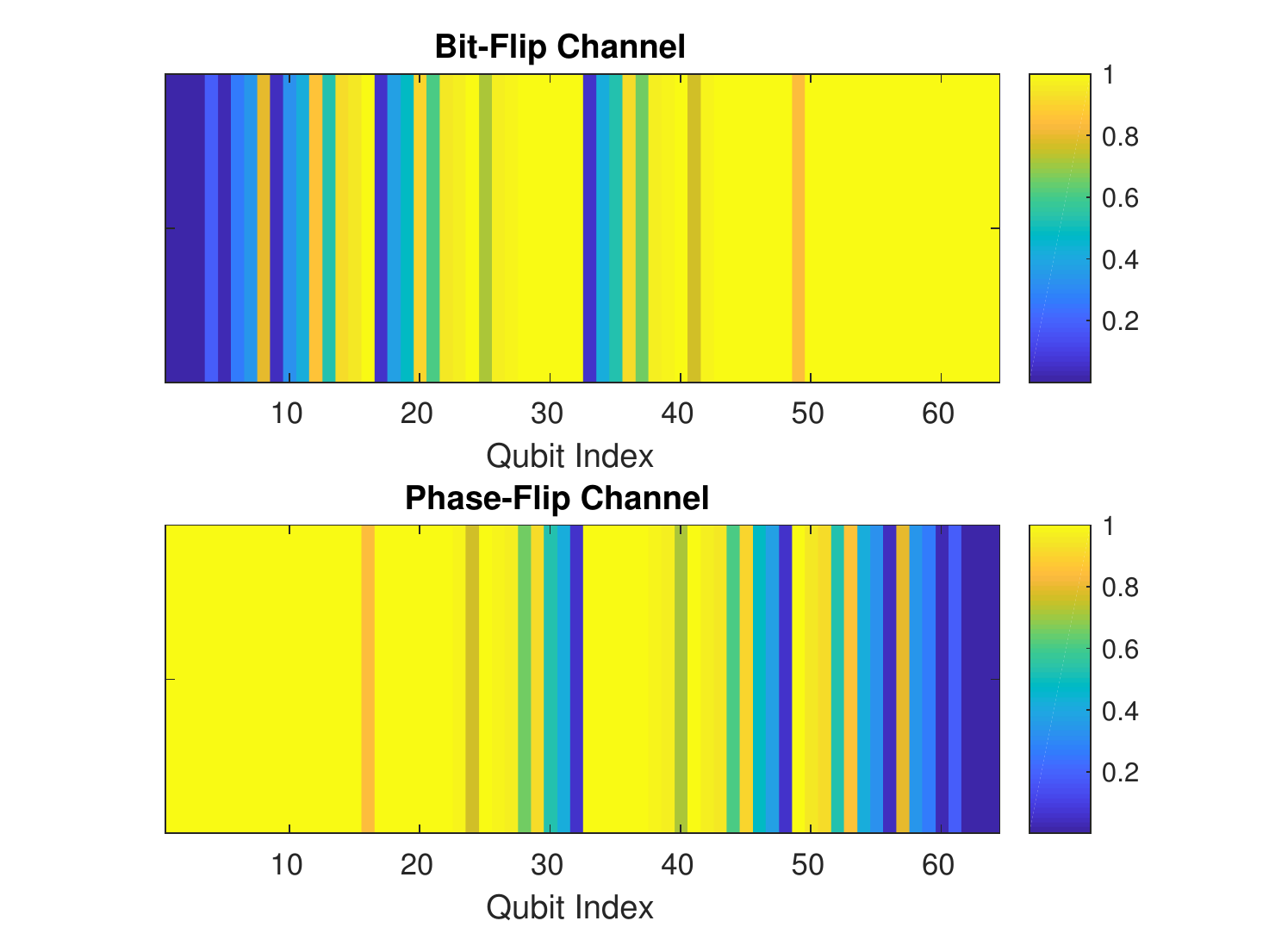}
                \caption{Mutual information intensity map for bit-flip and phase-flip channels at a depolarizing probability of $p = 0.06$, when a polar code of length $64$ is used.}
         \label{fig:intensity-64}
                \end{center}
\end{figure}

\fref{fig:intensity-64} shows the mutual information intensity maps for the bit-flip and the phase-flip channels at a 
depolarizing probability of $p = 0.06$, when a polar code of length $64$ is used. Explicitly, the colormap in \fref{fig:intensity-64} represents 
the mutual information and the qubit indices are sorted based on the mutual information of the bit-flip channel. As discussed in \fref{fig:encoder1}, the mutual information intensity map for the phase-flip channel is the same as that of the bit-flip channel, but it is flipped from right to left.

\subsection{Decoder} \label{sec:sec:Qdecoder}
\begin{figure*}[tbp]
        \begin{center}
                \includegraphics[width=0.9\linewidth]{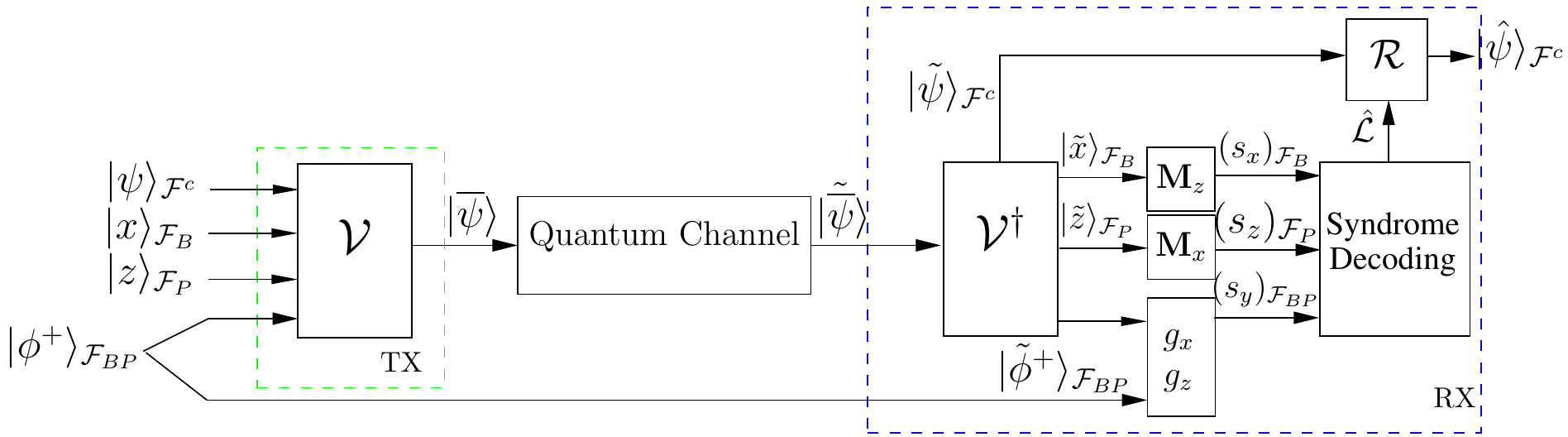}
                \caption{Schematic of a quantum communication system relying on a syndrome decoding based quantum polar code.}
         \label{fig:polar_system}
                \end{center}
\end{figure*}

\fref{fig:polar_system} shows the general schematic of a quantum communication system relying on a quantum polar code for 
protection against environmental decoherence.
Unlike the \ac{QSCD} based polar codes of~\cite{wilde_polar2013, 6472318, renes2012, wilde_polar2012, 6781023}, the 
system of \fref{fig:polar_system} invokes the syndrome-based classical decoding approach of~\cite{ferris2014tensor}. 
It is pertinent to mention here that quantum polar codes with syndrome-based classical decoding may not be capacity achieving for arbitrary
quantum channels, which require collective measurement over all channel uses. Nonetheless, we have adopted this approach because it is a more
direct application of the classical polar codes.
Recall from Section~\ref{sec:isomorphism} that a polar code polarizes the input quantum channels into four sets, which may  be denoted as:
\begin{itemize}
 \item $\mathcal{F}^c$: good bit-and-phase channels for transmitting arbitrary quantum information $\ket{\psi}$;
 \item $\mathcal{F}_P$: good bit-only channels for transmitting known Hadamard basis states $\ket{z}$, where $z \in \{+,-\}$;
 \item $\mathcal{F}_B$: good phase-only channels for transmitting known computational basis states $\ket{x}$, where $x \in \{0,1\}$;
 \item $\mathcal{F}_{BP}$: bad bit-and-phase channels for transmitting ebits.
\end{itemize}
Consequently, an $[N,k,c]$\footnote{We consistently use round brackets (.) for classical codes, while the
square brackets [.] are used for quantum codes.} polar code specified by the encoder $\mathcal{V}$ and the codespace $\mathcal{C}$ takes as input a $k$-qubit information word (logical qubits) $\ket{\psi}$ and maps it onto an $n$-qubit
codeword (physical qubits) with the aid of $(N-k-c)$ frozen qubits initialized to the known computational and Hadamard basis states and $c$ 
ebits, whose one qubit is pre-shared with the receiver. This may be mathematically expressed as:
\begin{equation}
 \mathcal{C} = \{\ket{\overline{\psi}} = \mathcal{V} \left( \ket{\psi}_{\mathcal{F}^c} \otimes \ket{x}_{\mathcal{F}_B} \otimes \ket{z}_{\mathcal{F}_P} \otimes \ket{\phi}^+_{\mathcal{F}_{BP}}\right) \}.
\end{equation}
The resulting encoded qubits $\ket{\overline{\psi}}$ are sent over a quantum depolarizing channel, which may inflict bit-flip, phase-flip or bit-and-phase-flip
errors, each with a probability of $p/3$. The received quantum information may be represented as:
\begin{equation}
 \ket{\tilde{\overline{\psi}}} = \mathcal{P} \ket{\overline{\psi}},
 \label{eq:ex-codewords}
\end{equation}
where $\mathcal{P}$ denotes the $n$-qubit Pauli error inflicted by the quantum depolarizing channel. At the receiver, the received information of \eqr{eq:ex-codewords}
is passed through the inverse encoder $\mathcal{V}^{\dagger}$, which yields:
\begin{align}
 \mathcal{V}^{\dagger}  \ket{\tilde{\overline{\psi}}} &= \mathcal{V}^{\dagger} \mathcal{P} \ket{\overline{\psi}} = \mathcal{V}^{\dagger} \mathcal{P} \mathcal{V} \left( \ket{\psi}_{\mathcal{F}^c} \otimes \ket{x}_{\mathcal{F}_B} \otimes \ket{z}_{\mathcal{F}_P} \otimes \ket{\phi}^+_{\mathcal{F}_{BP}}\right) \nonumber \\
 & = \mathcal{L} \ket{\psi}_{\mathcal{F}^c} \otimes \mathcal{S}_x \ket{x}_{\mathcal{F}_B} \otimes \mathcal{S}_z \ket{z}_{\mathcal{F}_P} \otimes \mathcal{S}_y \ket{\phi}^+_{\mathcal{F}_{BP}} \nonumber \\
 & = \ket{\tilde{\psi}}_{\mathcal{F}^c} \otimes \ket{\tilde{x}}_{\mathcal{F}_B} \otimes \ket{\tilde{z}}_{\mathcal{F}_P} \otimes \ket{\tilde{\phi}}^+_{\mathcal{F}_{BP}}.
\label{eq:rx1}
 \end{align}
where $\mathcal{V}^{\dagger} \mathcal{P} \mathcal{V} \equiv (\mathcal{L}\otimes \mathcal{S}_x \otimes \mathcal{S}_z \otimes \mathcal{S}_y)$ and $\mathcal{L}$
denotes the $k$-qubit error inflicted on the information word $\ket{\psi}$, while $\mathcal{S}_x$, $\mathcal{S}_z$ and $\mathcal{S}_y$ represent the errors
imposed on the frozen qubits, i.e. the computational, the Hadamard and the entangled qubits, respectively. As mentioned earlier, we assume that only the first ebit of Bell states
experiences noise, while the second ebit is pre-shared over a noiseless channel. Finally, the corrupted computational and Hadamard basis states,
i.e. $\ket{\hat{x}}$ and $\ket{\hat{z}}$, are measured in the computational and Hadamard basis, respectively, for the sake of determining the errors inflicted on these
qubits. Explicitly, the former reveals information about any bit-flips imposed on the frozen states $\ket{x}$, while the latter provides information about  
phase-flips inflicted on the frozen states $\ket{z}$. Let us denote the outcomes as $s_x$ and $s_z$, respectively, which are classical bits. 
Next, we have to determine the error imposed on the
ebits transmitted over the quantum channel. Recall that ebits are frozen in both the computational and the Hadamard basis, since the corresponding channels
are bad from the perspective of bit-flips as well as phase-flips. Consequently, we have to find the bit-and-phase-flip errors imposed on the ebits. This
may be achieved by using the Pauli operators $g_x = \mathbf{XX}$ and $g_z = \mathbf{ZZ}$, where the first Pauli operator acts on the ebit transmitted over the quantum
channel, while the second Pauli operator acts on the pre-shared noiseless ebit. Since ebits were created in the Bell states of \eqr{eq:ebit}, both  
Pauli operators $g_x$ and $g_z$ constitute the stabilizers. Furthermore, a bit-flip error on the first ebit will yield an eigenvalue of $-1$ for the stabilizer
$g_z$, while a phase-flip error will yield an eigenvalue of $-1$ for $g_x$. Hence, both bit-flip as well as phase-flip errors acting on the first ebit can be determined,
which may be denoted as $s_y$. The error patterns $s_x$, $s_z$ and $s_y$ acting on the frozen qubits are then fed to a syndrome-based polar decoder for the sake of estimating
the logical error $\hat{\mathcal{L}}$ experienced by the information word $\ket{\psi}$. Finally, a recovery operation $\mathcal{R}$ is applied to $\ket{\hat{\psi}}$
based on the estimated error pattern $\hat{\mathcal{L}}$; hence, recovering the transmitted information.

\begin{figure}[tbp]
        \begin{center}
                \includegraphics[width=\linewidth]{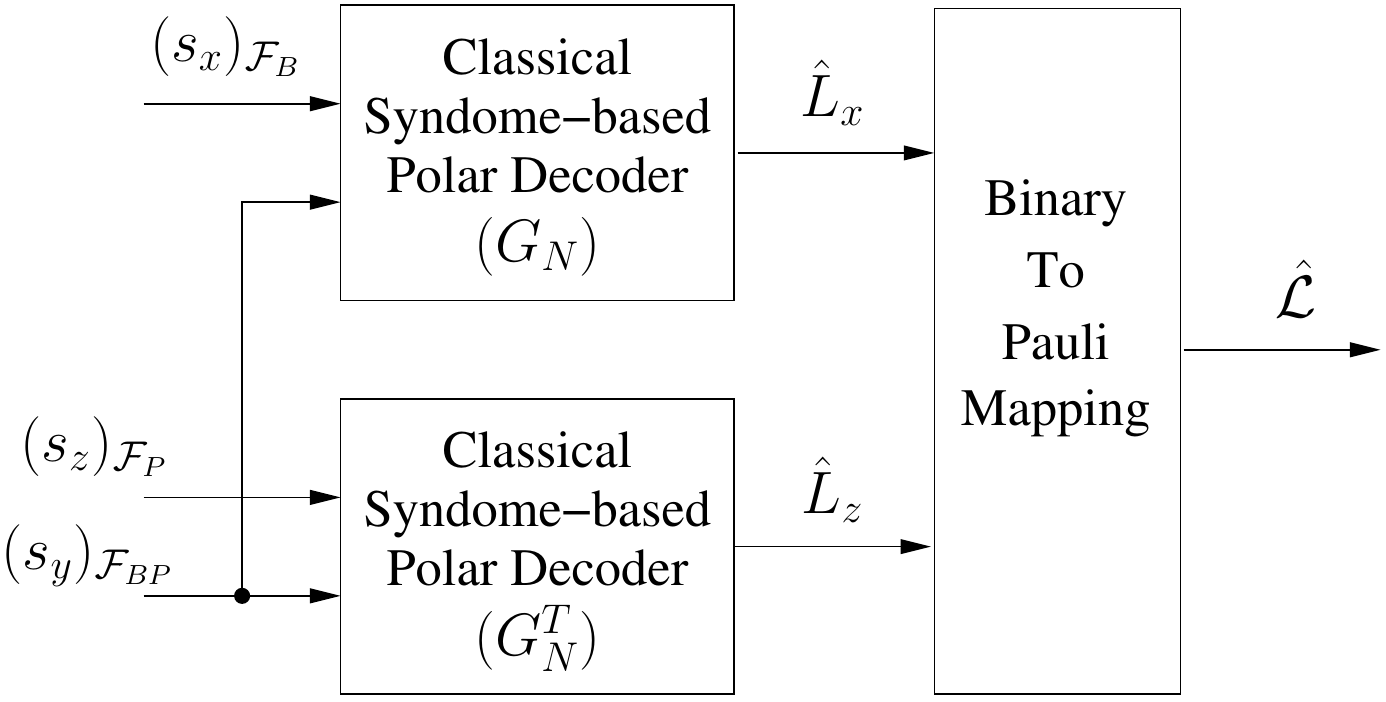}
                \caption{Schematic of quantum polar decoder.}
         \label{fig:Syndrome-dec}
                \end{center}
\end{figure}

Let us now elaborate on the syndrome decoding block of \fref{fig:polar_system}, which takes as input the errors experienced by the frozen qubits 
and estimates the error $\hat{\mathcal{L}}$ imposed on the logical qubits. The bit-flip and phase-flip errors constituting $\hat{\mathcal{L}}$ can be 
estimated independently as shown in \fref{fig:Syndrome-dec}, assuming the quantum depolarizing channel is approximated as two independent \acp{BSC}. 
More specifically,
the quantum polar decoder of \fref{fig:Syndrome-dec} consists of two independent classical syndrome-based polar decoders used for estimating
bit-flip and phase-flip errors, denoted as $\hat{L}_x$ and $\hat{L}_z$, respectively. 
The former decoder exploits the classical polar encoder $G_N$, while the latter relies on the encoder $G_N^T$. 
It is pertinent to mention here that the syndrome-based polar decoder already exist in the classical regime in the context of distributed source 
coding~\cite{korada2010polar2,onay2012polar}, where it is used to find the source information. However, when exploited from the perspective of 
channel coding, the syndrome-based polar decoder finds the most likely error on the information word. This is in contrast to the 
the conventional polar decoder, referred to as the
codeword-based polar decoder, which aims to find the most likely information word.
Such a syndrome-based polar decoder is obtained from the codeword-based polar decoder
of \fref{example.fig} by setting the values of the frozen bits according 
to the observed syndrome values, rather than the actual values of the frozen bits. Similarly, the channel \acp{LLR}
are replaced by the channel error \acp{LLR}, i.e. the probability of encountering channel error on the coded sequence. In case of depolarizing channel,
or equivalently a \ac{BSC} channel, the channel error \acp{LLR} are set according to the marginalized depolarizing probability $2p/3$.
The rest of the decoding process is same as that of the codeword-based decoding.
Finally, the estimated error patterns  $\hat{L}_x$ and $\hat{L}_z$ are mapped onto the corresponding Pauli operators using the binary-to-Pauli
mapping of \tref{tab:b2p}; hence, yielding the Pauli error $\hat{\mathcal{L}}$.
\begin{table}[tbp]
 \centering
 \begin{center}
\begin{tabular}[tbp]{|l|l|l|}
\hline
$(\hat{L}_z, \hat{L}_x)$& $\hat{\mathcal{L}}$& Estimated Error\\  \hline
$(0 , 0)$ & $\mathbf{I}$& No Error \\
$(0 , 1)$ & $\mathbf{X}$& Bit-flip Error\\
$(1 , 0)$ & $\mathbf{Z}$& Phase-flip Error\\
$(1 , 1)$ & $\mathbf{Y}$&Bit-and-phase-flip Error \\ \hline
\end{tabular}
 \end{center}
 \caption{Binary to Pauli mapping for estimating the Pauli error.}
\label{tab:b2p}
\end{table}
\section{Conclusions \& Future Directions} \label{sec:conclusion}
To conclude, Arikan's polarization phenomenon of Section~\ref{sec:ch-polarization} has paved the way to provably achieve the Shannon's capacity
at low encoding and decoding complexity. In particular, the polar code has a very structured encoder and decoder, as discussed in 
Section~\ref{sec:encoder} and Section~\ref{sec:decoder}, respectively; hence,
dispelling the notion that randomized coding structures are more apt for achieving the Shannon's capacity.
Furthermore, polar codes intrinsically support rate-adaptation, since the coding rate can be varied by only changing the number of
frozen bits, while retaining the same encoder and decoder. 

The attractive features of polar codes have stimulated a wave of interest in the research community as well as in the industry. Consequently, in just within a decade of its inception,
polar codes have already replaced their decades-old contemporaries in the $5$G~\ac{NR} for the control channels of the \ac{eMBB} use-case. 
However, the high decoding latency of polar codes is a major concern restricting the widespread application of polar codes, as discussed
in Section~\ref{sec:decoder}, where we have reviewed the major polar decoders with slow-paced tutorial examples. More specifically,
recall from \tref{tab:DecComp} that the different polar decoding schemes, namely the \ac{SC} of Section~\ref{sec:sec:SC},
the \ac{SSC} of Section~\ref{sec:sec:SSC}, the \ac{SCL} of Section~\ref{sec:sec:SCL}, the \ac{SCS} of Section~\ref{sec:sec:SCS},
the \ac{BP} of Section~\ref{sec:sec:BP} and the \ac{SCAN} of Section~\ref{sec:sec:SCAN}, entail a compromise between the imposed complexity,
space requirements, performance and the tendency for fully-parallel implementation. 

Another concern related to polar codes is their channel-specific nature, which necessitates the optimization of polar codes
for the channel under consideration. Explicitly, the optimization of polar codes entails selecting the right indices for the frozen 
bit-channels, which constitute the set $\mathcal{F}$. In Section~\ref{sec:polarcodedesign}, we detail the polar code design principles with 
particular emphasis on the \ac{BEC}-approximation, Monte-Carlo and \ac{GA}-\ac{DE} methods for estimating the reliabilities of the bit-channels.
Furthermore, we construct frozen bit-channel sequences for the \ac{AWGN} channel using these three methods and compare their performance. It is demonstrated 
that the sequences constructed using the Monte-Carlo and \ac{GA}-\ac{DE} are equally good, while those designed using the \ac{BEC}-approximation
have inferior performance. Perhaps, this is because a \ac{BEC}  does not truly depict the transmission over an \ac{AWGN} channel. 
We further demonstrate that, while the design \ac{SNR} is an important design parameter, limited variations in the design \ac{SNR} only slightly
affects the \ac{BLER} performance.

Polar codes have also been warmly welcomed by the quantum coding community, since the notion of channel polarization readily extends to the
quantum channels. Interestingly, Arikan's polar encoder is capable of concurrently polarizing the bit-flip and phase-flip quantum channels,
when the \ac{XOR} gate are replaced by the quantum \ac{CNOT} gates, as exemplified in Section~\ref{sec:isomorphism}. However,
the \ac{QSCD} is not a direct extension of the classical \ac{SC} decoder. The relevant contributions in this context are briefly summarized in
Section~\ref{sec:Qpolarcode}. Nonetheless, there exists quantum polar codes for quantum Pauli channels, which are more directly linked to
the classical polar codes, since they invoke the classical syndrome based polar decoders. The encoder and decoder of this class
of quantum polar codes are reviewed in Section~\ref{sec:sec:Qenc} and
Section~\ref{sec:sec:Qdecoder}, respectively.

As surveyed in this paper, intensive research efforts have been invested in the polar coding paradigm over the last decade for the sake of bringing it at par
with its contemporaries, namely the turbo and \ac{LDPC} codes. Nonetheless, there is a great potential
to explore this coding paradigm further, since it is still in its infancy. Some of the potential research directions are discussed below:
\begin{enumerate}
\item \textbf{Non-Arikan Polar Codes:} As discussed in Section~\ref{sec:sec:NonArikan}, efforts have been made to design multi-dimensional as well
as non-binary kernels for polar codes. However, these kernels have not been able to replace the Arikan's kernel owing to the associated encoding
and decoding complexities. This is still an open research area. In particular, the non-binary kernels are important from the perspective of
source coding.
 \item \textbf{Low-Latency, Power-Resource-Efficient \& Flexible Polar Decoders:} Polar decoder continues to be a major concern; hence preventing the adoption of
 polar codes for the data channel of $5$G \ac{NR}. This includes a range of open research problems both from the algorithmic perspective as
 well as from the implementation. In particular, the existing polar decoders
 incur a high latency, which is a primary concern for \ac{URLLC} applications. Recall from \tref{tab:DecComp} that only the \ac{BP} decoder supports fully-parallel implementation. However,
 it pays the price in terms of the performance, and the computational and space complexity. Efforts have been made to partially parallelize 
 the other polar decoders, for example in~\cite{6327689,6737143, 6876199,6803952,7756324}, 
 but their latency is still higher than that of the turbo and \ac{LDPC} codes, which lend themselves to a 
 fully-parallel implementation~\cite{rob_FPTD,8639065,chandrasetty2011fpga}. Furthermore, there is a need to explore more practical polar decoders as well as to  
 develop further the hardware implementations of the existing polar decoders to bring them
 at par with the turbo and \ac{LDPC} codes, particularly from the perspective of latency (or throughput), power efficiency, resource
 efficiency as well as flexibility. From the algorithmic perspective, soft-in soft-out polar decoding is also a promising research
 avenue, since the existing soft \ac{SCL} decoder is only applicable to systematic polar codes, while the performance of the \ac{SCAN}
 decoder is not at par with that of the \ac{SCL} decoder. So, a more general soft \ac{SCL} decoder is required for concatenated frameworks,
 for example concatenated coding schemes or joint detection and decoding schemes, invoking iterative decoding.
 \item \textbf{Stochastic Polar Decoders:} Stochastic \ac{LDPC} and turbo decodes are known to provide attractive benefits in terms of 
 fault-tolerance (to timing errors) as well as latency (or equivalently throughout)~\cite{zuo2016improving, perez2016stochastic}.
 However, stochastic polar decoders have not attractive much attention, except for 
 in~\cite{xu2014successive, yuan2015successive, liang2015efficient, yuan2016belief, xu2016stochastic, han2018bit}.
 Since the latency of polar decoders is already a prime concern, in this context it is worth investigating the area of stochastic
 polar decoders.
 \item \textbf{Universal Polar Codes:} Perhaps another limitation of polar codes is their channel-specific nature, which necessitates
 code optimization for the required channel characteristics. The impact of this is not very significant in practical scenarios. Hence, despite
 the channel-specific nature of polar codes, they have been adopted for the control channels of the $5$G \ac{NR}. However, this still remains
 a concern, especially from the theoretical perspective. More specifically, polar codes with \ac{SC} decoding
 incur a capacity loss over compound channels\footnote{Compound channels model the transmission scenario where the exact channel is unknown
 and only a set of channels to which the actual channel belongs is known.}, as demonstrated in~\cite{hassani2009compound}. This capacity loss
 is due to the sub-optimal \ac{SC} decoding and may be alleviated by invoking the optimal \ac{ML} decoding, as theoretically shown 
 in~\cite{sasoglu2011polar}. Explicitly, it was demonstrated in~\cite{sasoglu2011polar} that polar code optimized for a \ac{BSC}
 is universal under \ac{ML} decoding and hence is optimal for any channel of equivalent capacity. Unfortunately, \ac{ML} decoding of polar codes
 is not feasible. For the sake of overcoming this issue, universal polar codes
 were conceived in~\cite{hassani2014universal, csacsouglu2016universal}, which essentially retain the low encoding and decoding complexity (per bit)
 of classic polar codes. However, this universality is achieved by increasing the length of polar codes, which in turn imply a higher overall
 complexity and longer latencies (or equivalently lower throughputs). This is a promising research avenue and must be explored further
 to find improved approaches for achieving universality for binary and non-binary classical polar codes as well as for quantum polar codes.
 \end{enumerate}



%

\bibliographystyle{ieeetr}

\end{document}